\documentclass[amsmath,amssymb,prb,superscriptaddress,preprint,showpacs]{revtex4-1}
\usepackage{graphicx}
\usepackage{dcolumn}
\usepackage[colorlinks=true,linkcolor=blue,citecolor=blue,urlcolor=blue]{hyperref}
\usepackage{stmaryrd}
\usepackage{esvect}
\usepackage{multirow}

\usepackage{amsmath}
\usepackage{graphicx}
\usepackage{xfrac}
\usepackage{multirow}
\usepackage{subfigure}
\usepackage{mathtools}

\usepackage{lineno}
\usepackage{bm}
\usepackage{color}

\normalsize

\begin{document}

\title{ Theoretical investigation of antiferromagnetic skyrmions in a triangular monolayer }
\author{Zhaosen Liu\footnote{Email: liuzhsnj@yahoo.com}}
\affiliation{College of Physics and Electronic Engineering, Hengyang
	Normal  University, Henghua Road 16,  Hengyang  421002, China.}
\author{Manuel dos Santos Dias}
\author{Samir Lounis}
\affiliation{Peter Gr\"unberg Institut and Institute for Advanced Simulation, Forschungszentrum J\"ulich \& JARA, 52425 J\"ulich, Germany.}
\vspace{0.2cm}

\begin{abstract}
The chiral spin textures of a two-dimensional (2D) triangular system, where both  antiferromagnetic (AF) Heisenberg exchange and chiral  Dzyaloshinsky-Moriya  interactions co-exist, are investigated   numerically  with   an  optimized   quantum Monte Carlo method  based on mean-field theory.  We find  that: helical, skyrmionic and vortical AF crystals can be formed when  an external magnetic field is applied perpendicular to the 2D monolayer; the sizes of these  skyrmions and vortices   change abruptly  at several  critical points of the external magnetic field; each of these AF   crystals can be decomposed into three periodical ferromagnetic (FM) sublattices.  The quantum ingredient implemented  into the   theoeretical framework helps to track the existence of AF skyrmion lattices  down to low temperatures. 
\end{abstract}


\maketitle

\section{Introduction}
 Skyrmions in ferromagnetic materials have been intensely studied   in recent years. \cite{Bogdanov94,Bogdanov99,Bogdanov01,Pappas,Yu10,Banerjee,Yi,Buhrandt,Huang,Romming15}
Their small size,  in a range  of 1 nm $\sim$ 100 nm, and the very weak electronic current, around 10$^6$ Am$^{-2}$,  required to drive them to motion, \cite{Iwasaki}   make them ideal candidates in future data storage  and  racetrack memory  devices. Unfortunately,  the Magnus force,  \cite{Tomasellor14,Everschor11,Everschor17}  that acts  transversely   on the   ferromagnetic (FM) skyrmions  by the applied electric current,  eventually pushes them  off  the nano-track edge, which greatly hinders  their applications in  the aforementioned memory devices.\cite{Fert13}

In contrast, an  antiferromagnetic (AF) skyrmion  in a square lattice  can be  decomposed into two   almost identical FM  sublattices.  Thus, the Magnus forces acting on the  two FM sublattices  can be completely cancelled, \cite{Barker16}    so that  the  AF skyrmion can  move much faster straightly along the direction of the applied electric  current. \cite{Barker16,Velkov16}

AF  skyrmions have been investigated  theoretically in 2D square and triangular   antiferromagnets.
\cite{Okubo,Keesman,Rosales}   In their Monte Carlo study,  Keesman and his colleagues discovered that  an isolated skyrmion  can be stabilized by a strong external magnetic field   at nonzero temperatures, but only in a very tiny finite-sized $8 \times 8$ square antiferromagnet. \cite{Keesman}
For  triangular lattices, Rosales et al.~\cite{Rosales}observed  with classical  Monte Carlo (CMC) simulations that when AF Heisenberg exchange (HE) and  Dzyaloshinsky-Moriya (DM) interactions co-exist, AF skyrmionic lattices  can be induced   by an  external  magnetic field  applied normal to the lattice plane  in  a  very low temperature range. However, they found no evidence of AF skymrion lattice (SL) formation  in the AF chiral square lattice.  The latter aspect was addressed by one of us\cite{LiuIan19} utilizing a quantum simulative method, namely the SCA approach, where a  self-consistent algorithm is used in the frame of quantum theory.  \cite{liujpcm,liupssb,Liu14phye,LiuIan16,LiuIan16cpl,LiuIan17,LiuIan18, LiuIan19, LiuIan19-2} It was found  that  both  Bloch  and N\'eel types AF SLs   could be induced  in a 2D   lattice of square crystal structure by a very strong external magnetic field exerted normal to the monolayer plane.\cite{LiuIan19} We expect that quantum effects, missing in CMC simulations, to be important in AF materials, which motivates the current study by reinvestigating  AF SLs formed in the 2D triangular lattices with a method built upon quantum theory.

 We utilize here an optimized  quantum  Monte Carlo (OQMC) method  based on a mean-field approach. The  Metropolis algorithm is employed, and a simple trick is adopted  to improve the computational efficiency. Astonishingly, by using   this   method we are able to plot, for examples, the well  periodic and symmetric AF vortical lattices (VLs), AF skyrmionic lattices (SLs) of both Bloch and N\'eel types, which can appear  at any temperatures in 2D AF chiral magnets, just with the computational results obtained from the last iteration. In contrast, if CMC method is applied to calculate a spin configuration at a finite temperature,  averages   must be made  over thousands of last iterations.

Using the OQMC method, we simulate  and investigate in detail   the chiral spin textures of a   2D triangular system  where   the AF  HE and chiral  DM interactions are both present. Consequently, we find that, within   an external magnetic field   applied perpendicular to the 2D monolayer, AF HLs (helical lattices),  SLs  and VLs   can be induced at much higher temperatures than   predicted by previous authors   with CMC simulations; \cite{Rosales} the  SL  states prevail   in a broad $T-H$ phase area; the sizes of the   skyrmions and vortices formed in the spin crystals or lattices  changes abruptly   when modifying the external magnetic field; each of these AF   SL and VL  can be decomposed into three FM sublattices that are in fact FM SLs or FM VL, respectively. To check the accuracy of the results obtained with the OQMC method, we carry out simulations using the SCA approach for a particular case  with a fixed field strength,   and find the resulting AF skyrmionic lattices from both theoretical approaches to be identical.

\section{The Quantum Model and Computational Algorithm}

The  two-dimensional  triangular  spin-lattice is considered to be  in the $xy$-plane,  and    its corresponding  Hamiltonian can be expressed as
{\small
\begin{eqnarray}
{\cal H} = &  -\frac{1}{2}\sum_{i,j}\left[{\cal J}_{ij}{ \vec{S}_i
\cdot }\vec{S}_j
- {\vec{D}_{ij}\cdot(\vec{S}_i\times}\vec{S}_j)\right] 
-K_A\sum_i\left(\vec{S}_i\cdot \hat{n}  \right)^2 - {\vec
H }\cdot\sum_i{\vec S}_i\;.
 \label{hamil}
 \end{eqnarray}
Here, the  first two  terms represent the HE and DM interactions  between a pair of spins at the $i$-th and $j-$th lattice sites, respectively.  In practice, we restrict the interactions to the nearest neighboring spins. The third term stands for the uniaxial magnetic  anisotropy assumed to be normal to the 2D monolayer, which is usually termed as the perpendicular magnetic anisotropy (PMA), and the last one denotes the Zeeman energy of the magnetic system placed in an external magnetic field.  The  strengths of the two-spin HE, DM  and     single-spin PMA interactions are   ${\cal J}_{ij}$,  $ \vec{D}_{ij}$,  $K_A$, respectively.  If  ${\vec D}_{ij}$  is along the ${\vec r}_{ij}$ direction  (the vector ${\vec r}_{ij}$ connecting both spin sites), the  induced skyrmion is  of Bloch-type; whereas if ${\vec D}_{ij}$ lies  in the film plane and is perpendicular to ${\vec r}_{ij}$, the   skyrmion is  of  N\'eel-type.

In light of quantum theory, the spins appearing  in Eq.(\ref{hamil}) are quantum operators,
\cite{liujpcm,liupssb,Liu14phye,LiuIan16,LiuIan16cpl,LiuIan17,LiuIan18, LiuIan19, LiuIan19-2} rather than classical vectors.
When $S$ = 1   as it is assumed in the present work, the matrices  of the three spin components are:

\begin{eqnarray}
S_x = 
\frac{1}{2}\left(
\begin{array}{ccc}
0 &  \sqrt{2}  & 0\\
\sqrt{2} & 0 & \sqrt{2}  \\
0 & \sqrt{2}  & 0\\
\end{array}
\right)\;, & S_y =\frac{1}{2i}\left(
\begin{array}{ccc}
0 &  \sqrt{2}  & 0\\
-\sqrt{2} & 0 & -\sqrt{2}  \\
0 & \sqrt{2}  & 0\\
\end{array}
\right)  \;,
& S_z =    \left(
\begin{array}{ccc}
1 &  0  & 0\\
0 & 0 & 0 \\
0 & 0  & -1\\
\end{array}
\right)\;,
\end{eqnarray}
respectively, in the Heisenberg representation.   The thermal expectation value of a  physical observable $A$  at temperature $T$ can be evaluated with
\begin{equation}
\langle A\rangle = \frac{{\rm Tr}\left[\hat{A}\exp(\beta{\cal H}_i)\right]}{{\rm Tr}\left[\exp(\beta{\cal H}_i)\right]}\;,
\label{avq}
\end{equation}
where $\hat{A}$ is the  operator of the observable $A$, and $\beta = -1/k_BT$.

 When the quantum Monte Carlo (QMC) method was proposed,    Metropolis algorithm was also conventionally  employed. \cite{Liu14phye,LiuIan18} That is, in every simulation step, a spin ${\vec S_i}$ is randomly selected from the considered magnetic system, then rotated randomly within a narrow spatial cone. Afterwards, a   random number $r$ is generated to compare with $p = \exp(-\Delta E_i/k_BT)$, where $\Delta E_i$  is the energy change  caused by the rotation. If $r \leq p$, the operation is accepted, otherwise discarded.  All   simulations  are  started from  random magnetic configurations  above  the magnetic transition temperatures, then carried out stepwise down   to very low temperatures with a reducing step  $\Delta T < 0$.

 A  simple   trick has been   taken to  optimize the QMC method: Once the rotated state of ${\vec S_i}$ is accepted,  the states of its   neighboring spins have to be updated immediately, since their states are also affected by the operation. Otherwise, errors in the total energy of the magnetic system will be accumulated,  so that  the computational program may converge to an incorrect spin configuration.

 The  above technique taken to optimize QMC  method  seems very easy and simple. However,  it   has been proved to be very effective, leading to quick convergence of the simulations. More specifically,  OMQC method makes it possible for  the computational results obtained  in the   final iteration after convergency to accurately describe the complicated and detailed spin textures of   considered magnetic system  at all  temperatures.   For instance,  with such  results we are able  to depict  well symmetric and periodic   FM and AF SkLs of the Bloch- and N\'eel-types in a broad  $T-H$ phase area in 2D magnetic systems. 

\section{Computational Results}

Each lattice site of the AF monolayer  is considered to be occupied by an $S$ = 1 spin, both  HE and DM interactions are limited to the nearest neighboring spins,  and their strengths   assigned to ${\cal J}_{ij} = {\cal J}$ = -1, $D_{ij} = D$ = 1, respectively. That is, all parameters and other physical quantities  are scaled with $|{\cal J}|$. Since these two interactions   are  comparable in magnitude,  the ground state  of the magnetic system is neither simply antiferromagnetic  nor vortical. To generate isolated skyrmions or SLs, an external magnetic field ${\vec H}$ that is   normal to the   monolayer  has to be considered. \cite{Keesman,Rosales}   A    $30\times 30$ square lattice  is chosen in most our   simulations,   and periodical boundary conditions are imposed to mimic the infinite size of the 2D monolayer and to facilitate  formations  of chiral spin lattices.

 \subsection{ Phase Diagram}
  From our simulated results, we find  that well symmetric and periodical AF SLs   can be generated  when  the external magnetic field falls in  a wide range $1 \leq H \leq 7.4$, as shown in the phase diagram plotted in Figure 1(a).   To get this diagram,  the applied magnetic field strength is varied from 0.4 to 7.8;  and  in an external  field of fixed strength, the simulation is initiated from a random spin configuration in  the paramagnetic phase, then temperature is lowered  with a reducing step $\Delta T$ = -0.1 or -0.05.
\begin{figure*}[htb]
\centerline{
 \includegraphics[width=0.45\textwidth,height=6cm,clip=]
 {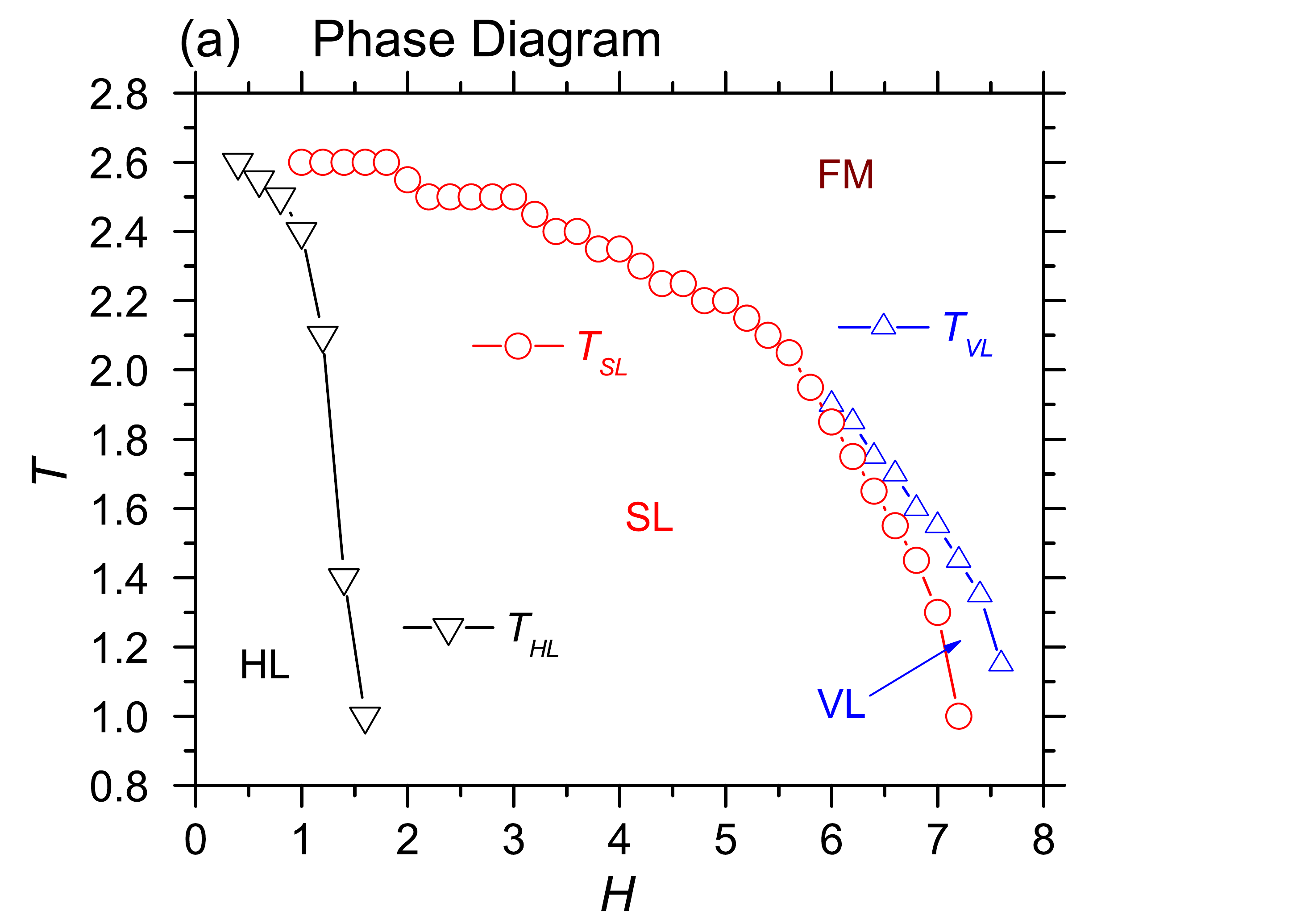}
 \includegraphics[width=0.45\textwidth,height=6cm,clip=]
 {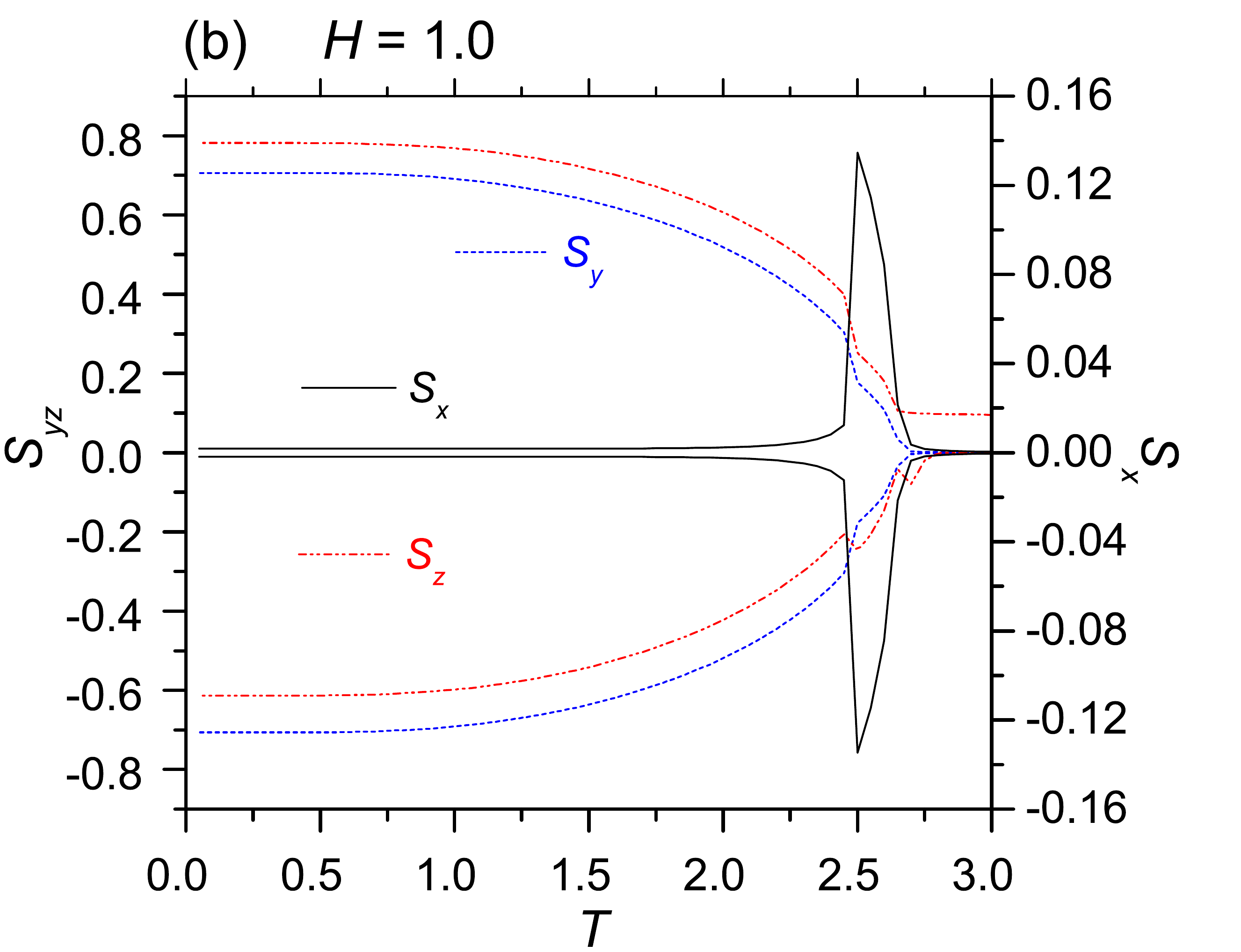}
 }
 \centerline{
 \includegraphics[width=0.45\textwidth,height=6cm,clip=]
 {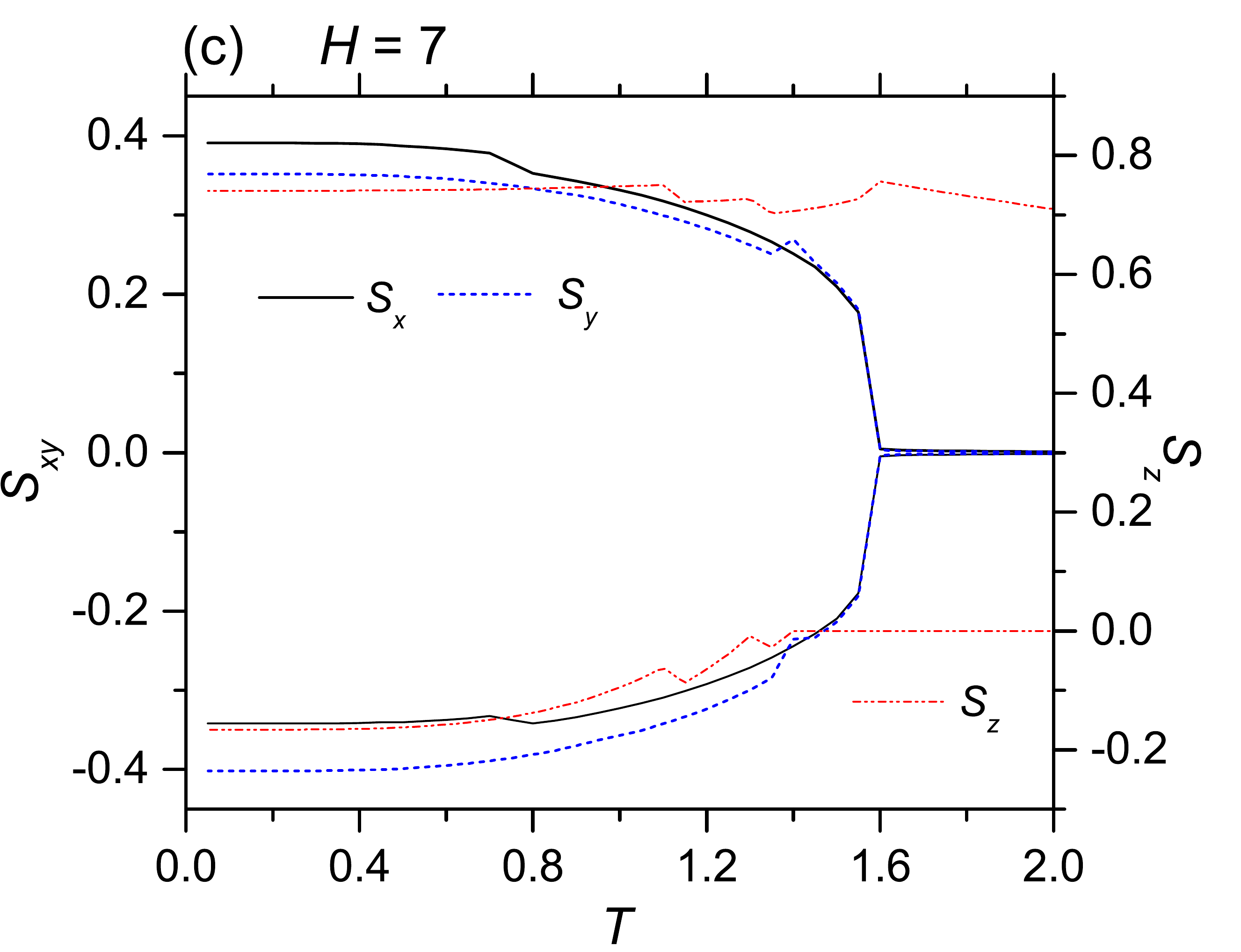}
 \includegraphics[width=0.45\textwidth,height=6cm,clip=]
 {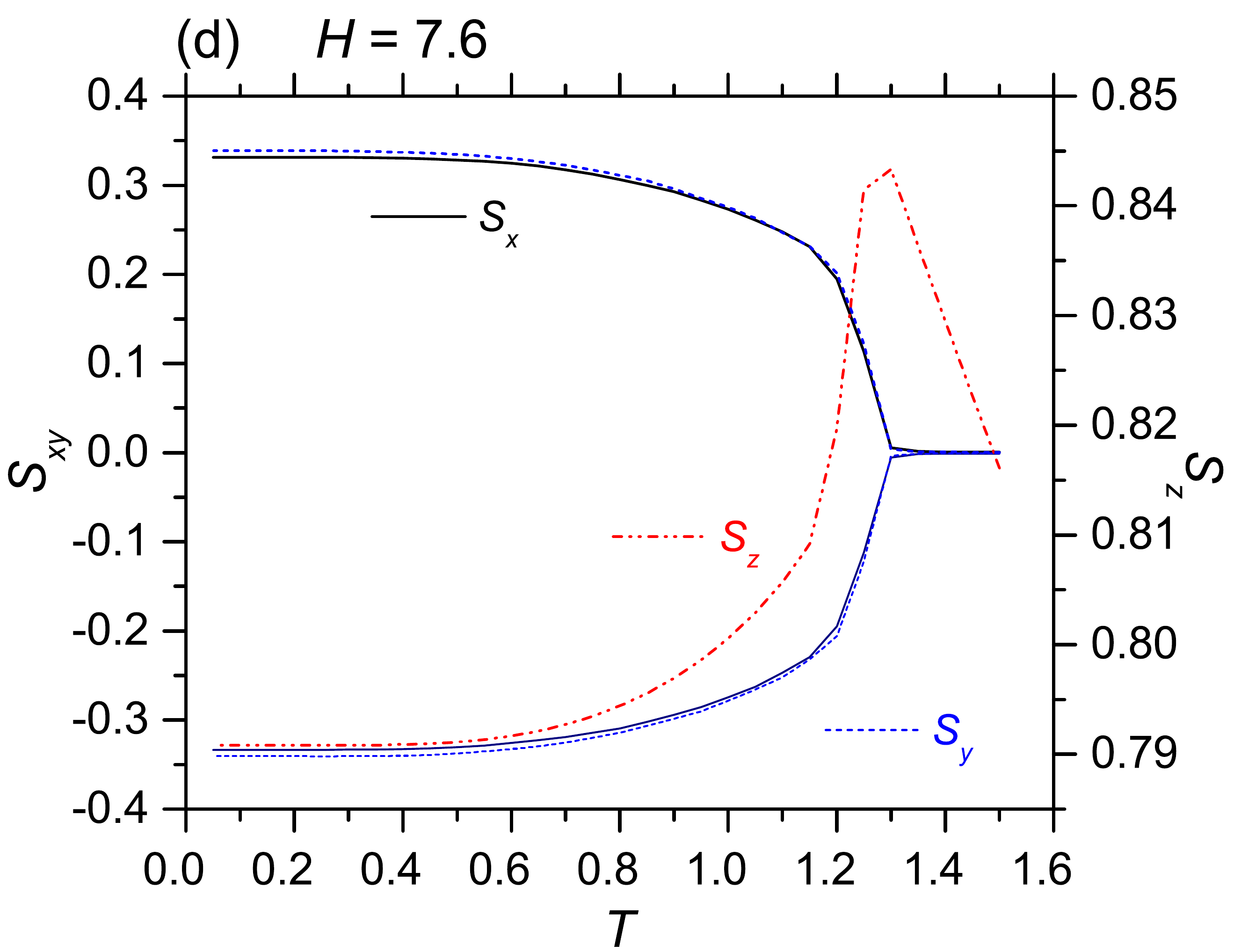}}
\caption{Calculated   phase diagram  (a),  $S_{x}$, $S_y$ and $S_z$ curves for  $H$ =  1.0 scaled with ${\cal J}$ ($S_{x}$ is displayed in a different scale)  (b), $H$ =  7  (c),  and  $H$ =  7.6  (d), versus  temperature.}		
\end{figure*}

In Figure 1(a),  $T_{HL}$, $T_{SL}$, $T_{SV}$   represent    the  transition temperatures when the magnetic  system condenses to  the AF helical,   skyrmionic  and vortical  lattices, respectively. Therefore, the area below  the curve $T_{\alpha}$ and above the curve $T_{\beta}$ is  in the $\alpha$ phase; whereas the region below the lowest curve $T_{\gamma}$ is in the $\gamma$ phase. For example,  as $  6.0 \leq H \leq 7.4$, the area between the $T_{VL}$ and $T_{SL}$ curves is the VL  phase, whereas the region below the $T_{SL}$ curve is  in the  SL phase;  while in the field range $  1.0 \leq H \leq 1.6$,   the area between the $T_{SL}$ and $T_{HL}$ curves is the SL  phase, whereas the region below the $T_{HL}$ curve is   the  HL phase.  In  a wide field range 1.8 $\leq H \leq$ 5.8, only    AF SLs appear  below the $T_{SL}$ curve; while  for   0.4 $\leq H \leq$ 0.8,  only HL states prevail below the $T_{HL}$ curve.
As expected,  at high temperatures  the system is completely polarized by the applied magnetic field and becomes ferromagnetic.

Figures 1(b,c,d) display  the   $\langle S_x\rangle$,  $\langle S_y\rangle$ and $\langle S_z\rangle$ curves  of the  AF   system  when it is placed in external magnetic fields of different strengths. In the first case  shown in Figure 1(b),  $H$ = 1.0 which is in the unit of ${\cal J}$.   Please note here that $\langle S_x\rangle$ and  $\langle S_y\rangle$ are displayed in different scales.  Above $T$ = 2.8,  the spins are completely polarized to the $z$-direction by the external magnetic field.  When temperature drops to $ T $ = 2.6 and 2.5,  the positive and negative branches of $\langle S_x\rangle$ and  $\langle S_y\rangle$, though very weak, are comparable in magnitude, thus AF SLs can be generated. However, as temperature drops further, the magnitudes of $\langle S_x\rangle$ and  $\langle S_y\rangle$  differ  considerably,  while those of $\langle S_y\rangle$ and  $\langle S_z\rangle$ are comparable, so that vertical AF HL state  dominates  the whole low temperature range.

 In the second case shown in Figure 1(c), $H$ =7.   As $T  >$ 1.6, only $\langle S_z\rangle$ is sizable. That is, the whole system is almost completely polarized by this strong external magnetic field. Below $T  =$ 1.6, the positive and negative branches of both $\langle S_x\rangle$ and  $\langle S_y\rangle$ grow gradually with decreasing  temperature, and later become comparable in magnitude, but they are  shifted relatively with each other in the  vertical direction, so that  VLs  can be formed  within temperature rangle 1.55 $\geq T >$ 1.3,  whereas SLs  are  induced below $T_{SL}$ = 1.3.

  It is easy to understand that, when   $H$  = 7.6 as shown in Figure 1(d), only VL can be observed below $T_{VL}$ =1.15, since  all spins have been rotated by the external magnetic field toward the $z$-direction.    Especially,  when $H$  is increased to  7.6, the size of each vortex suddenly grows, and we have to use a 60 $\times$ 60 lattice to generate these VL structures.

\subsection{ FM Sublattices of Helical, Skyrmionic and Vortical Lattices}

The co-presence of AF HE and   DM interactions gives rise to very complicated spin configurations, so that the AF spin lattices,   though they are well  periodical and symmetric, look quite puzzling if the $xy$ projection and $z$-contour of such a spin crystal are plotted together. Fortunately,  each of these  spin  textures can be decomposed into three FM sublattices.
\begin{figure*}[htb]
\centerline{
 \includegraphics[width=0.45\textwidth,height=6cm,clip=]
 {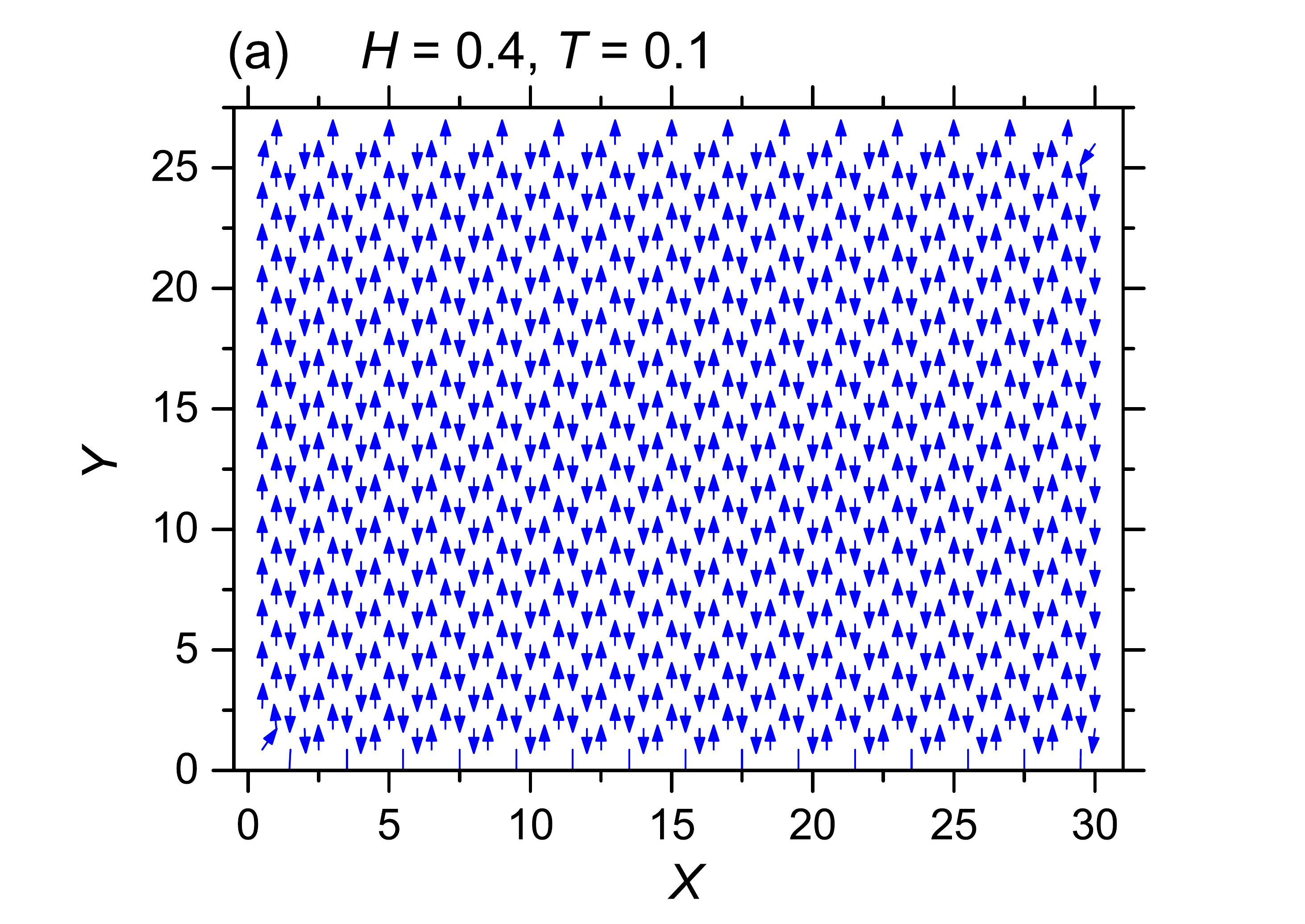}
 \includegraphics[width=0.45\textwidth,height=6cm,clip=]
 {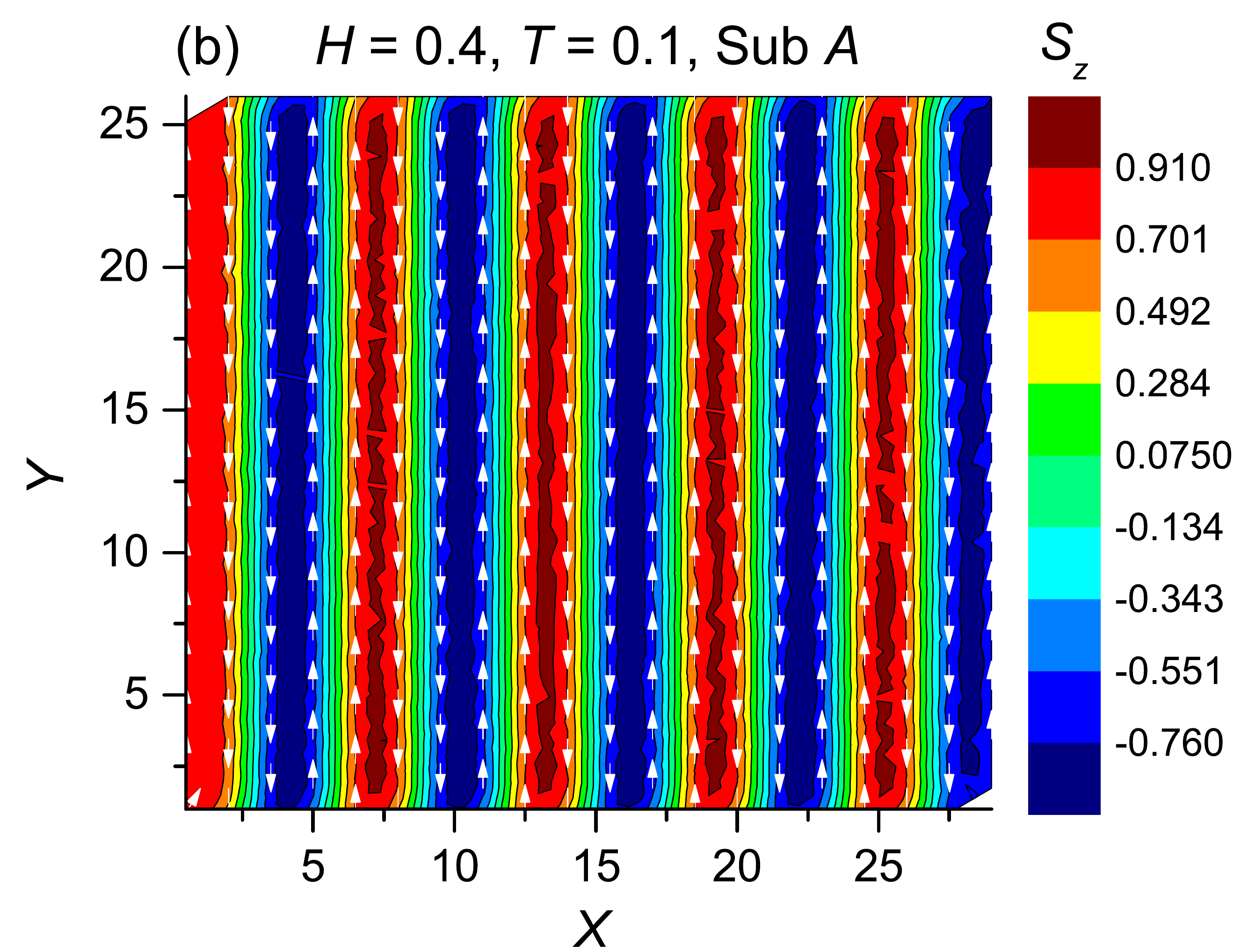}
} \centerline{
 \includegraphics[width=0.45\textwidth,height=6cm,clip=]
 {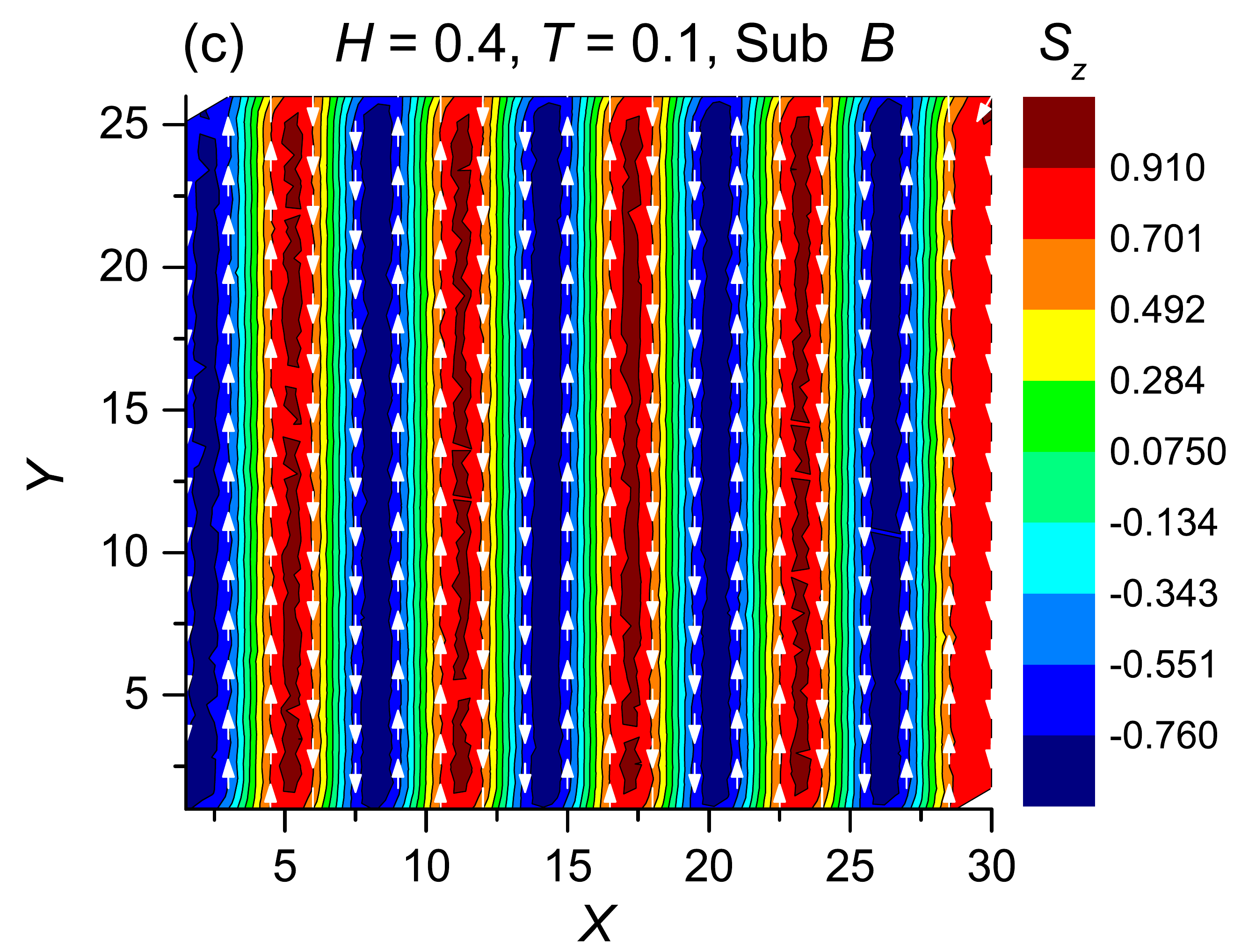}
 \includegraphics[width=0.45\textwidth,height=6cm,clip=]
 {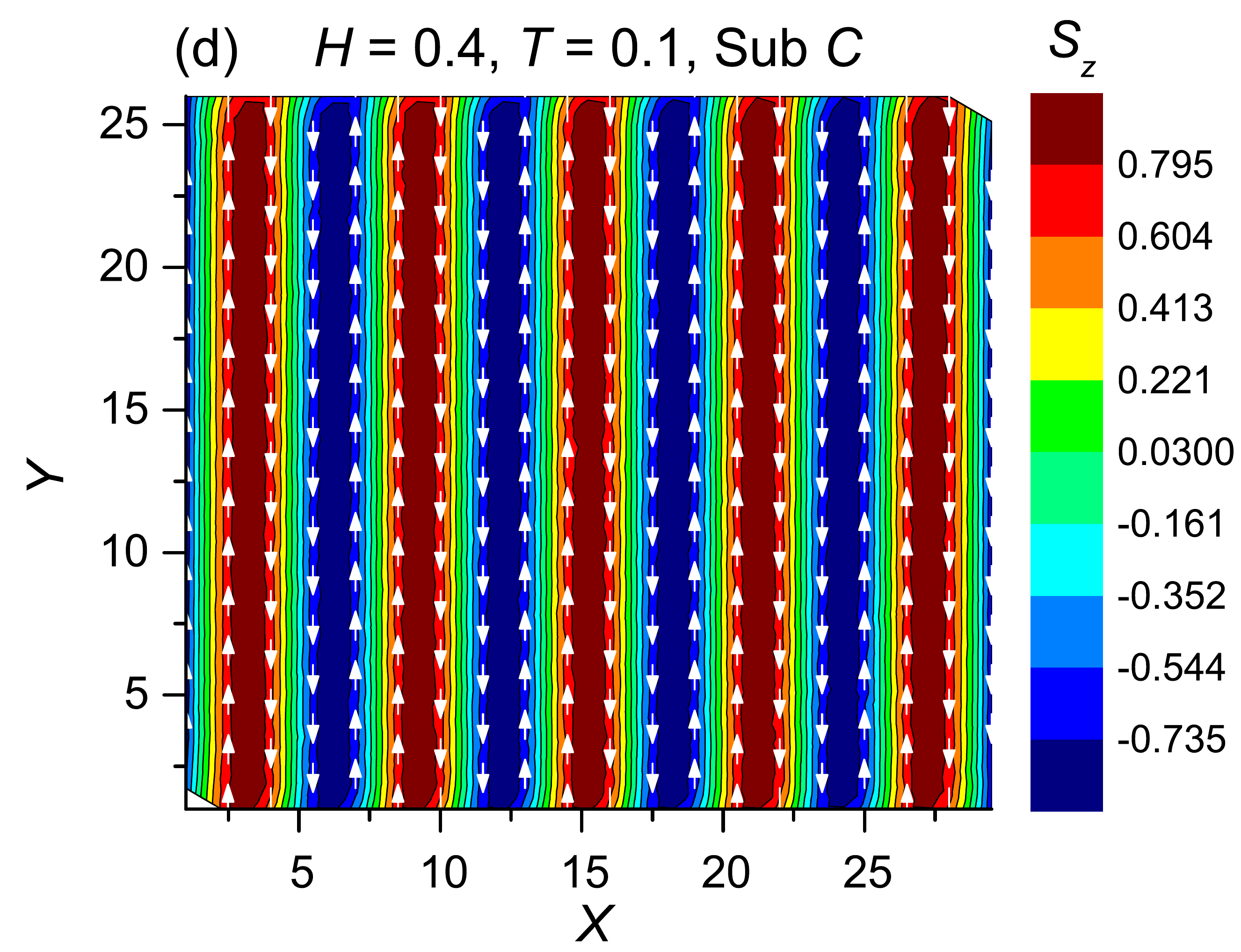}
 }
 \caption{ Calculated  $xy$-projection  (a),  and  three FM sublattices (b,c,d) of the AF  spiral   when $H$ = 0.4 and $T$ = 0.1. Sub \emph{A}, \emph{B} and \emph{C} are the three sublattices.}	
\end{figure*}

Figures 2 (a-d) display the $xy$-projection and the three FM  sublattices of the helix obtained at $H$ = 0.4 and $T$=0.1. The helical lattice is vertical, i.e.,   the $x$-component of each spin is negligible, whereas other two components are  dominant. In each   sublattice, the periodical wavelength  $\lambda$ = 6$a$ in the $x$-direction; and the three sub-HLs are staggered  with each other in the same direction.

\begin{figure*}[htb]
\centerline{
 \includegraphics[width=0.45\textwidth,height=6cm,clip=]
 {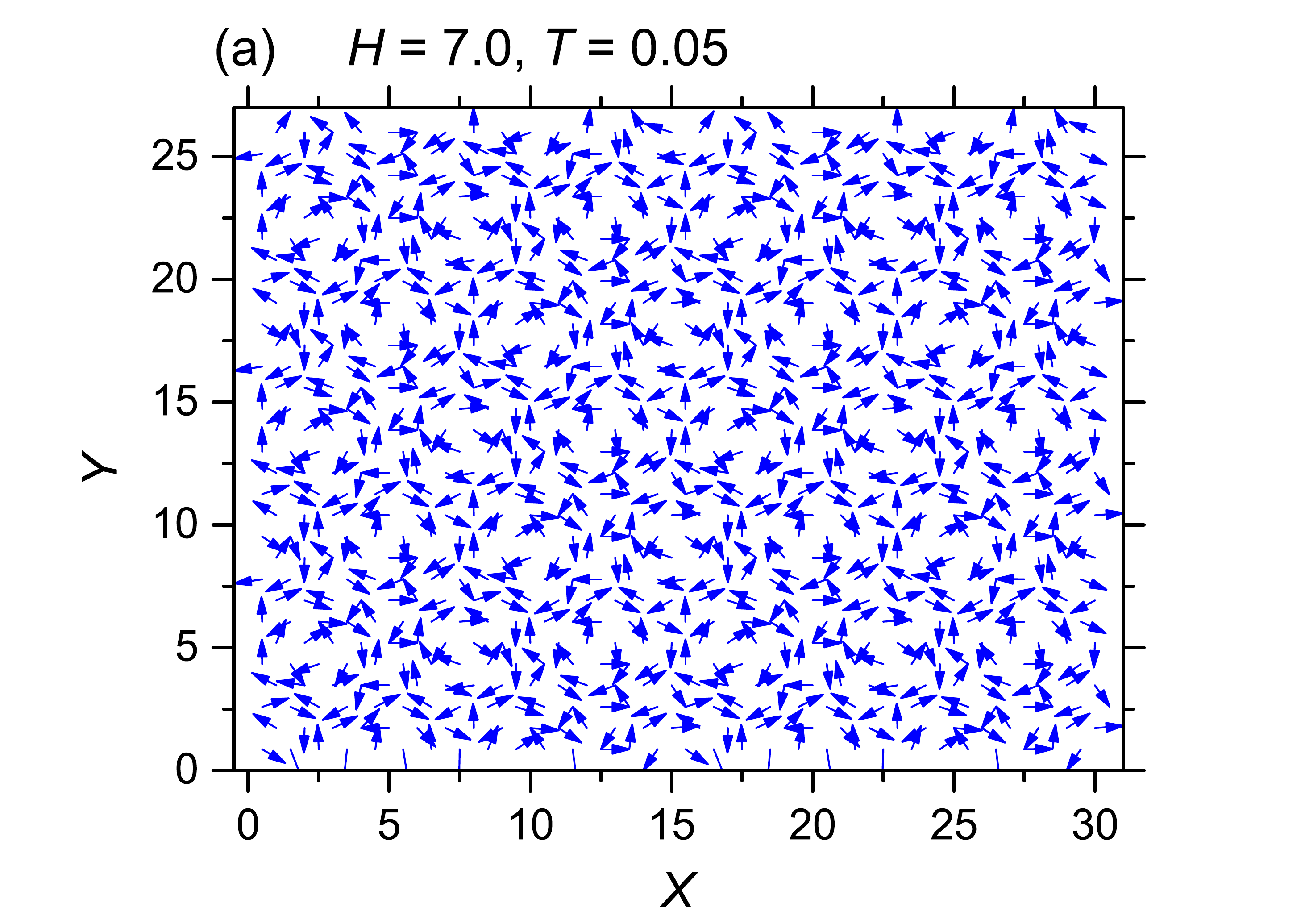}
 \includegraphics[width=0.45\textwidth,height=6cm,clip=]
 {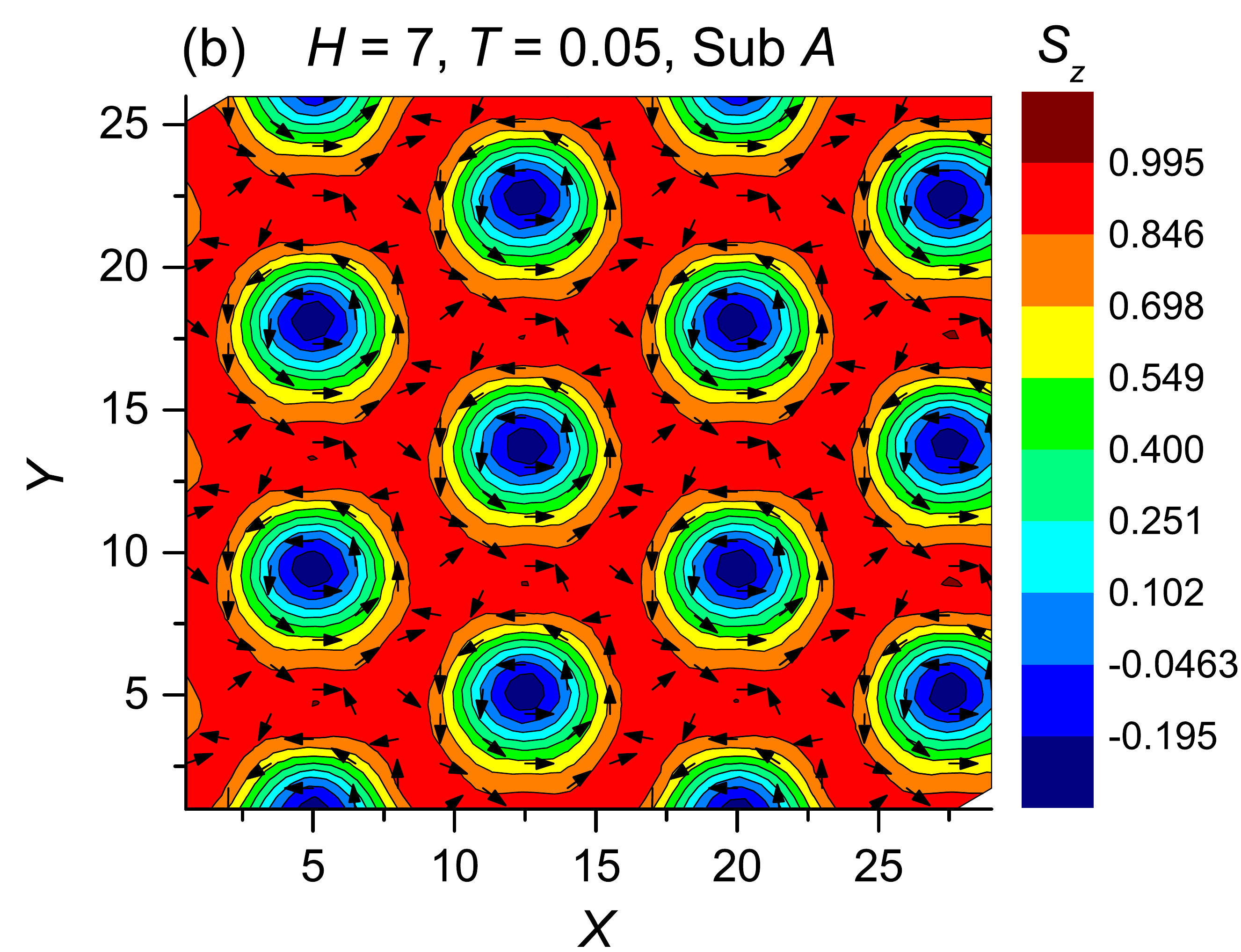}
 }
 \centerline{
 \includegraphics[width=0.45\textwidth,height=6cm,clip=]
 {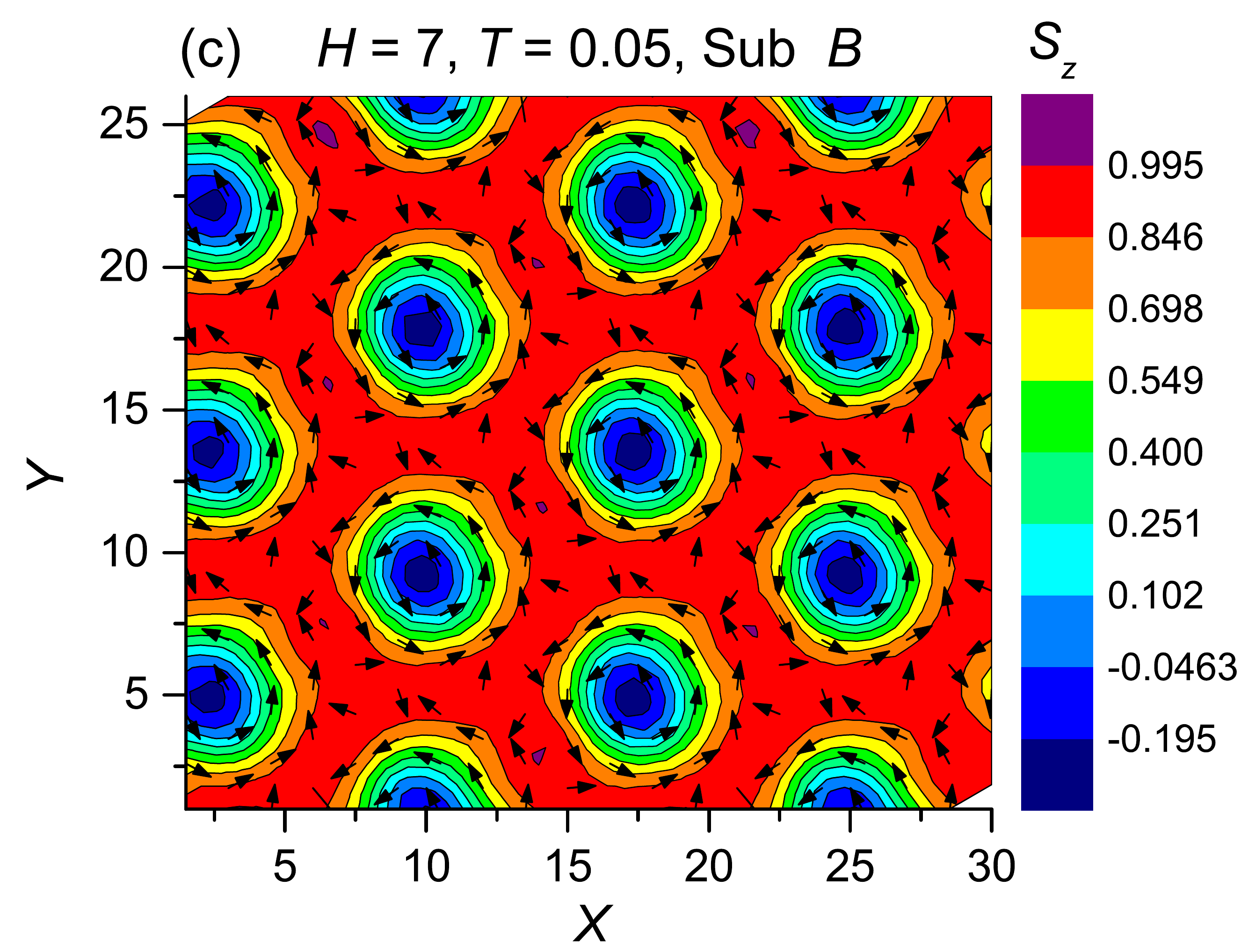}
 \includegraphics[width=0.45\textwidth,height=6cm,clip=]
 {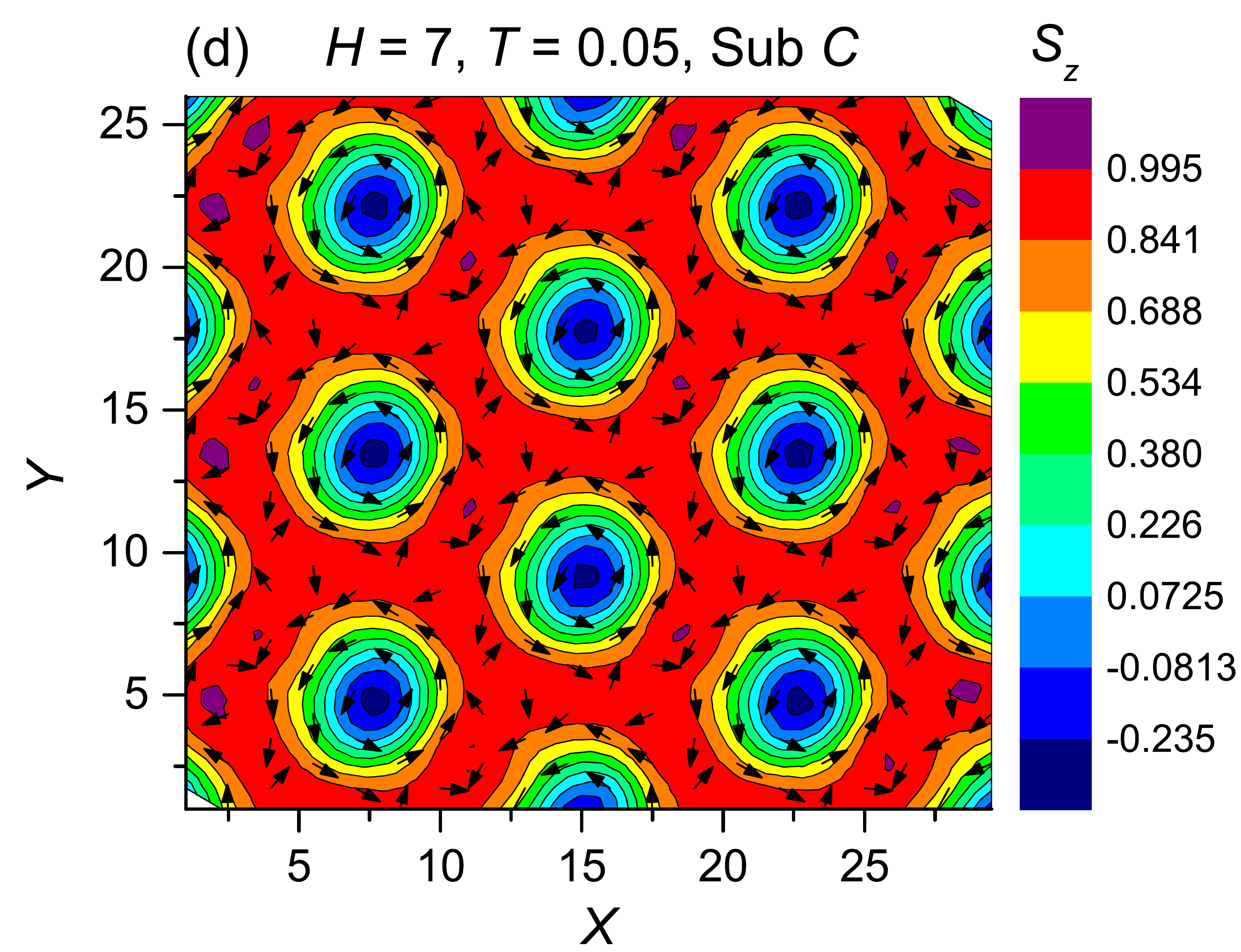}
}
\caption{Calculated  $xy$-projection   (a)  and  three FM sublattices   (b,c,d)  of the AF SL when $H$ =7, $T$ = 0.05.}
\end{figure*}
Figure 3   displays the $xy$-projection and the three sublattices of an AF SL calculated at $H$ = 7, $T$=0.05. In Figure 3(a)   we can find that AF SL shows a hexagonal   pattern, so are its three FM sublattices. But as displayed in Figure 3(b-d),  the latter three  have been  rotated about $\pi/4$ clockwise.  Interestingly, each of the three sublattices is an FM SL  consisting of 12 FM skyrmions inside the 30 $\times$ 30 lattice. All these FM skyrmions are left-handed since $D/{\cal J} <$   0.  The three FM SLs are staggered with each other,  the skyrmions  in these sublattices  are elongated along slightly different   directions, so that the FM  sublattices are  not identical. In each SL, the spins in the interstitial region, which has a  size comparable  to that of  the skyrmion, do not form vortex, and some of the neighboring spins align approximately antiparallel  owing to spin frustration.
 The spin textures in these interstitial areas seem quite 'random' in the $xy$-plane, but they are  astonishingly still periodic in each FM sublattice. These  'disordered' textures look   completely different from those obtained in CMC simulations, \cite{Rosales}  and as a  result, give rise to   increased  magnetic  entropy, $S_M$,  and reduced  total free energy, $F$, of the whole system, since $F=E - TS_M$, so that the AF SLs are better stabilized.

\begin{figure*}[htb]
\centerline{
 \includegraphics[width=0.45\textwidth,height=6cm,clip=]
 {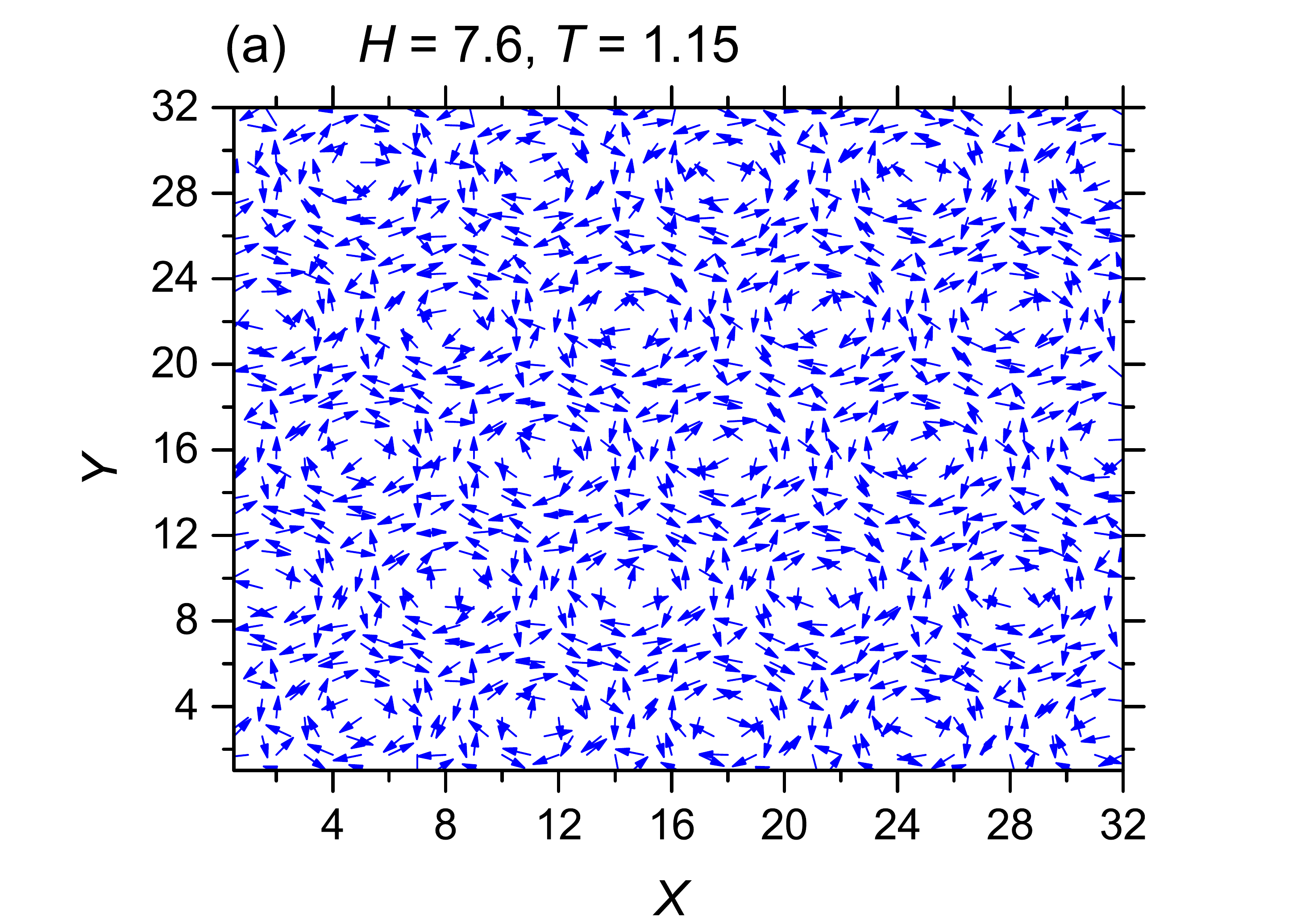}
 \includegraphics[width=0.45\textwidth,height=6cm,clip=]
 {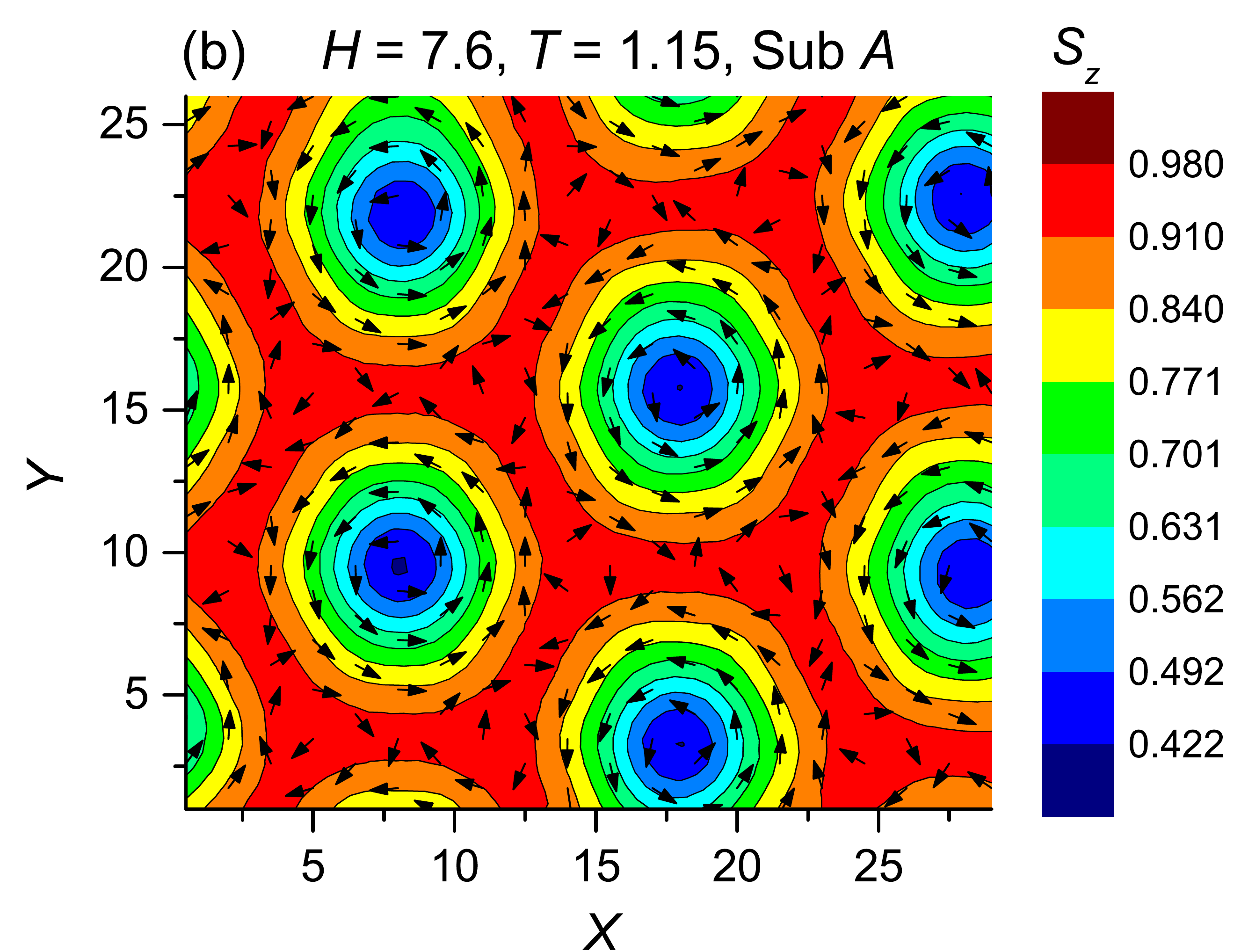}
 }
 \centerline{
 \includegraphics[width=0.45\textwidth,height=6cm,clip=]
 {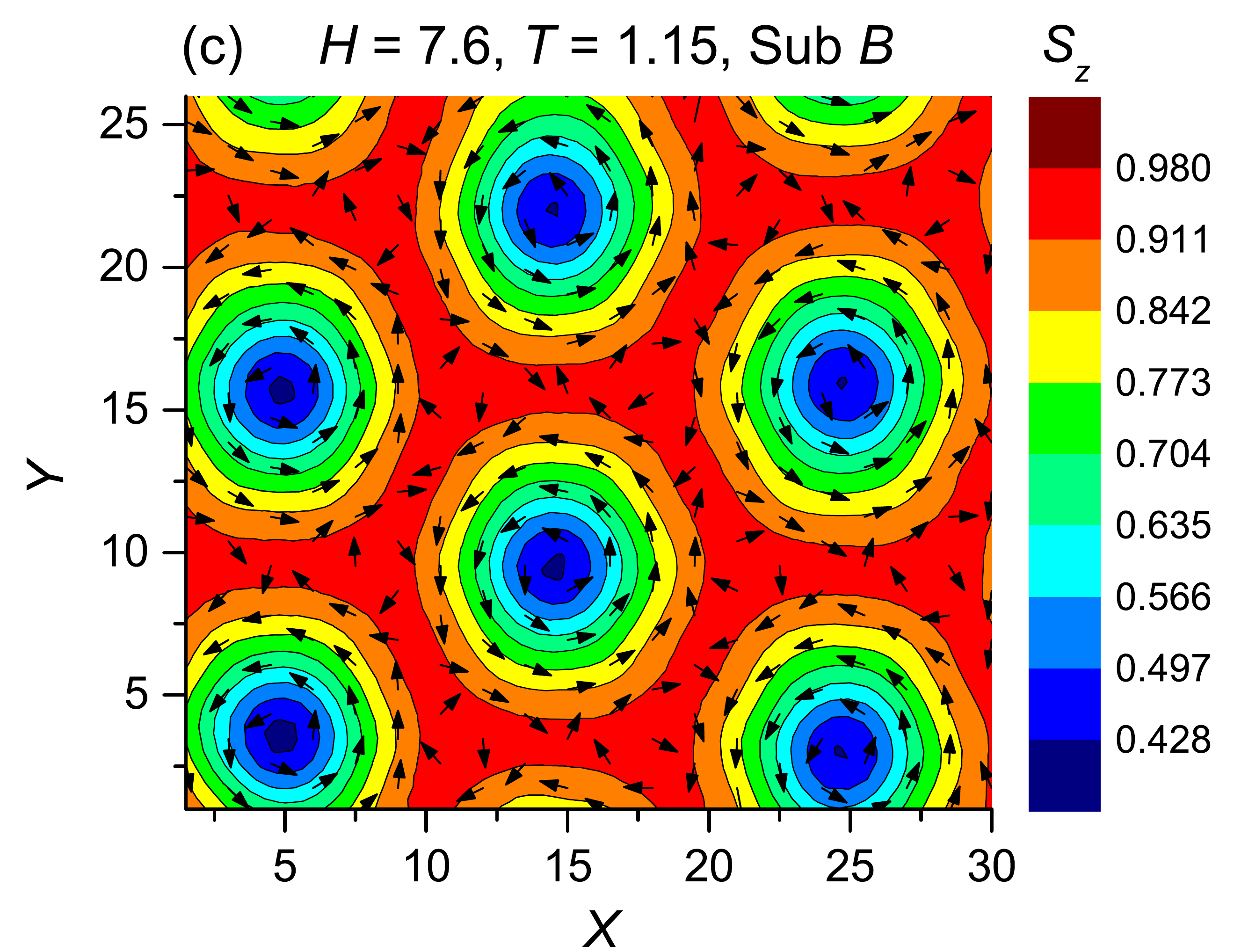}
 \includegraphics[width=0.45\textwidth,height=6cm,clip=]
 {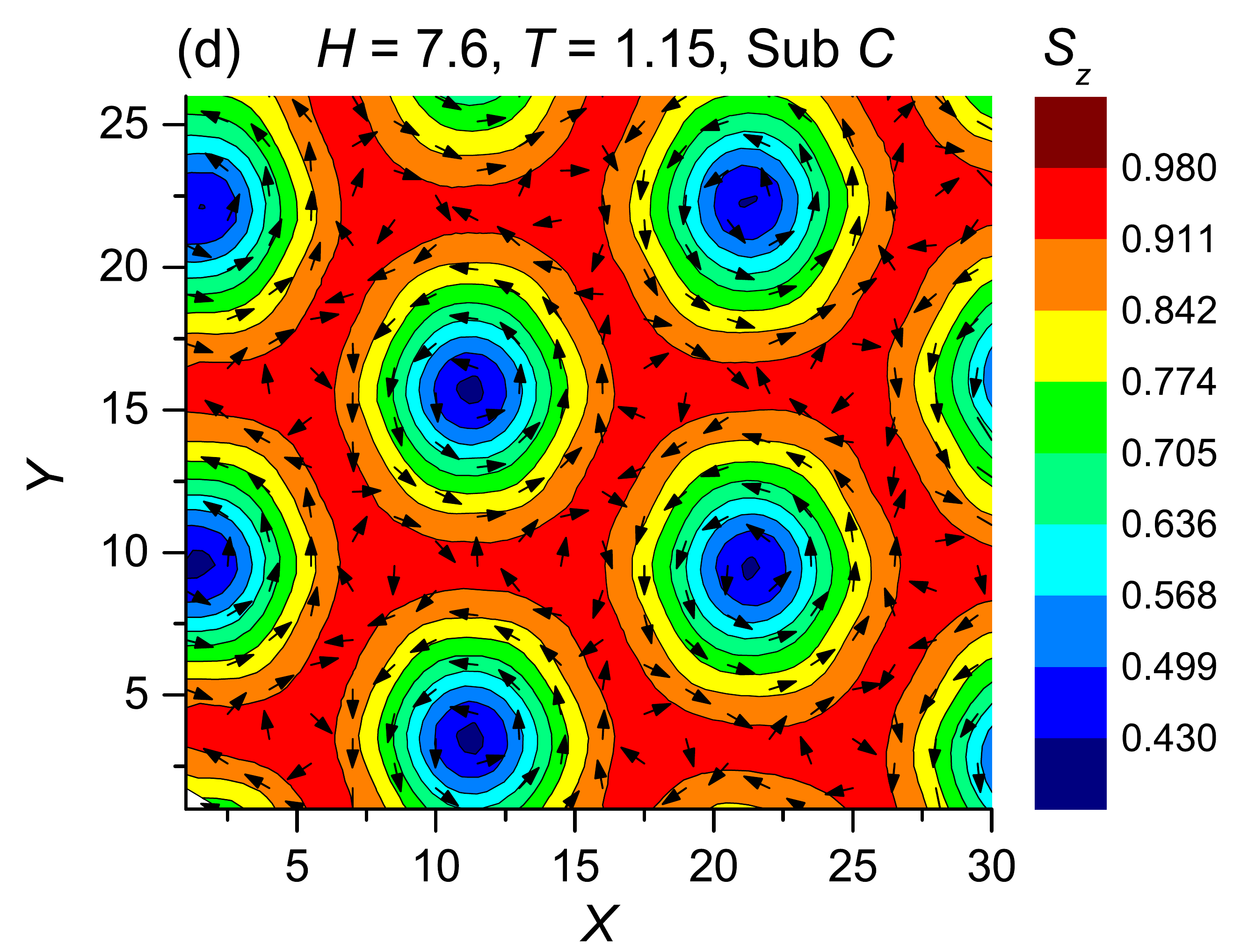}
}
\caption{Calculated  $xy$-projection   (a),  and  three FM sublattices  (b,c,d)  of the AF  SL as $H$ =7.6, $T$ = 1.15.}
\end{figure*}
When $H$ = 7.6   as shown in Figure 4,   the $z$-components of all spins become positive,  so only VL can be formed below $T_{VL}$ = 1.15.  The AF vortices form regular hexagonal  VL,  whose three   FM sublattices show the same pattern but all rotated around the $z$-axis. More interestingly,  the size of each vortex increases suddenly,  so that we have to use a 60 $\times$ 60 lattice to produce the VL texture. For comparison,  a 30 $\times$ 30 portion is displayed in Figure 4, therein 6  FM vortices are observed in each FM sublattice. Once again, the three sublattices are   FM VLs, they are nonidentical, and staggered with each other. Nevertheless, each FM VL shows excellent periodicity and symmetry.

\begin{figure*}[htb]
\centerline{
 \includegraphics[width=0.45\textwidth,height=6cm,clip=]
 {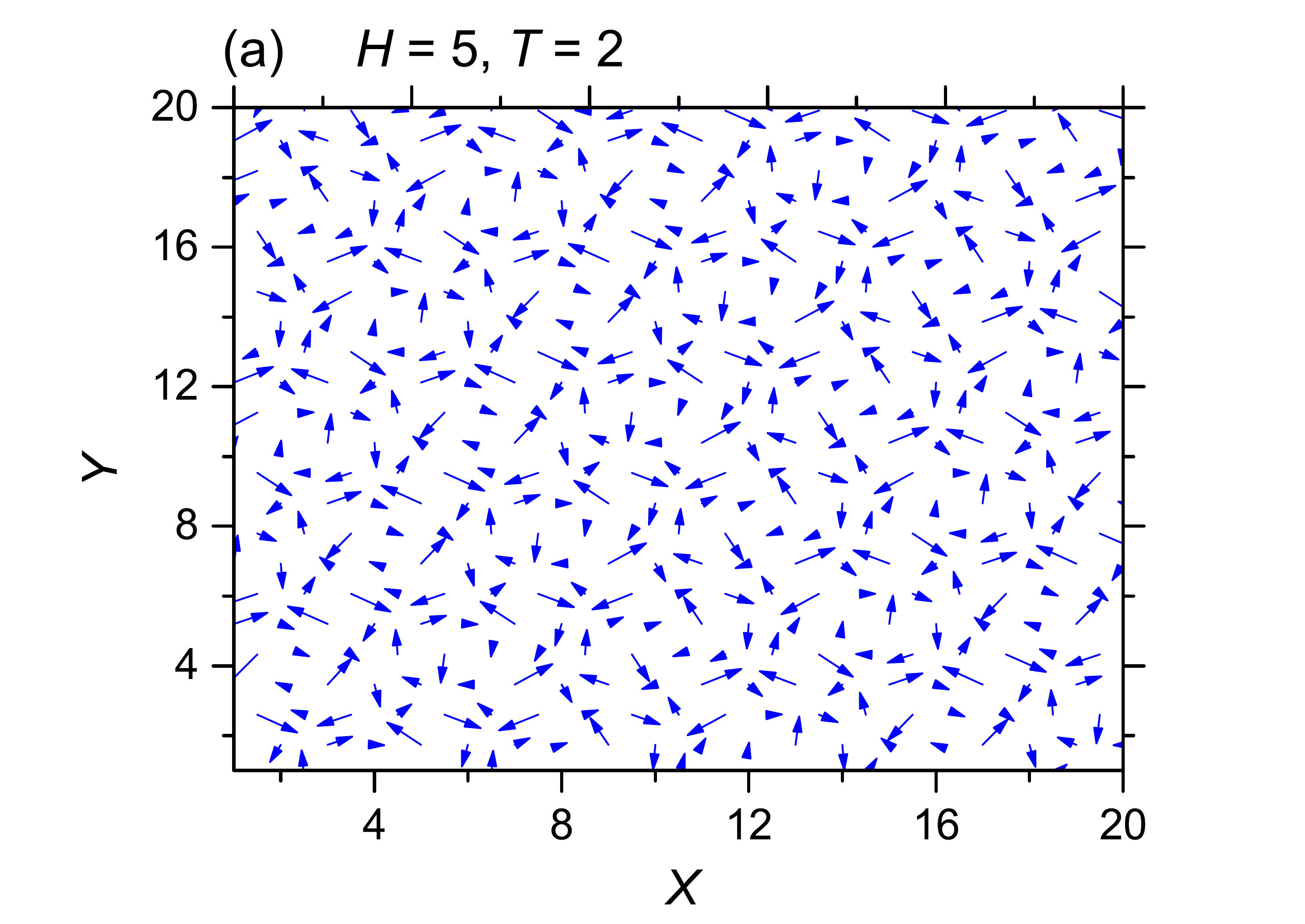}
 \includegraphics[width=0.45\textwidth,height=6cm,clip=]
 {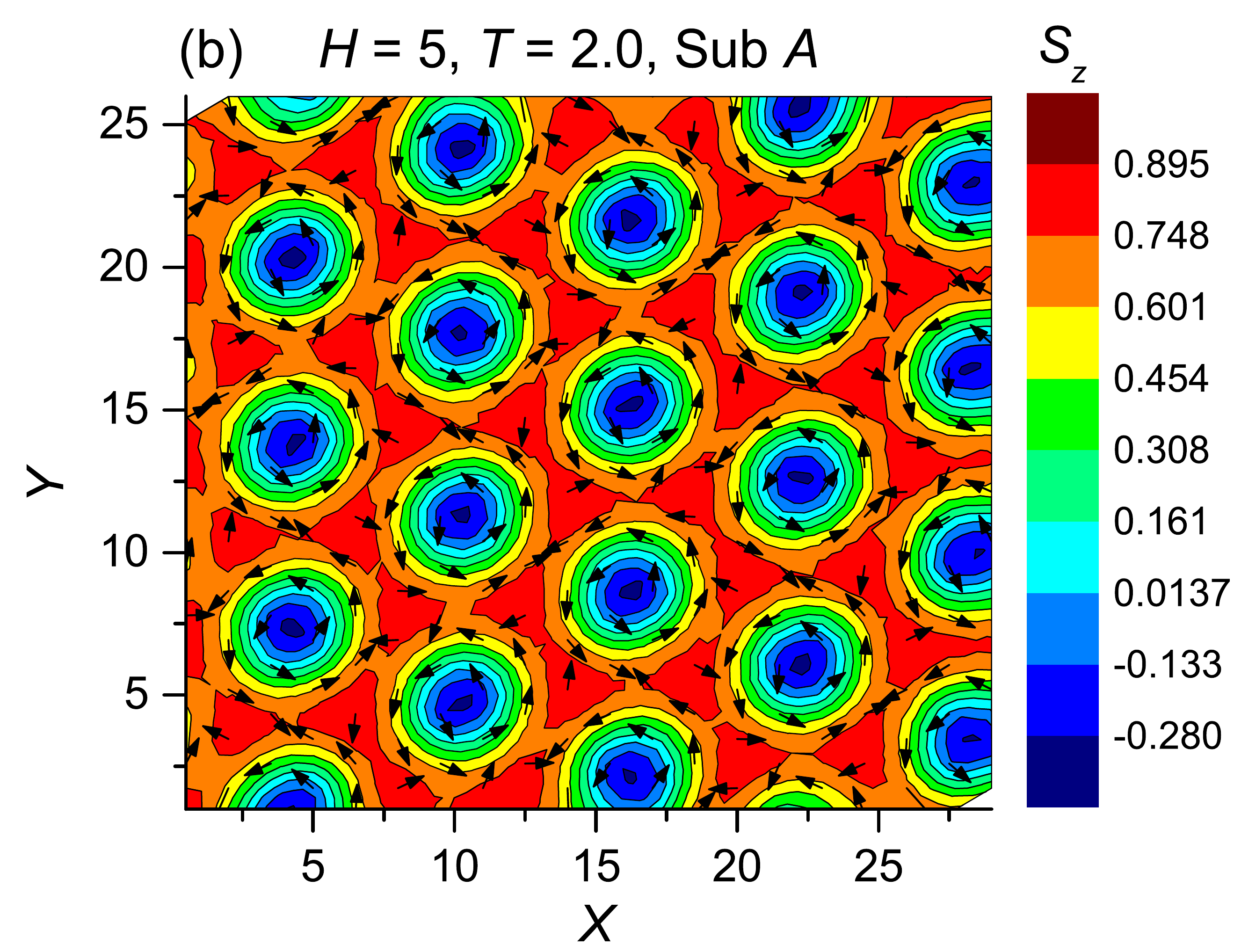}
 }
 \centerline{
 \includegraphics[width=0.45\textwidth,height=6cm,clip=]
 {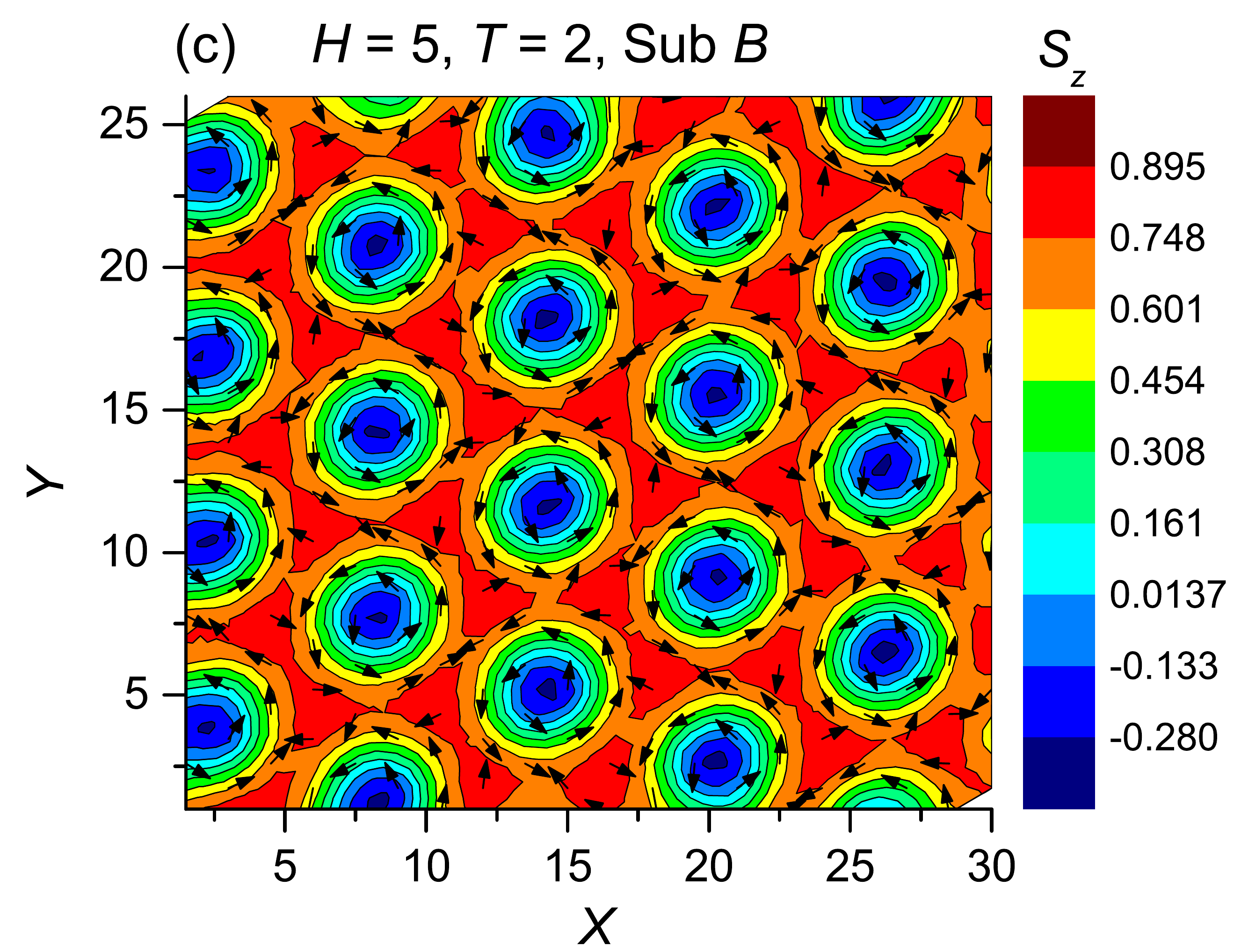}
 \includegraphics[width=0.45\textwidth,height=6cm,clip=]
 {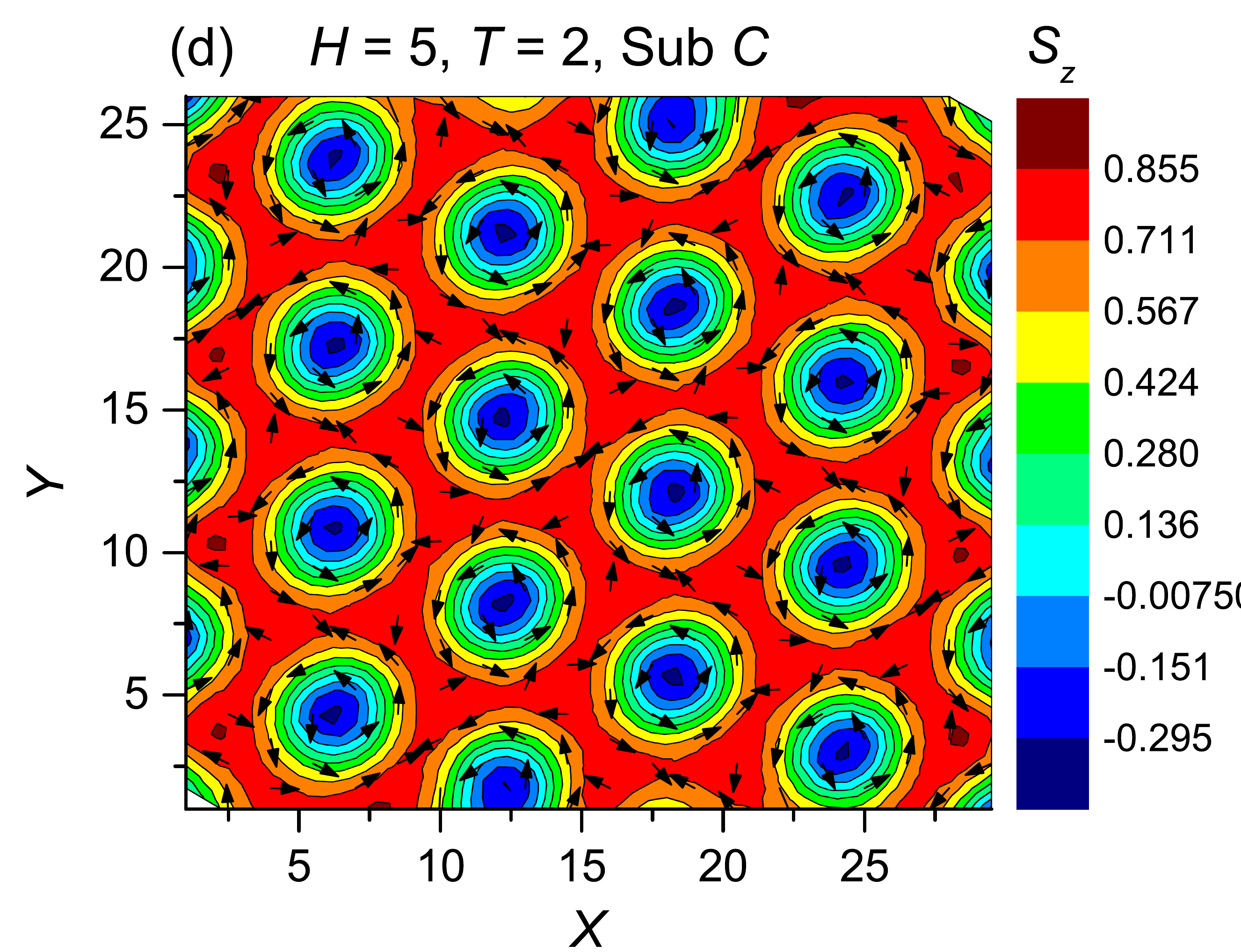}
}
\caption{Calculated    $xy$-projection (a),  and  three FM sublattices  (b,c,d) of the AFM SkL as $H$ =5, $T$ = 2.}
\end{figure*}

When $H$ is reduced to  6.6,  each skyrmion in the AF SL shrinks considerably.  So we can find in Figure 5, obtained at $H$ = 5, $T$ = 2, each FM SL contains 20 skyrmions in the rotated HCP pattern. In addition, in the (a) sub-figure, two  inclined spin strips  emerge with their lower ends terminating  around $x$ = 1  and  15, respectively. The two strips do spoil the periodicity of the AF SL shown in the first sub-figure, however, they have  seemly no evident effects upon the periodicity and symmetry of the three FM sublattices.

\subsection{Distribution of Topological Charge Density  of  Each  Sublattice}

For a  discretized   model, the topological charge can be expressed with \cite{Rosales,Anderson,Ohgushi}
\begin{equation}
\chi_Q=\frac{1}{4\pi}\left\{\sum_{\vec{r}_i}A_{\vec{r}_i}^{(12)}{\rm sign}[\chi_{L,\vec{r}_i}^{(12)}]
+A_{\vec{r}_i}^{(34)}{\rm sign}[\chi_{L,\vec{r}_i}^{(34)}]\right\}\;,
\end{equation}
where $A_{\vec{r}_i}^{(ab)}=||\left(\vec{S}_{\vec{r}_a}- \vec{S}_{\vec{r}_i}  \right) \times \left( \vec{S}_{\vec{r}_b}- \vec{S}_{\vec{r}_i}    \right)||/2$ is the local area of the surface spanned by three spins at $\vec{r}_i$, $\vec{r}_a$ and $\vec{r}_b$  of  an elementary  triangle,     $\chi_{L,\vec{r}_i}^{(ab)}=  \vec{S}_{\vec{r}_i}\cdot\left( \vec{S}_{\vec{r}_a}\times\vec{S}_{\vec{r}_b} \right)$  denotes the   local chirality,
and $\vec{r}_i$, $\vec{r}_1 \sim  \vec{r}_4$  are the sites involved in the calculation of $\chi_Q$ as described in Rosales et al.'s article.  \cite{Rosales}
\begin{figure*}[htb]
\centerline{
 \includegraphics[width=0.45\textwidth,height=6cm,clip=]
 {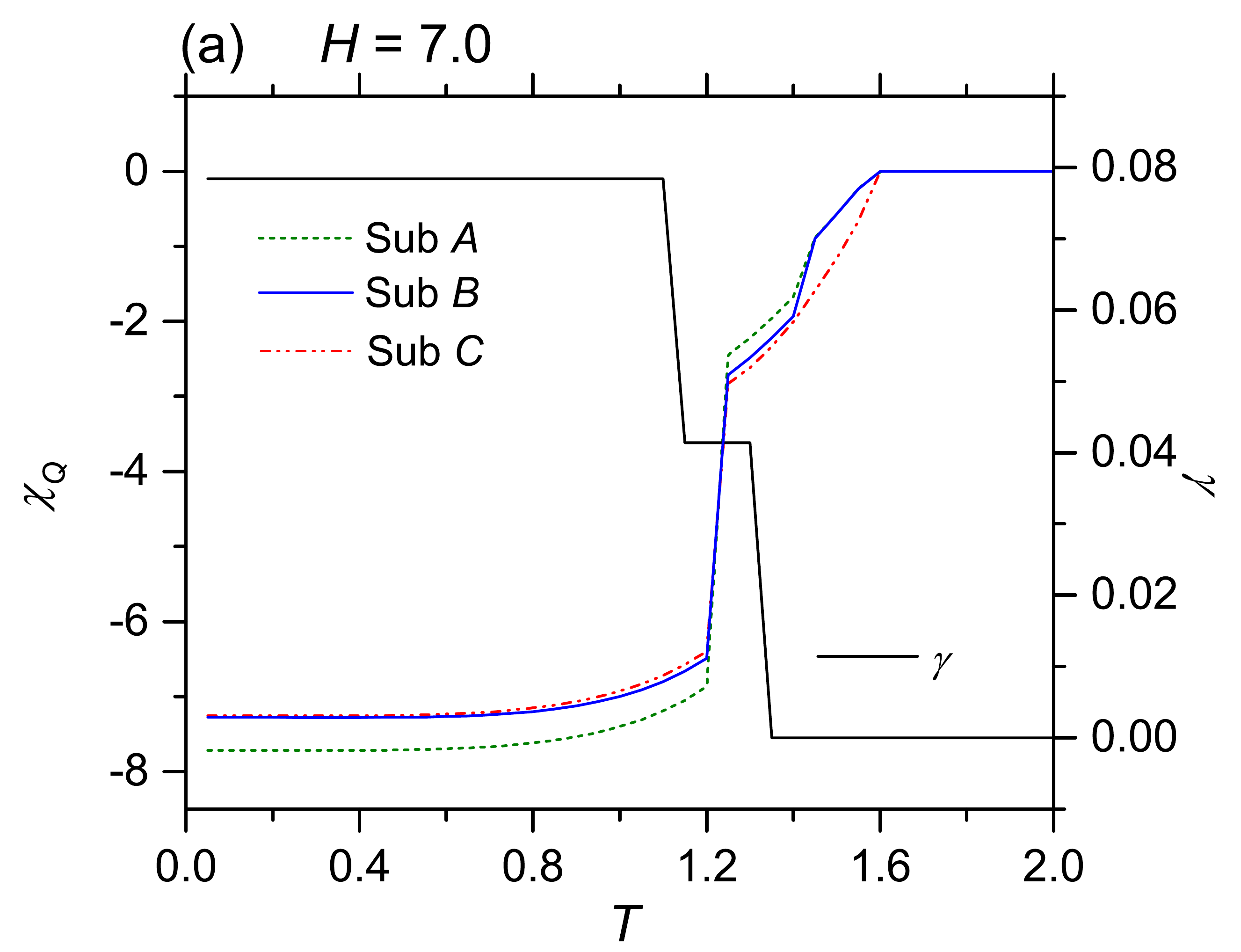}
 \includegraphics[width=0.45\textwidth,height=6cm,clip=]
 {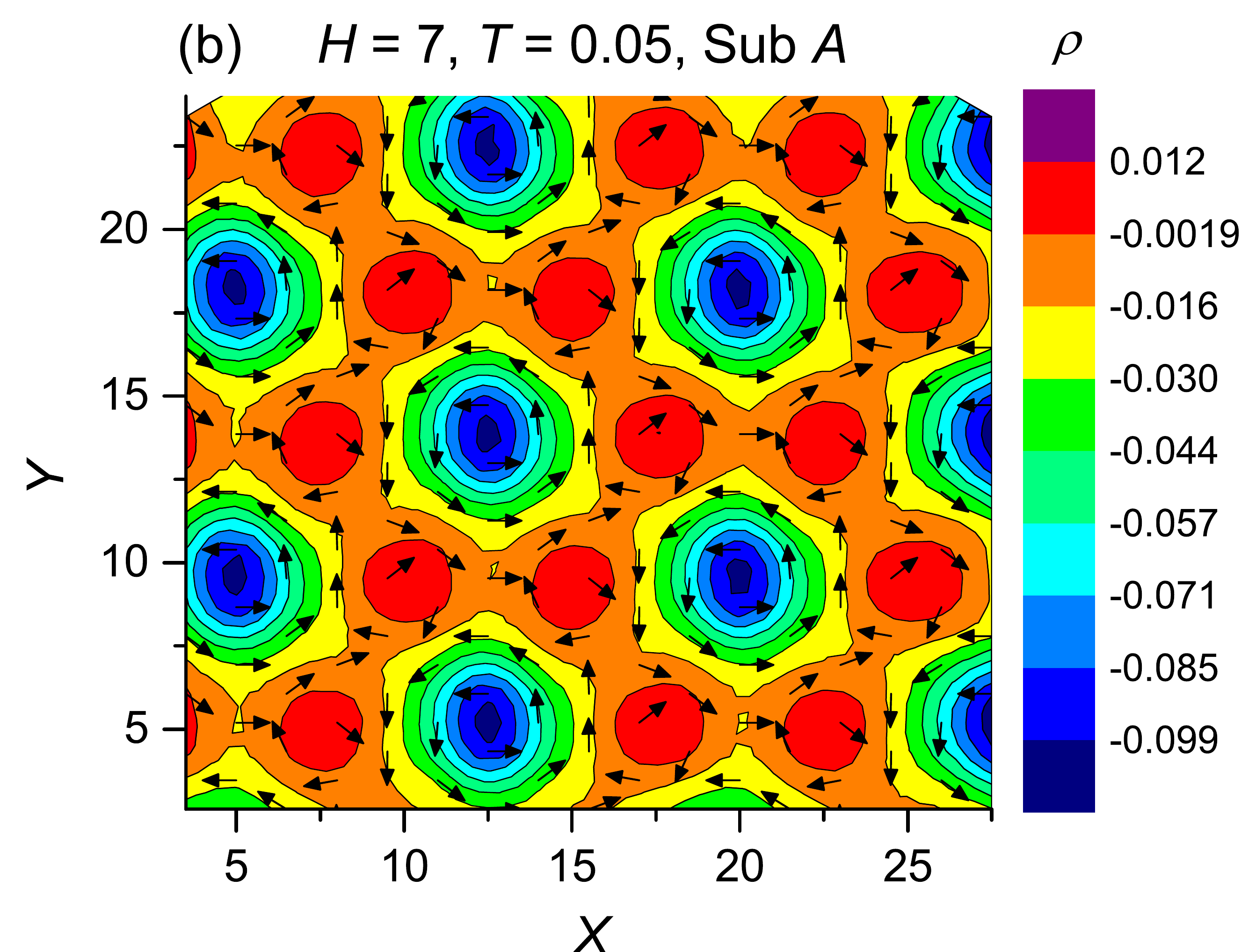}
 }
 \centerline{
 \includegraphics[width=0.45\textwidth,height=6cm,clip=]
 {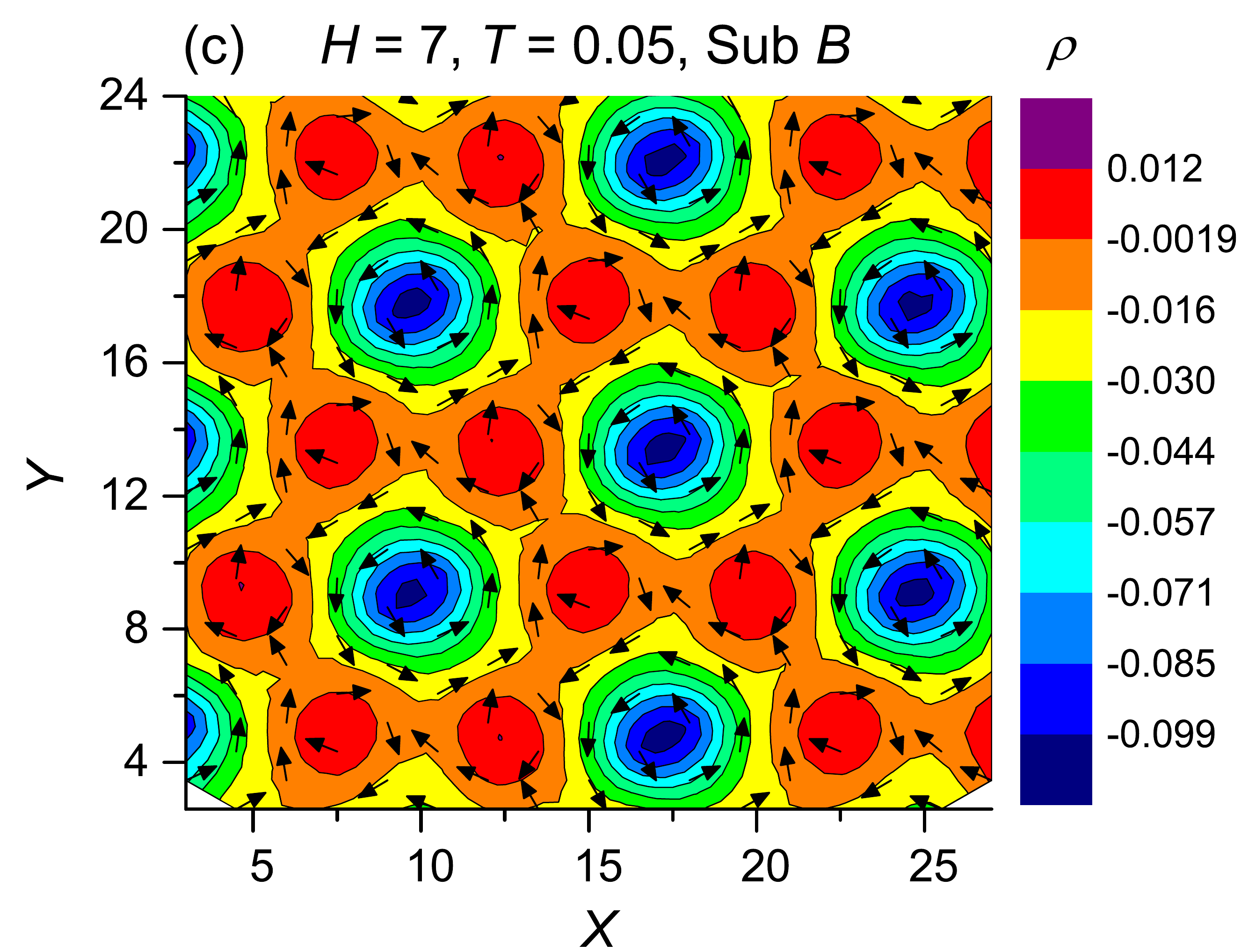}
 \includegraphics[width=0.45\textwidth,height=6cm,clip=]
 {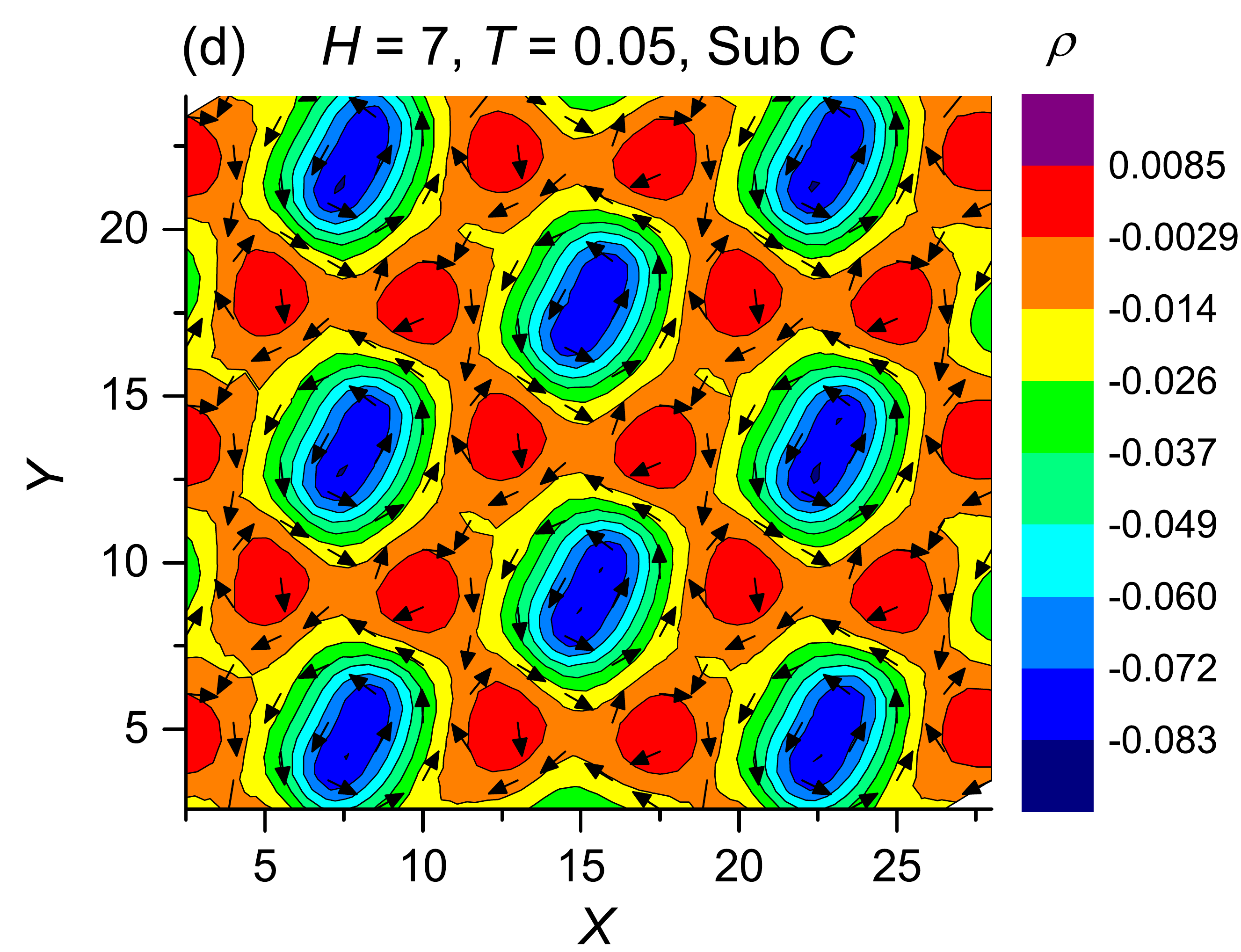}
}
\caption{ Total topological charges (a), and  charge density distributions of the three FM sublattices (b,c,d),  of the   AF  SL calculated at   $H$ =7, $T$ =0.05.}
\end{figure*}

\begin{figure*}[htb]
\centerline{
 \includegraphics[width=0.45\textwidth,height=6cm,clip=]
 {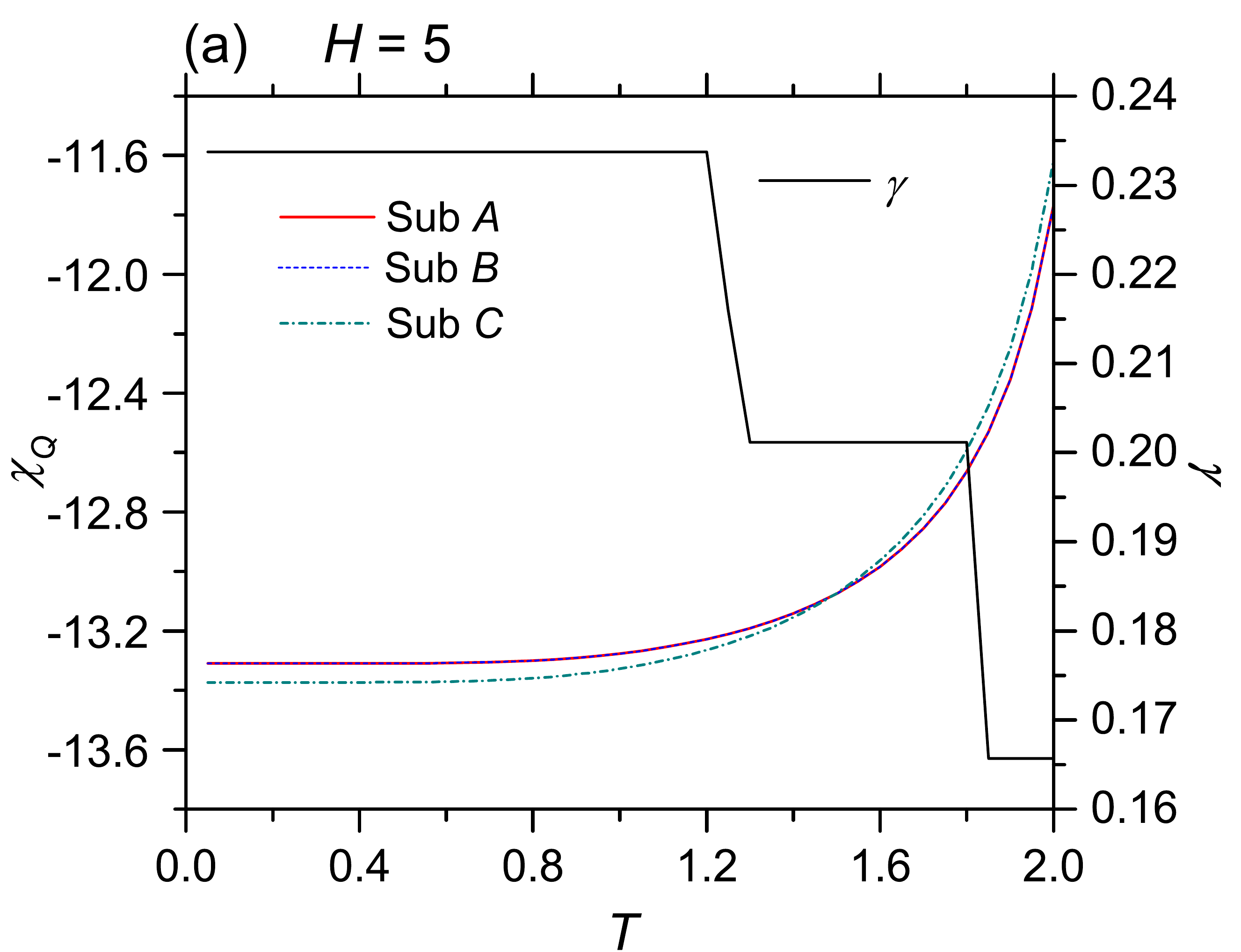}
 \includegraphics[width=0.45\textwidth,height=6cm,clip=]
 {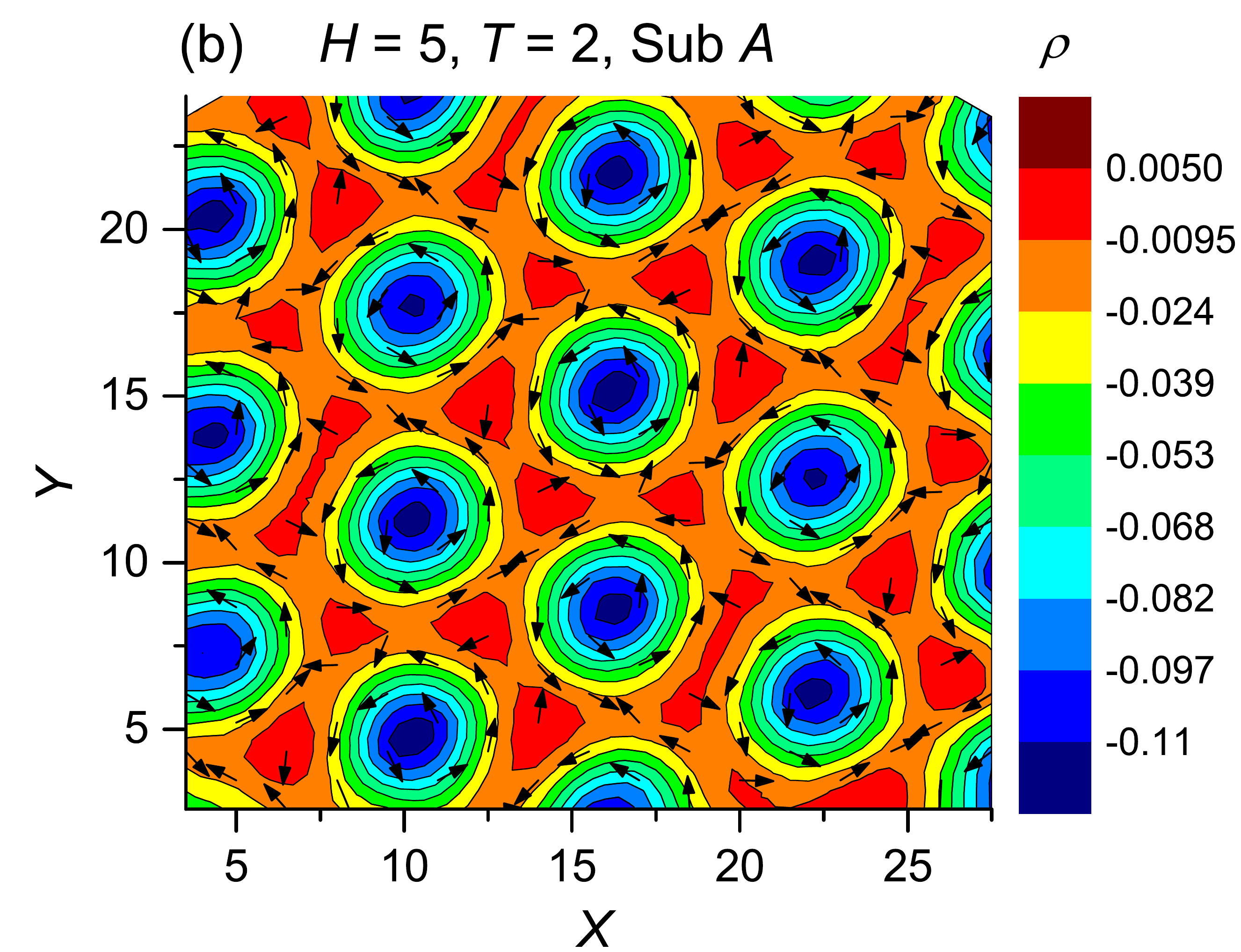}
 }
 \centerline{
 \includegraphics[width=0.45\textwidth,height=6cm,clip=]
 {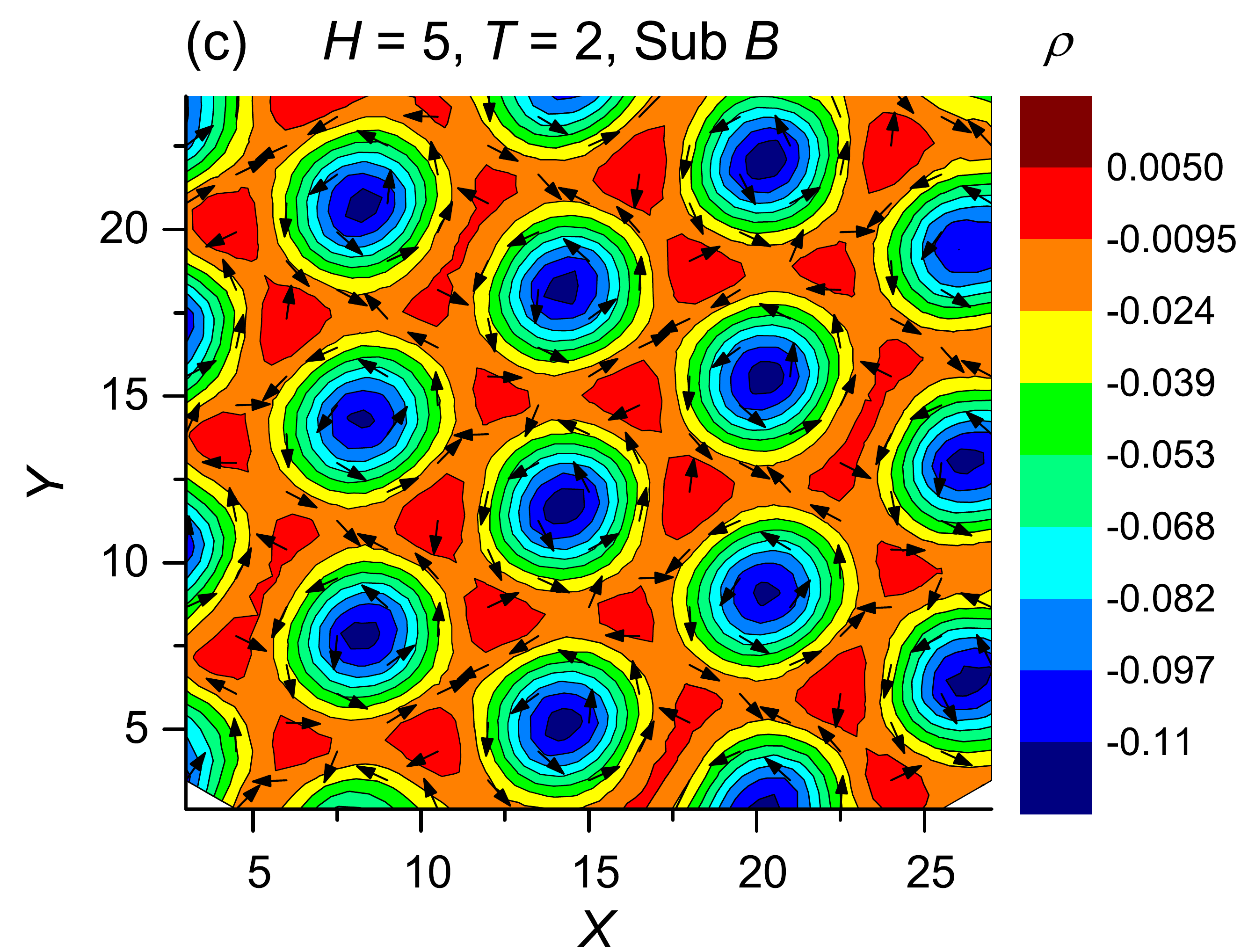}
 \includegraphics[width=0.45\textwidth,height=6cm,clip=]
 {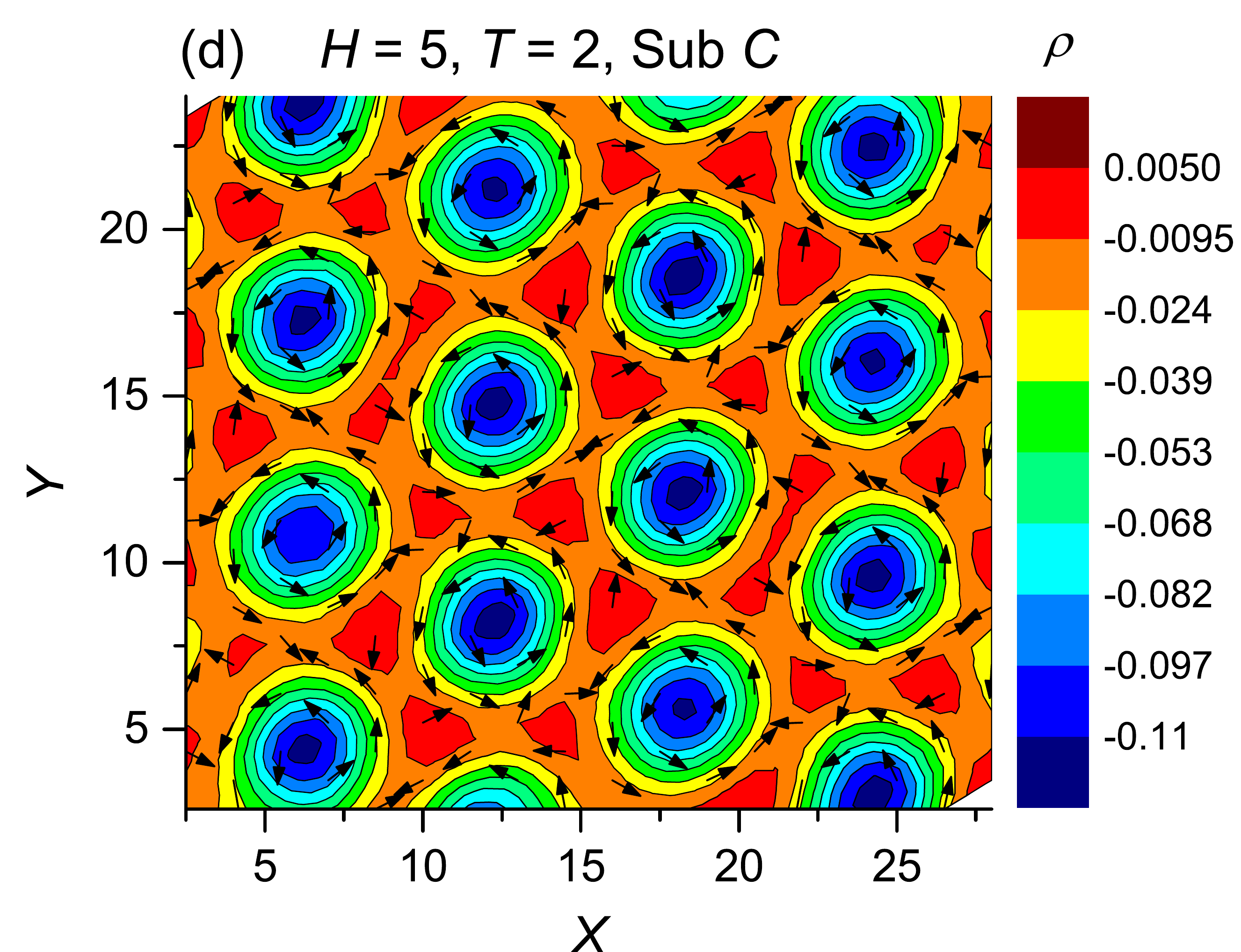}
}
\caption{Total topological charges (a), and  charge density distributions of the three FM sublattices (b,c,d),  of the   AF  SL calculated at $H$ =5, $T$ = 2.}
\end{figure*}
Using above formulas, we find that the calculated topological charge density  of each sublattice also forms periodical crystals, as shown in Figures 6 $\sim $ 7(b,c,d,), that are obtained at $H$ = 7, $T$ = 0.05, and $H$ = 5, $T$ = 2, respectively. Here, the rainbow-like contour   denotes the topological charge density, $\rho$,   at the lattice sites, and the arrows show the $xy$ projections of the FM spin sublattice  of the   AF SL. Clearly,  each   charge density crystal is almost exactly identical with the  corresponding FM spin sublattice. The three $\rho$ sublattices of an  AF SL differ from each other, and we observe much densely packed $\rho$ unit cells in Figure 7(b,c,d).

In Figure 6(a) and Figure 7(a), the $\chi_Q$ curves are plotted versus changing temperature.  There, $\gamma$ represents the ratio of the spin number of those with negative $\langle S_z\rangle$ to the whole spin number. In the first case, $H$ = 7, $T_{VL}$ = 1.55 and $T_{SL}$ = 1.3.  So above $T_{VL}$,  $\gamma$ and $\chi^{A,B,C}_Q$ are all equal to zero. From $T_{SL}$ down to $T$ = 1.10,  $\gamma$ increases from 0.0414 to 0.0784, then remains unchanged until $T$ = 0.05.  At all temperatures,   less than 8\% spins align opposite to the applied magnetic field, so the chiral spin textures are far away from  fully 'skyrmionic'.  Carefully inspecting each FM sublattice, we can see that the spin texture in each interstitial region exhibits  completely different features from the surrounding skyrmions, and the two sorts of spin textures are entangled together. Thus, at $H$ = 7 and $T$ = 0.05,   the averaged topological charge per skymionic complex, consisting of one skyrmion and its surrounding area, $Q_{av} \in$ (-0.643,   -0.606). In the second case, $H$ = 5 and $T$=2.0. Though the three spin sublattices are FM SLs,  $\gamma  $ is around  0.166,  the SLs are also not fully skyrmionic yet, thus  $Q_{av} \in$ (-0.588,   -0.581).

\subsection{Spin Structure Factors of  the Helical, Skyrmionic and Vortical Lattices}
To characterize  the  spin-textures,  the static spin structure factors in the reciprocal space have to be calculated, of which the  out  and in plane components $S_{\bot}(q)$  and $S_{//}(q)$ are defined as

\begin{eqnarray}
S_{\bot}(q) &=& \frac{1}{N}\left(|\sum_{\vec{r}_i} S_{\vec{r}_i}^x{\rm e}^{-i\vec{q}\cdot{\vec{r}_i} }|^2 + |\sum_{\vec{r}_i} S_{\vec{r}_i}^y{\rm e}^{-i\vec{q}\cdot{\vec{r}_i} }|^2  \right)\;,\\
S_{//}(q) &=& \frac{1}{N}\left(|\sum_{\vec{r}_i} S_{\vec{r}_i}^z{\rm e}^{-i\vec{q}\cdot{\vec{r}_i} }|^2   \right)\;,\nonumber
\end{eqnarray} respectively.

\begin{figure*}[htb]
\centerline{
 \includegraphics[width=0.45\textwidth,height=6cm,clip=]
 {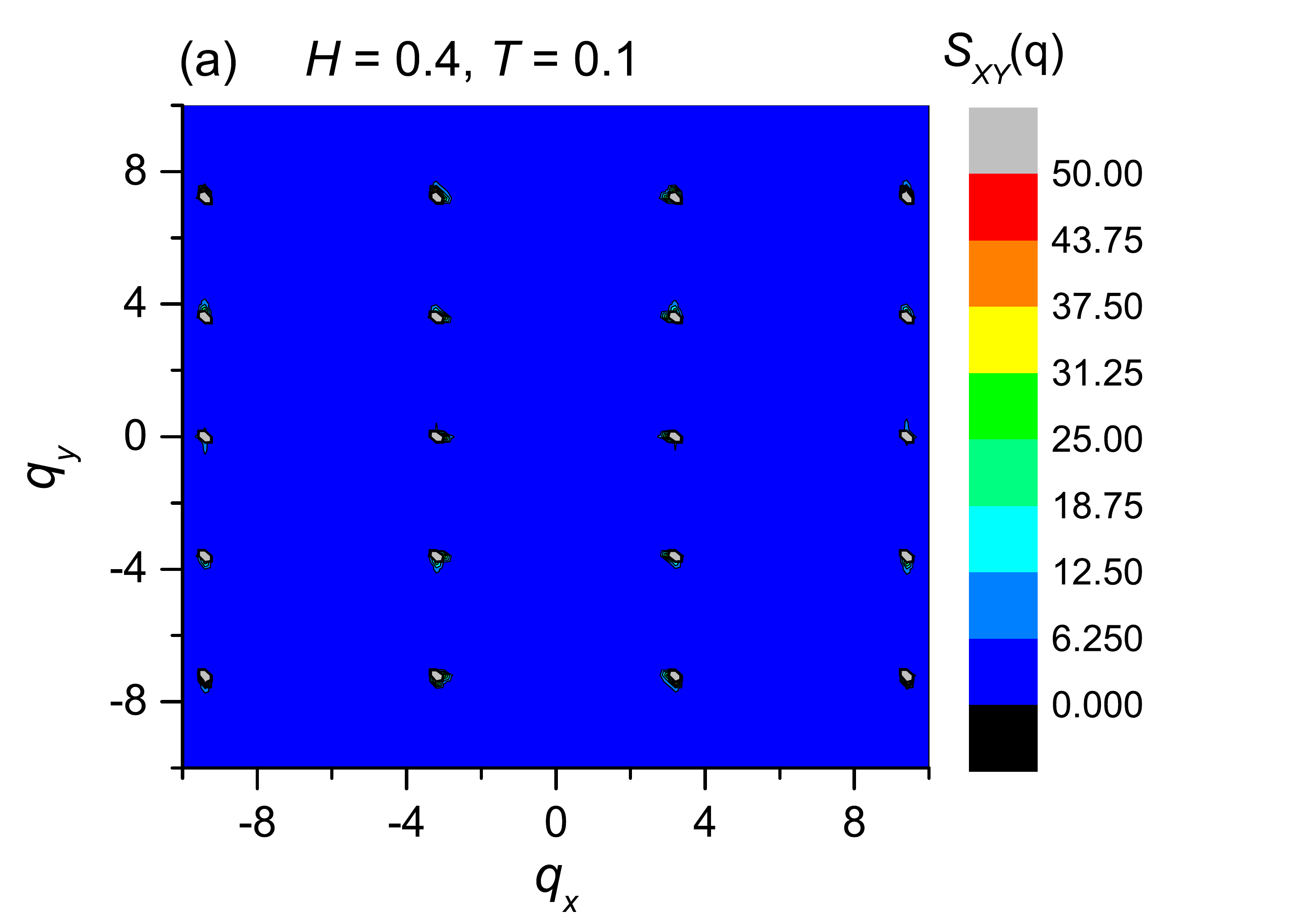}
 \includegraphics[width=0.45\textwidth,height=6cm,clip=]
 {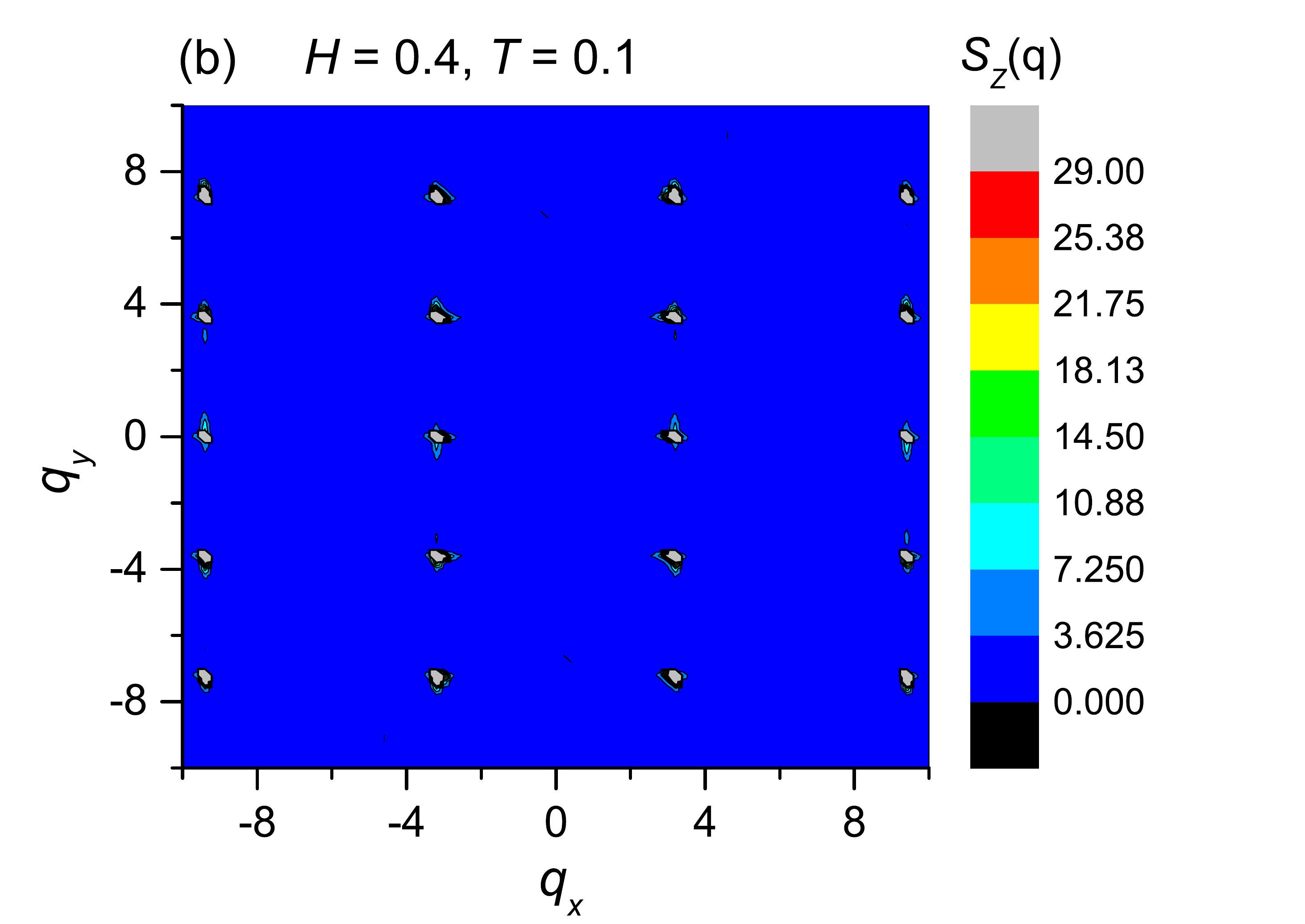}
 }
 \centerline{
 \includegraphics[width=0.45\textwidth,height=6cm,clip=]
 {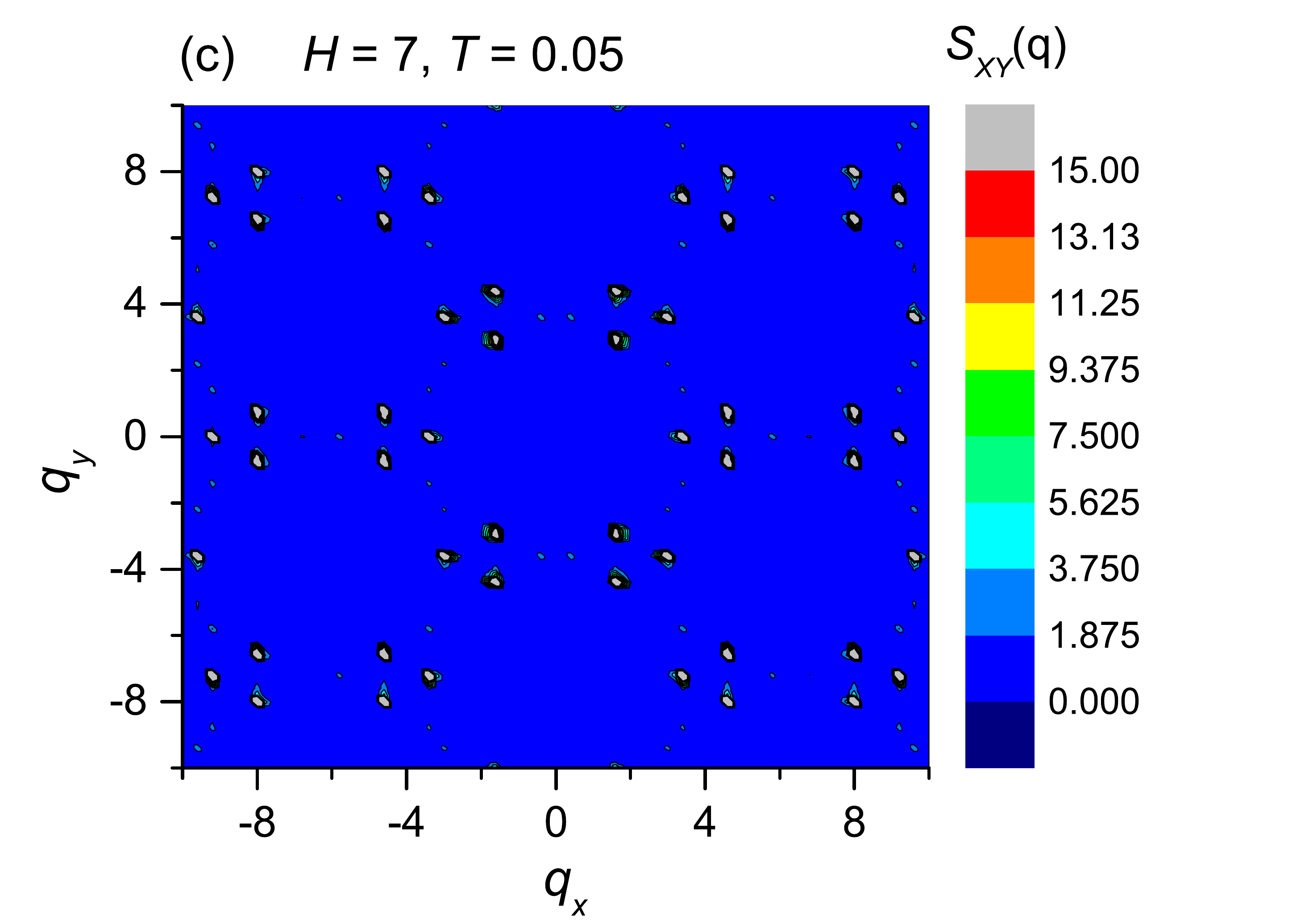}
 \includegraphics[width=0.45\textwidth,height=6cm,clip=]
 {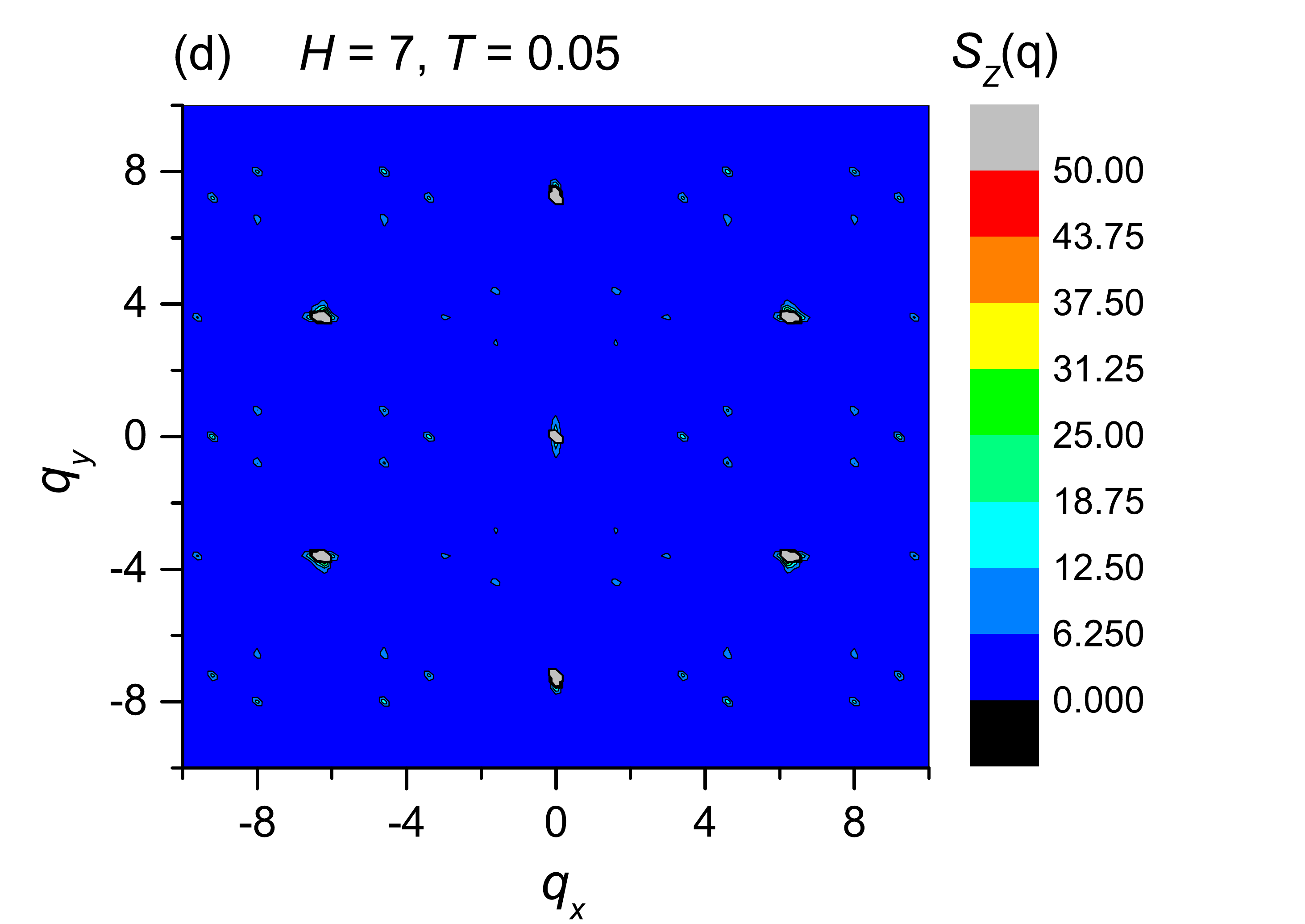}
}
\centerline{
 \includegraphics[width=0.45\textwidth,height=6cm,clip=]
 {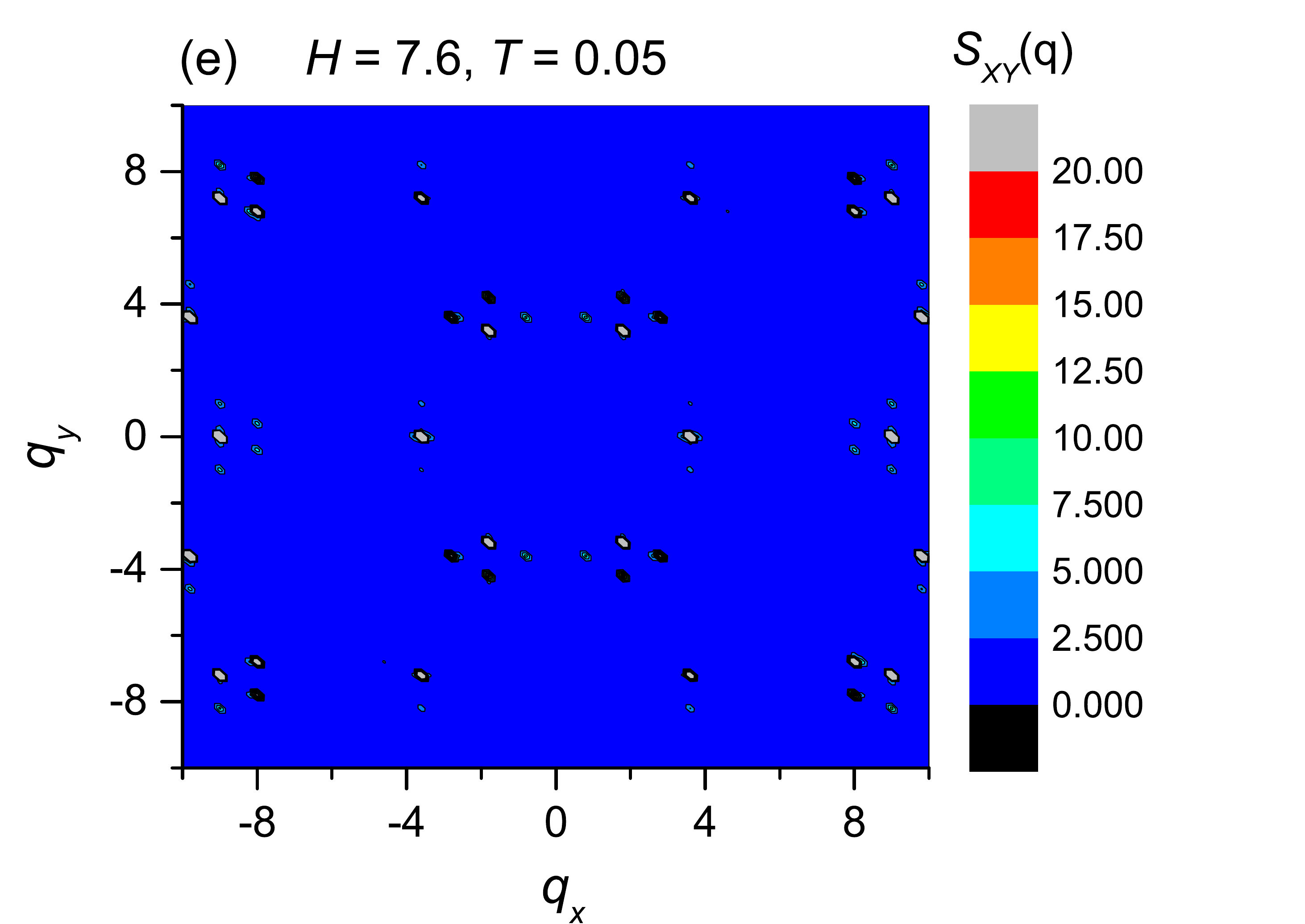}
 \includegraphics[width=0.45\textwidth,height=6cm,clip=]
 {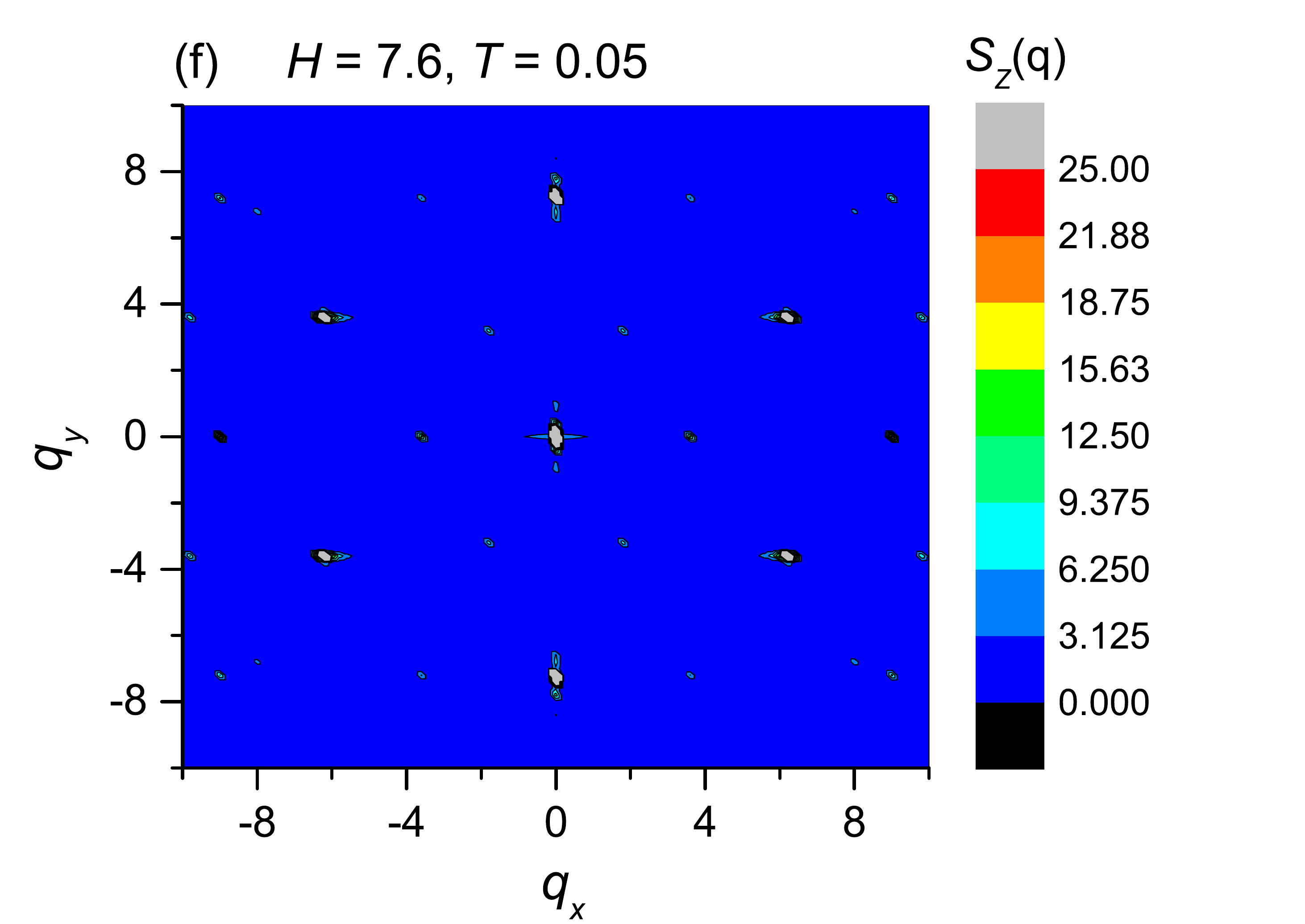}
 }
\caption{ Intensity patterns of the static spin structure
	factors  for the helical phase at $H $ = 0.4, $T$ = 1 (a,b);   AF SL
	phase  when  $H $ = 7, $T$ = 0.05 (c,d), and AF VL phase at  $H $ = 7.6, $T$ = 0.05  (e,f), repectively.}
\end{figure*}

Figure 8(a $\sim$ f) displays  the intensity of the spin texture factor  for the  AF HL,  AF SL  and AF VL, respectively. The colored spectral band on the right side of each sub-figure shows the scaled intensity of the spin texture factor. In both Figure 8(a) and(b), the two   spectra look the same, those bright spots form square pattern out of the blue background. Though the skyrmions and vortices shown in Figure 3 and 4 respectively are of very different sizes, the   brightest points of $S_{\bot}(q)$ in Figure 8(d,f)  exhibit the same hexagonal pattern if other less strong strong points are neglected. Large differences  are detected   in  Figure 8(c) and (e), where  the   $S_{//}(q)$ spectra of the   AF SL and VL are displayed.  However, each of them is symmetric about its center, and has a quite large  hexagon  in  the central region of the reciprocal 2D space.

\section{Comparison  with SCA Method}
 From now on,   other simulated   results will be  displayed in the appendixes  for compactness.

To check our  results, we repeat  the simulations by means of the SCA method.\cite{liujpcm,liupssb,Liu14phye,LiuIan16,LiuIan17,LiuIan18, LiuIan19, LiuIan19-2}  Astonishingly, as $H$ = 7, the $xy$-projections of the AF SLs obtained at different temperatures  with the two methods, when plotted together, overlap with each other exactly, except
 for  $T$ = $T_{VL}$ = 1.55, where the system  starts to condense  to the VL state,  only at a few sites the spins align in slightly different directions.  Figure S1  shows an example  at a very low  temperature, $T$ = 0.05,  in comparison with  the results, that are shown in Figure 3 and calculated with   the same set of parameters using the OQMC method. No doubt, if the $A$, $B$ and $C$ FM sublattices obtained with the two quantum approaches are  plot together  correspondingly, every pair will overlap with each  other almost exactly.

For the same purpose,   the calculated total topological charges versus varying temperature as $H$ = 7, and  charge density textures of the $A$, $B$ and $C$ FM sublattices   of the AF  SL obtained at  $T$ = 0.05  with   the SCA approach,  are displayed in Figure S2.  When  compared with the four pictures displayed  in Figure 6,  that are obtained  with  the OQMC method using the same set of  parameters,  we can hardly  detect any disparity in each pair of the  corresponding  sub-figures.

\section{Conclusions and Discussion}

In this work, we have investigated spin textures of a 2D triangular  lattice   in the presence of  AF HE and chiral DM interactions with a quantum Monte Carlo method  optimized recently.  It is observed   that within an external magnetic field   applied normal to the 2D plane, well periodic and  symmetric AF helical, skyrmionic and vortical lattices can be generated at temperatures  higher  than those predicted  by previous authors;\cite{Rosales}
  the  AF SL states  prevail  in a broad $T-H$ region in the phase diagram; the sizes of skyrmions and vortices  change discontinuously at a few critical points of the applied magnetic field;  each of these AF periodical SLs and VLs can be decomposed into three FM sublattices; and the topological charge density of each  FM sublattices of an AF SL   also forms  crystal which coincides with that FM SL almost exactly; the  intensity of  spin structure factor of each  AF lattice varies in a very broad range in the 2D reciprocal space, where   the bright points   in  the  $q_x \sim  q_y$  plane  form  symmetric patterns, but most  points have very low  intensity, which are connected continuously to form  a blue background.

 In fact, we have performed simulations with different $D/|{\cal J}|$ ratios, but only the results for $D/|{\cal J}|$ = 1  are presented in this manuscript for compactness. In general, the larger the ratio, the smaller periodicity of the AF SL; and our calculated $T_{VL}$ and $T_{SL}$ are usually several times larger than those estimated by the CMC method. \cite{Rosales} For instance, when $D/{\cal J}$ = -0.5  and   $H$ = 2.4,   our calculated  $T_{SL} \approx 2.0$, that is about 6 time larger than the value obtained from the CMC simulations.\cite{Rosales}

   According to the Metropolis algorithm, in every computing step,   a random number $r_i $  must be  generated and compared with    $p_i =   e^{-\triangle E_i/k_BT}$, where    $\triangle E_i$ denotes the energy change caused by  rotating spin $\vec{S}_i$. Near the transition temperature,  $|\triangle E_i|/k_BT \ll 1$.
   To correctly determine the new state of $\vec{S}_i$,   $\triangle E_i$ must be accurately evaluated.  In the system we currently consider, the co-existence of chiral DM  and especially AF HE interactions make it very difficult to calculate  $\triangle E_i$ accurately   if a classical method is   employed, since antiferromagnetic systems are intrinsically 'quantum'. On the other hand, in CMC simulations, thousands of spin configurations  after 'convergence' must be taken to calculate the averaged values of the spin vectors  to determine the spin textures at a given temperature.  { Thus, if   these spin configurations are not accurately determined, or because of the AF HE and DM interactions, the  orientations of any spin  can differ greatly or may be nearly opposite  in different iterations,  so the regular spin textures may be  smeared out by averaging over those non-well determined spin configurations  and finally become featureless, so that  the  transition  temperature  may be considerably  underestimated}. Or even worse, the AF chiral spin texture existing  at very low temperatures can not be detected with the CMC method.


\vspace{0.4cm}
 \centerline{\bf Acknowledgements}
\vspace{0.2cm}
Z.-S. Liu  acknowledges  Professor Stefan  Bl\"ugel for helpful discussion,  the financial support provided by National Natural Science Foundation of China  under grant No.~11274177, and by Peter Gr\"unberg Institut and Institute for Advanced Simulation, Forschungszentrum J\"ulich, Germany during the visit in last Summer.

\appendix

\section{Comparison of results: OQMC versus  SCA methods}

To assess the  results obtained with the OQMC mehtod, we have repeated the simulations by means of another quantum approach, namely the SCA  method since the self-consistent algorithm is employed there. \cite{liujpcm,liupssb,Liu14phye,LiuIan16,LiuIan17,LiuIan18, LiuIan19-2} The results obtained with ${\cal J}$ = -1, $D$ = 1, and $H$ = 7 are displayed in Figure S1 and S2 in comparison with those  depicted in Figure 3 and 6, correspondingly.

\begin{figure*}[htb]
\centerline{
 \includegraphics[width=0.45\textwidth,height=6cm,clip=]
 {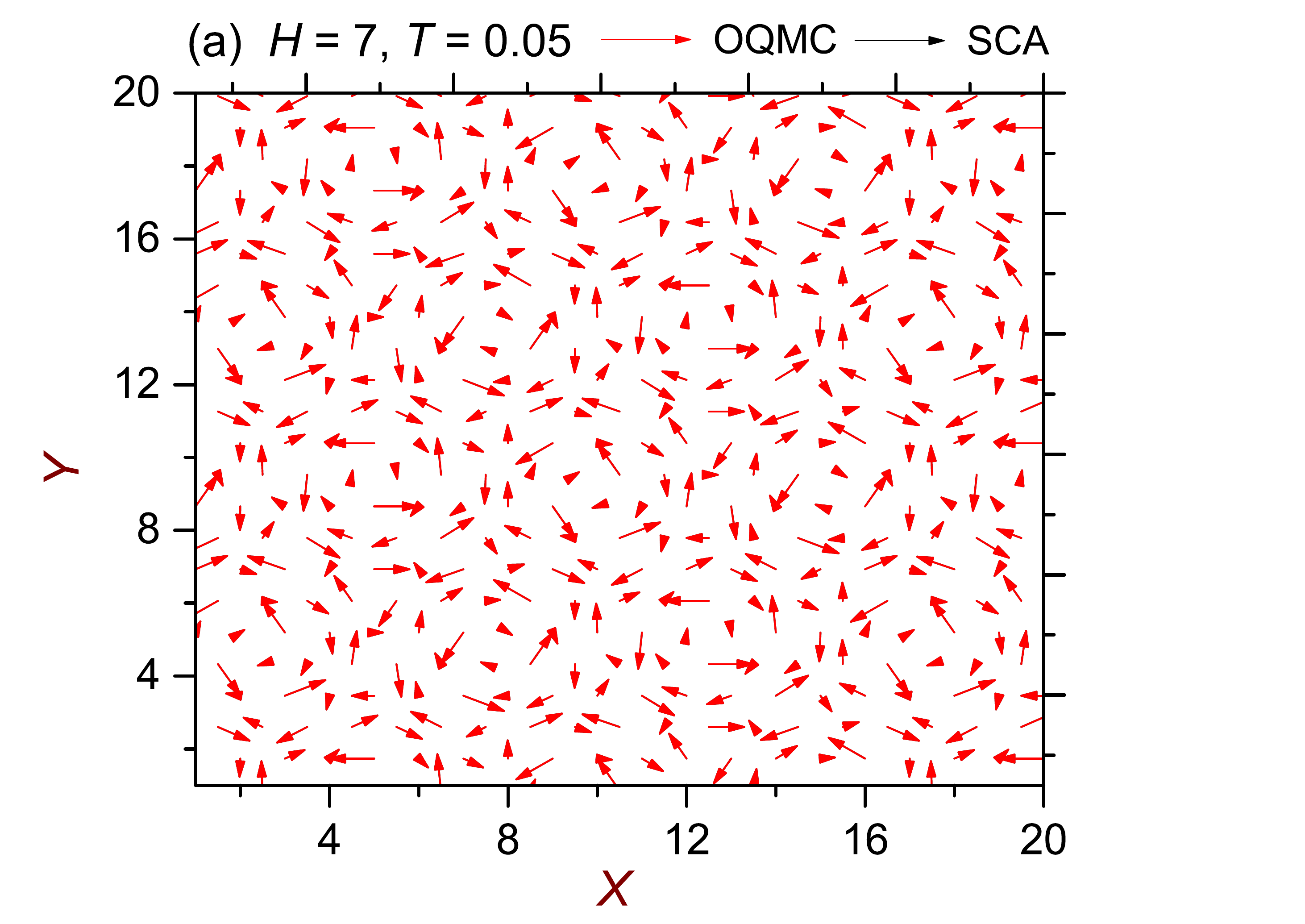}
 \includegraphics[width=0.45\textwidth,height=6cm,clip=]
 {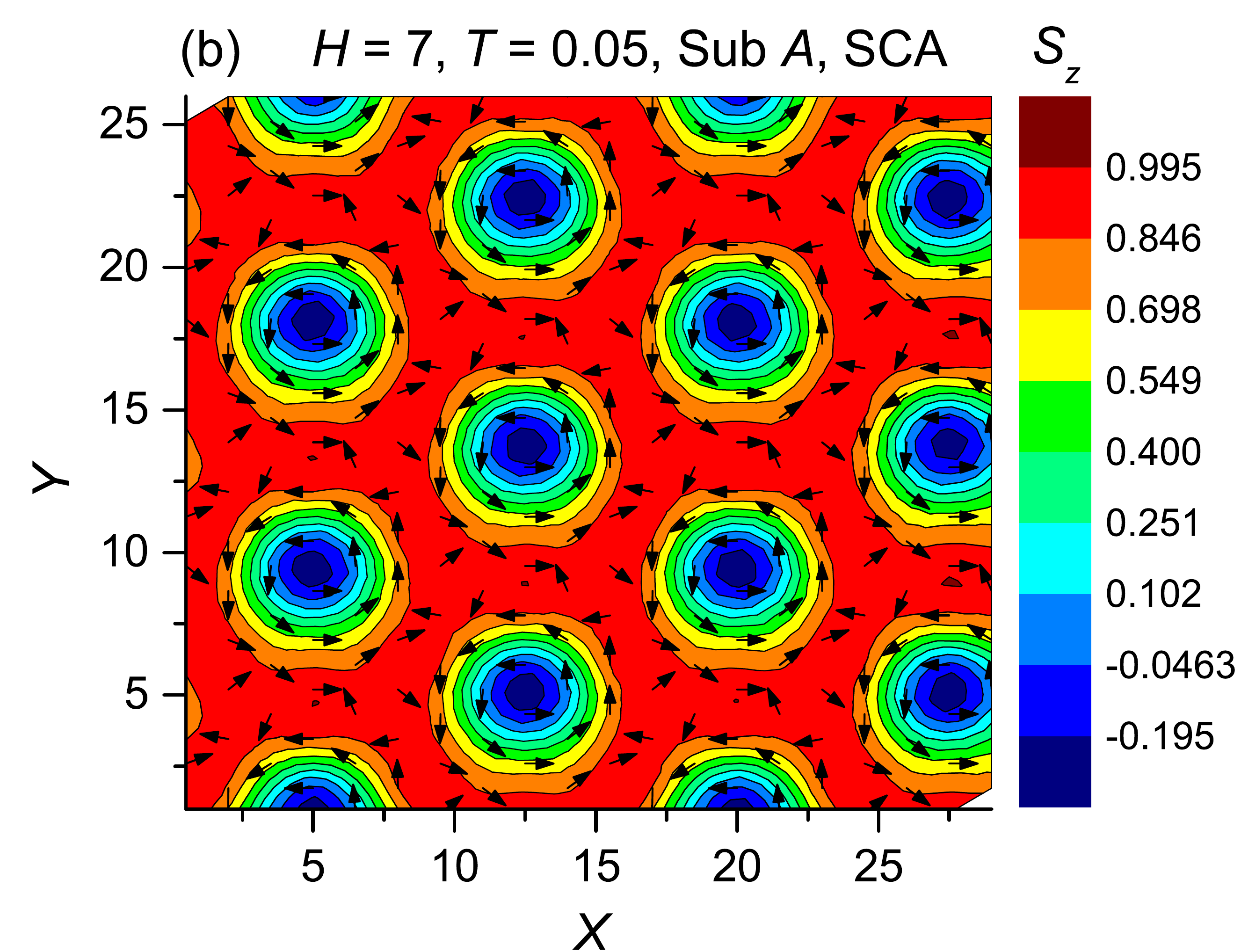}
 }
 \centerline{
 \includegraphics[width=0.45\textwidth,height=6cm,clip=]
 {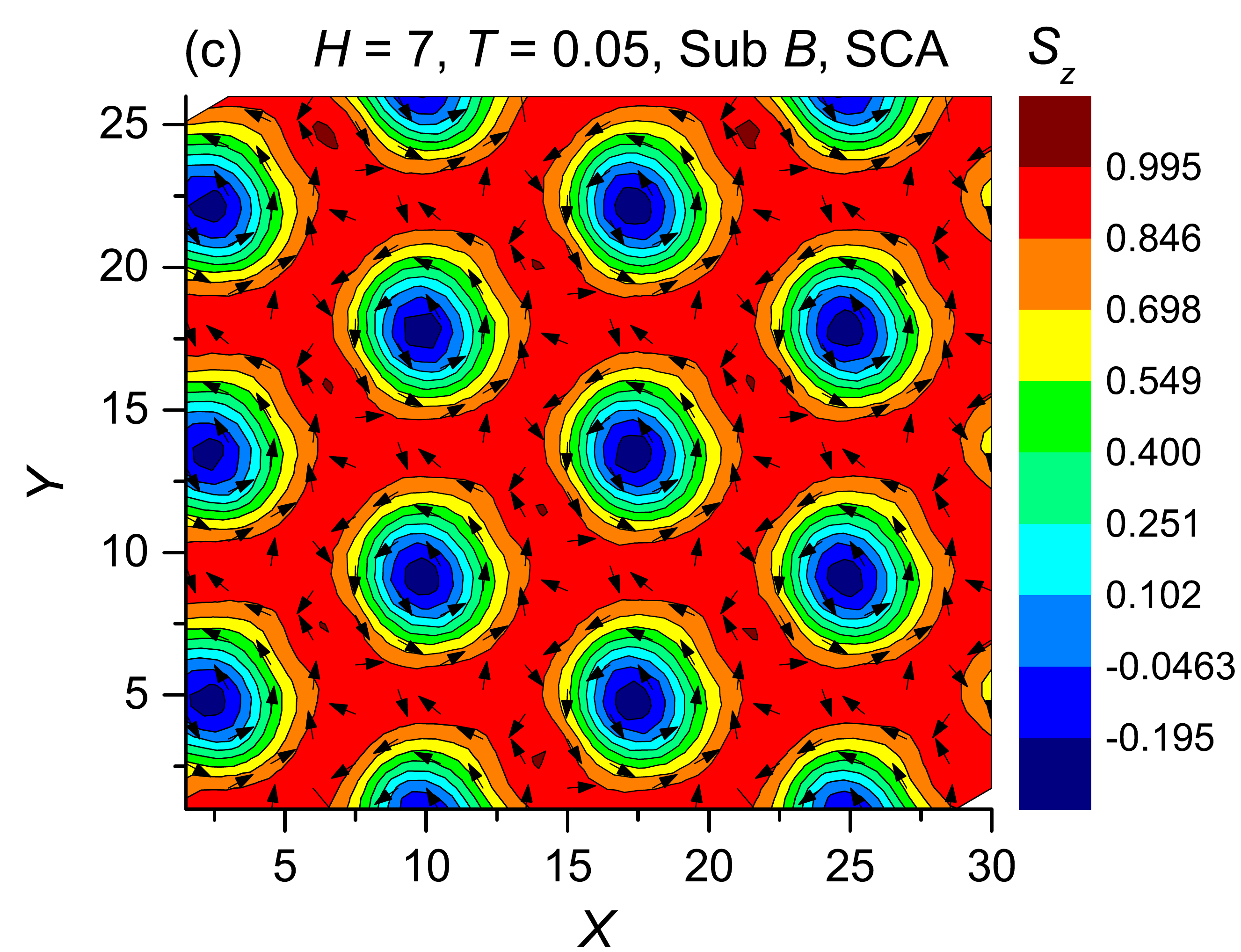}
 \includegraphics[width=0.45\textwidth,height=6cm,clip=]
 {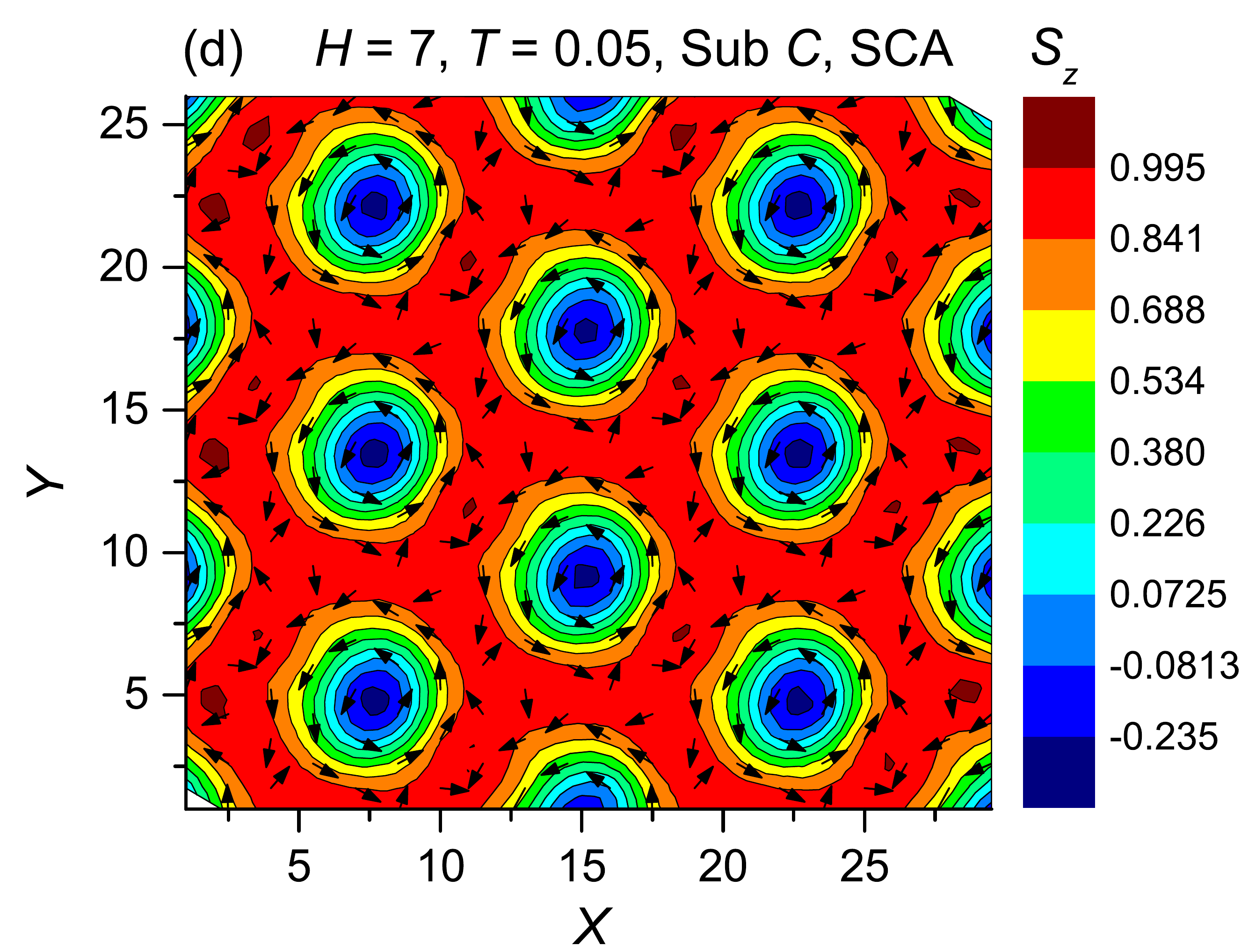}
}
\caption{Calculated   (a)    $xy$-projections,  and (b,c,d)  three FM sublattices for the AF SLs  at $H$ = 7, $T$ = 0.05 with the SCA methods.}
\end{figure*}

\begin{figure*}[htb]
\centerline{
 \includegraphics[width=0.45\textwidth,height=6cm,clip=]
 {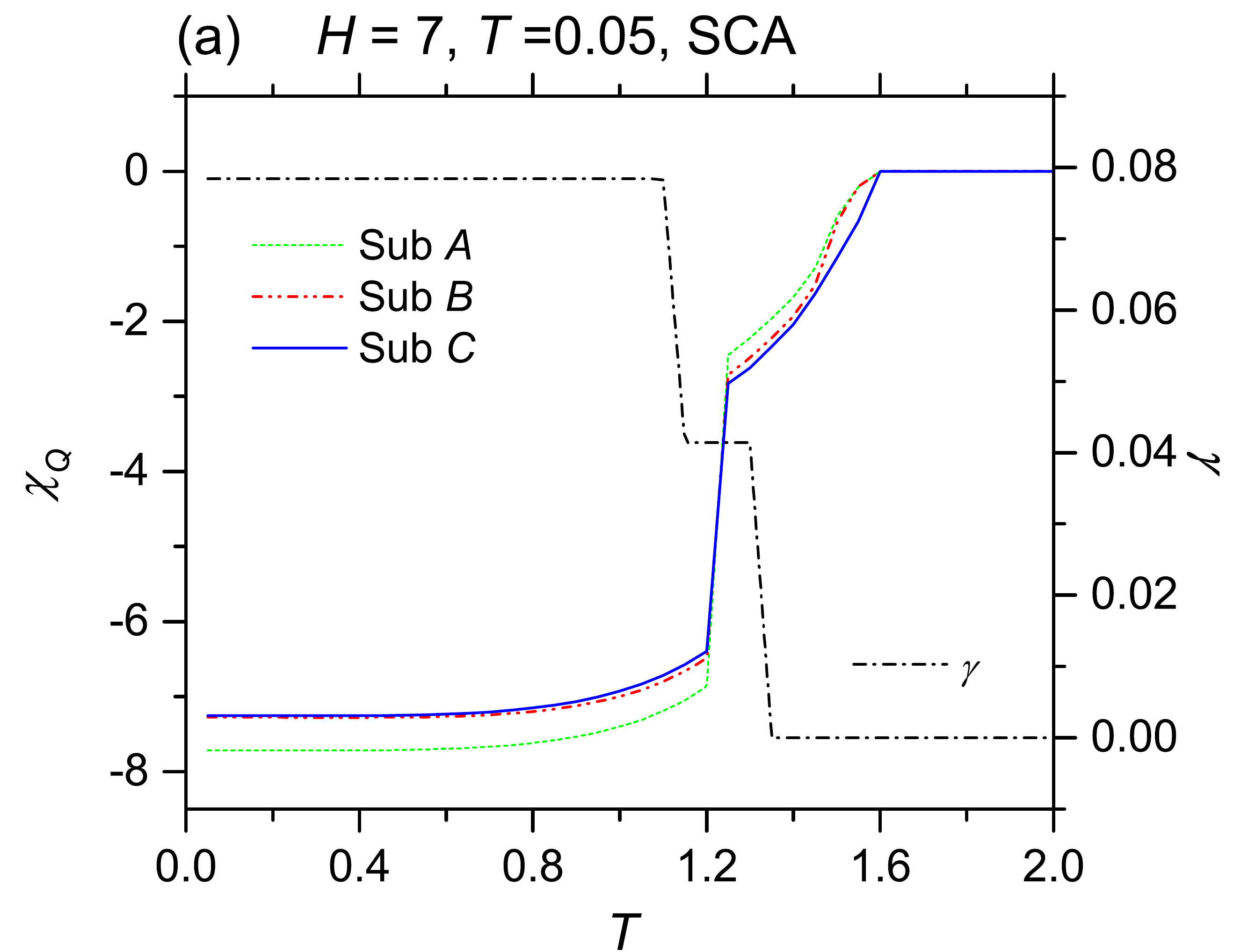}
 \includegraphics[width=0.45\textwidth,height=6cm,clip=]
 {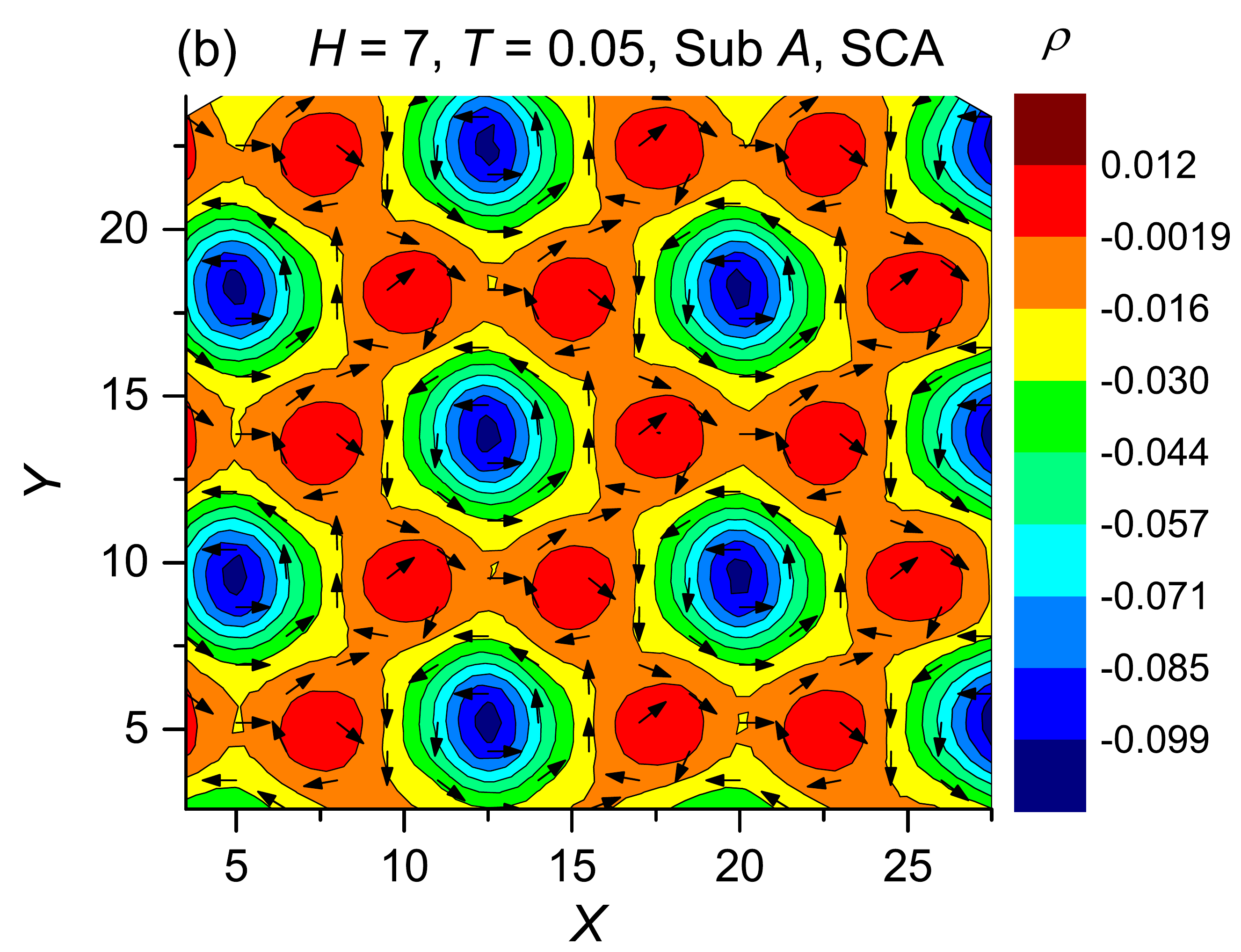}
 }
 \centerline{
 \includegraphics[width=0.45\textwidth,height=6cm,clip=]
 {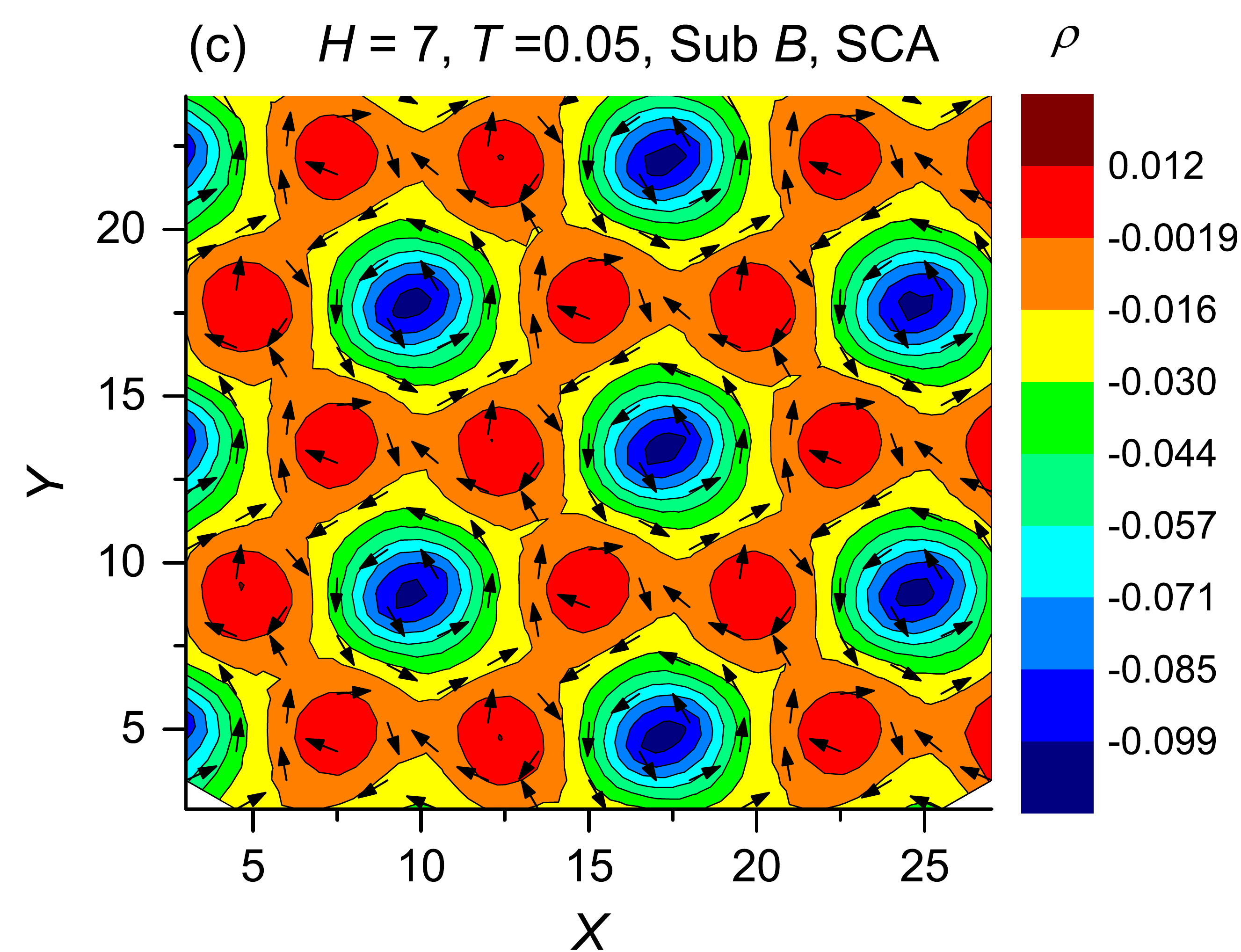}
 \includegraphics[width=0.45\textwidth,height=6cm,clip=]
 {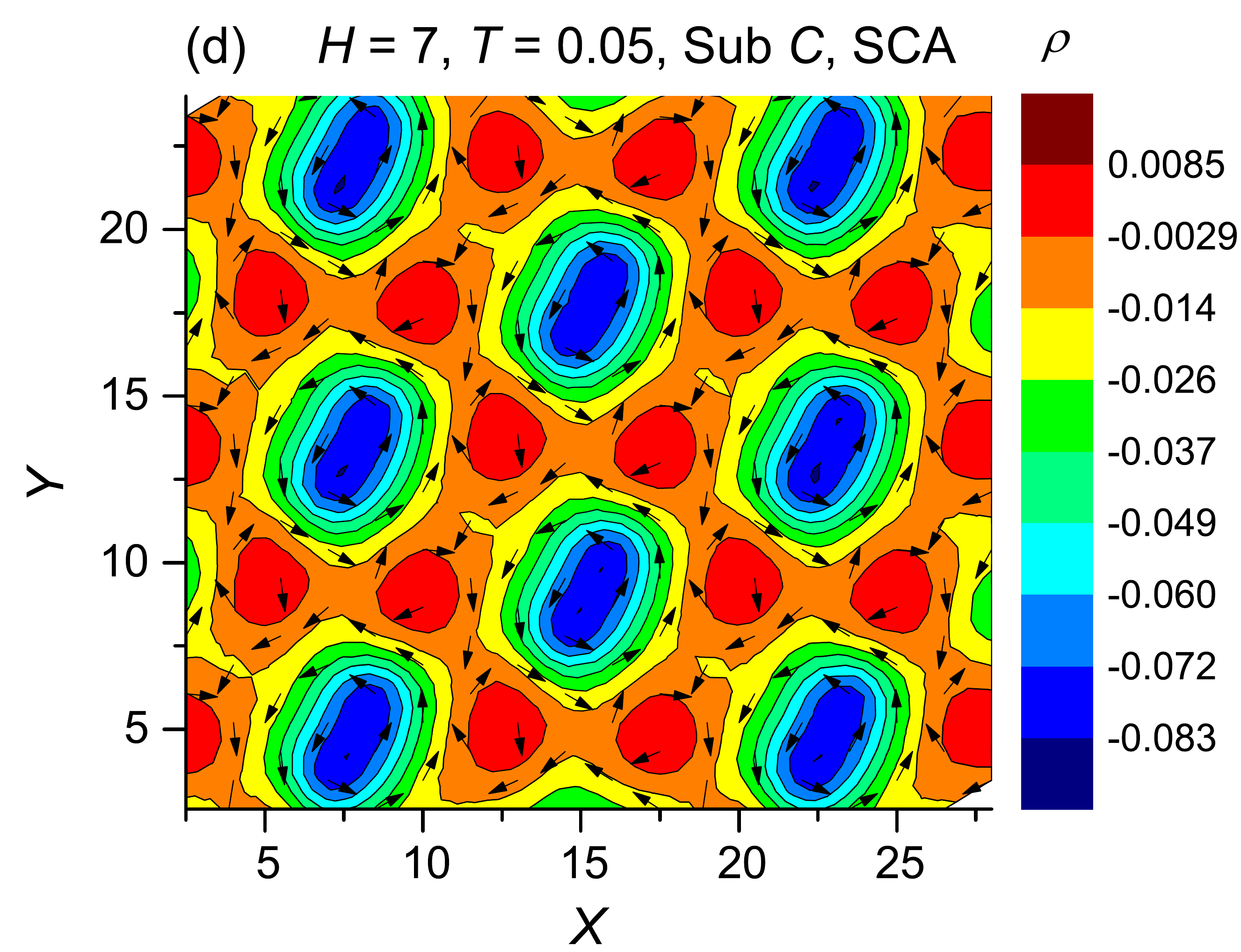}
}
\caption{The total topological charges (a) , and  charge density distributions of the three FM sublattices (b,c,d)   of the AF  SL at $H$ = 7, $T$ = 0.05 calculated with   the SCA  method.}
\end{figure*}
The   $xy$-projections of the AF SLs obtained at different temperatures  using the two methods, when plotted together, overlap with each other almost exactly, as shown in Figure S1(a). Just at $T$ = $T_{VL}$ (= 1.55), where the system starts to condenses to AF VL state, only  a few spins align in  different directions. If the  $A$, $B$ and $C$ FM sublattices obtained with the two quantum approaches are  plotted together, every   pair of them overlaps with each other almost exactly as well.

In  Figure S2 and Figure 6,   the curves of the total topological charges versus changing temperature,   the charge density distributions of the $A$, $B$ and $C$ FM sublattices of the AF  SL  at $T$ = 0.05, that are calculated with the two  methods,  show no   disparity.

\section{ N\'eel-Type  AF SL Simulated with OQMC Method }

\vspace{0.3cm}
By assigning ${\cal J}$ = -1, $D$ = 1, and $H$ = 7 which is perpendicular to the 2D antiferromagnet, OQMC  simulations are started from a paramagnetic state, then carried out down to a very low  temperature, $T$ = 0.05. Now, $\vec{D}$ is assumed to be in-plane and normal to $\vec{r}_{ij}$, so the induced AF SL shown in Figure S3 and S4 are of N\'eel-type.

\begin{figure*}[htb]
\centerline{
 \includegraphics[width=0.45\textwidth,height=6cm,clip=]
 {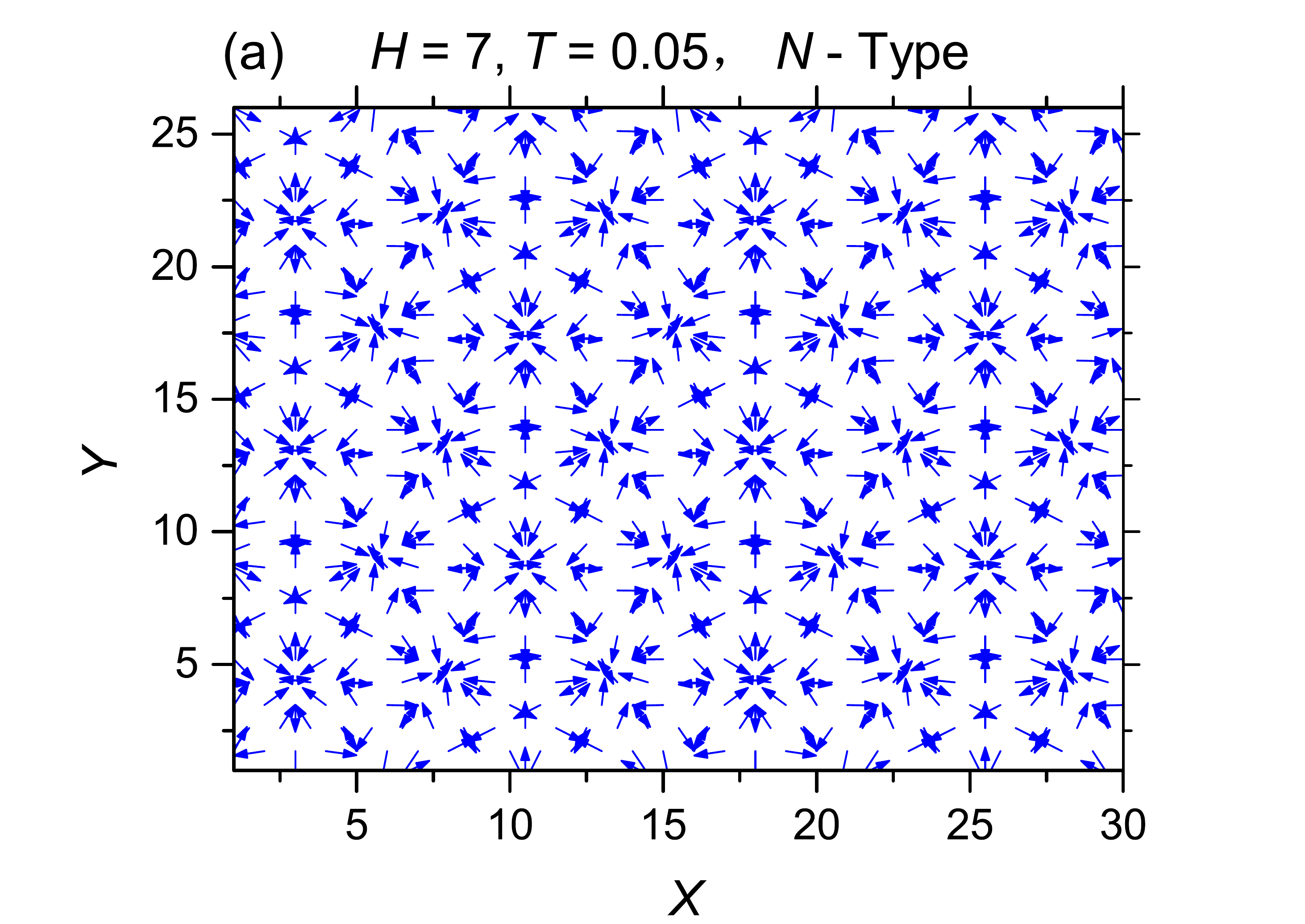}
 \includegraphics[width=0.45\textwidth,height=6cm,clip=]
 {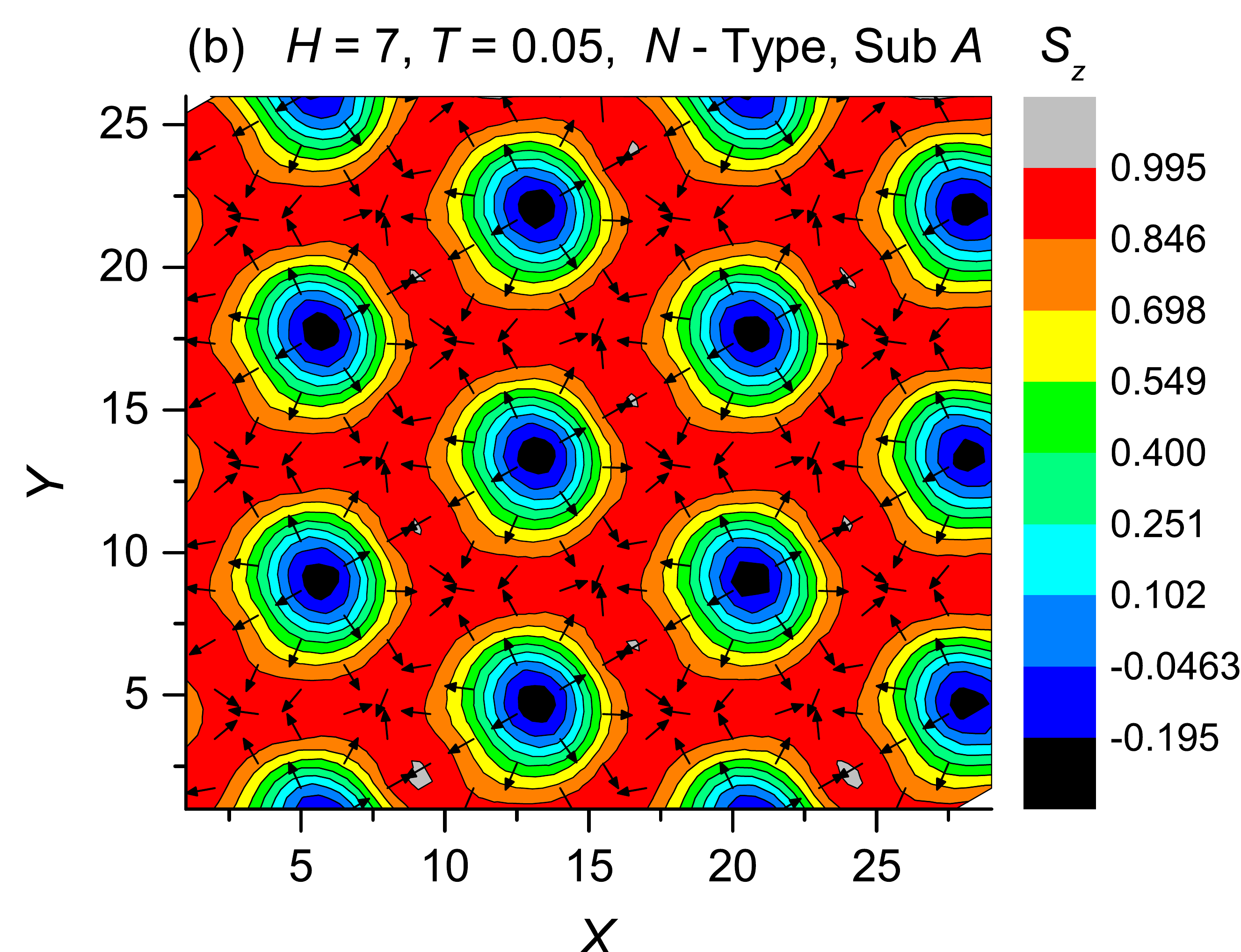}
 }
 \centerline{
 \includegraphics[width=0.45\textwidth,height=6cm,clip=]
 {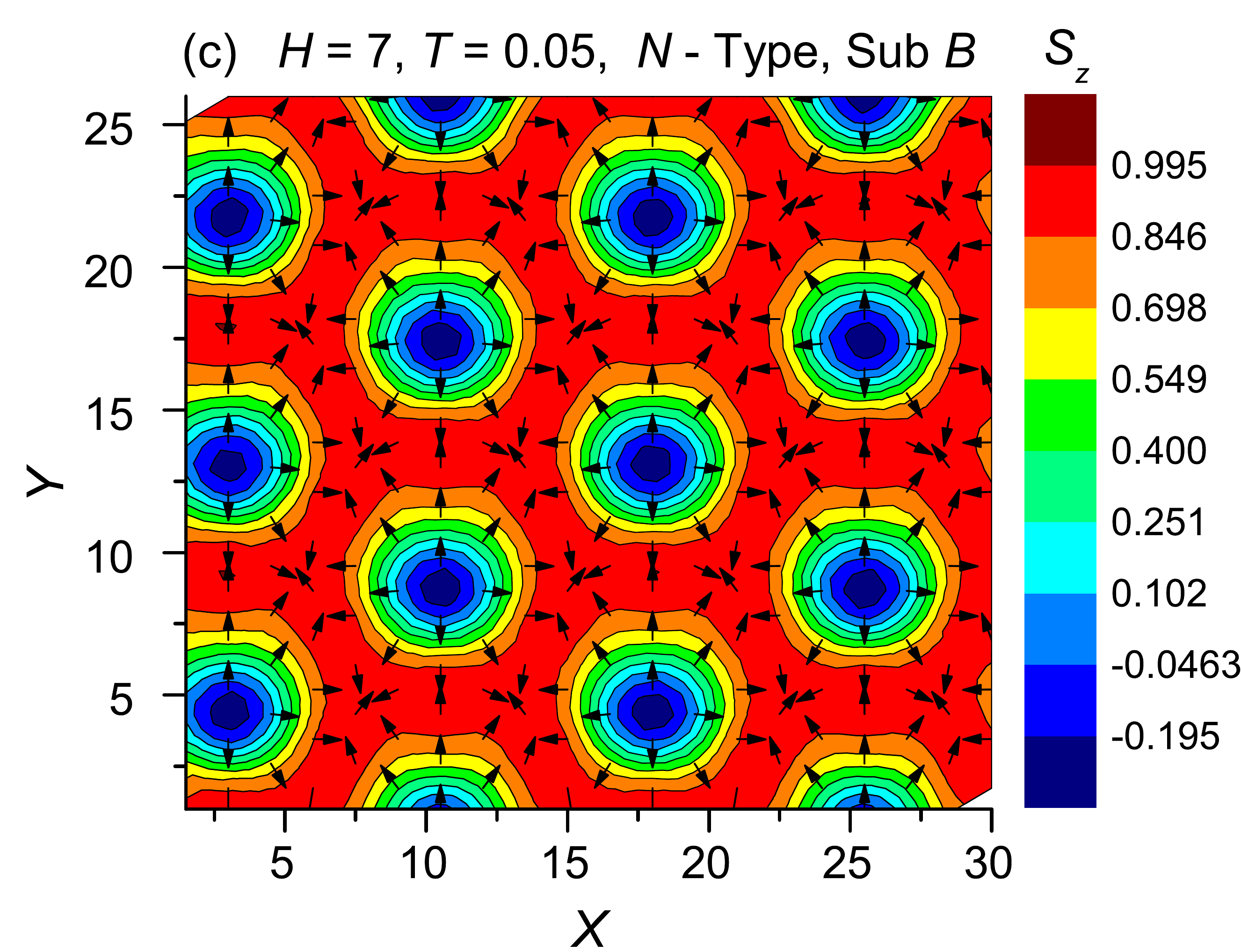}
 \includegraphics[width=0.45\textwidth,height=6cm,clip=]
 {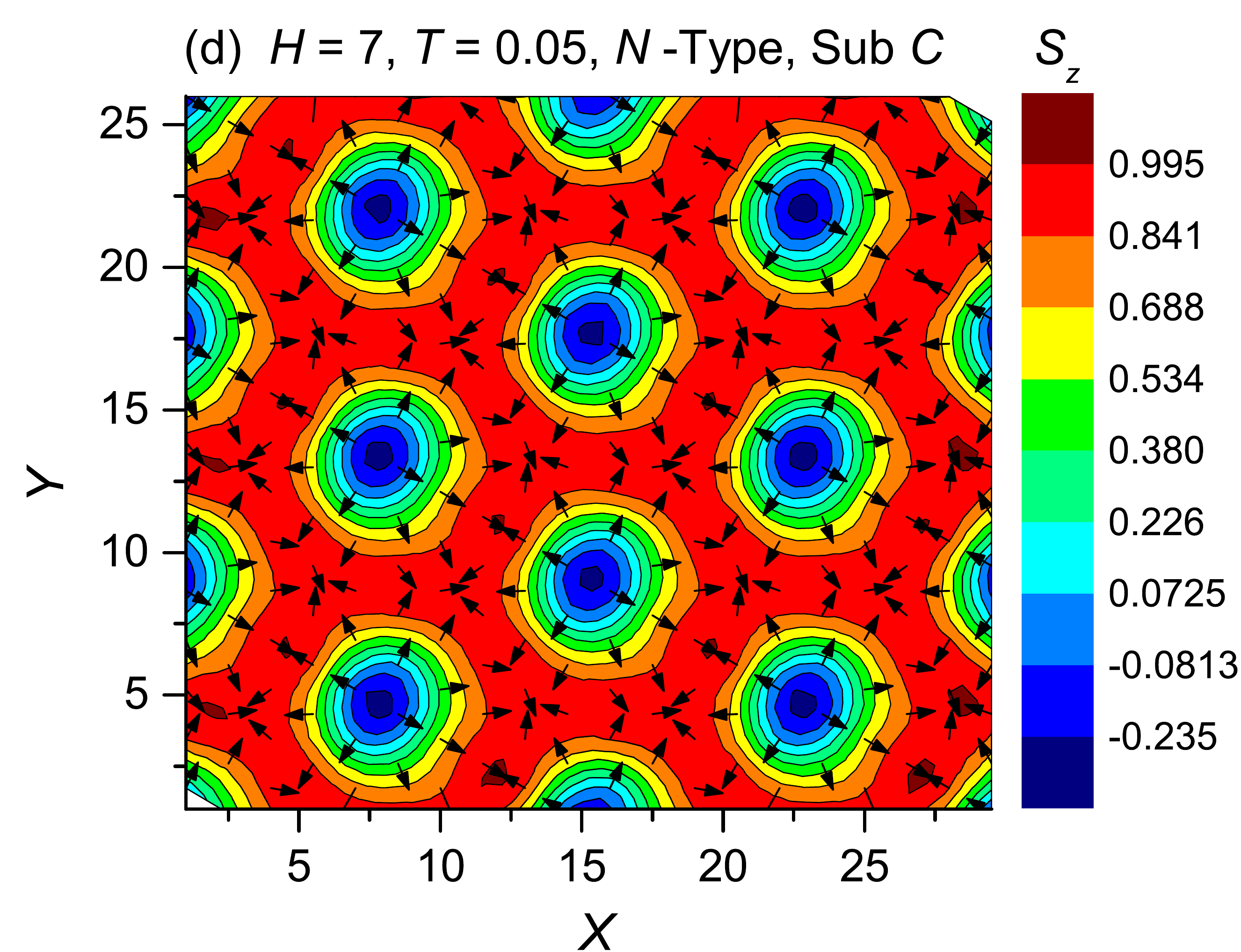}
}
\caption{The   $xy$-projection    (a),  and   three FM sublattices (b,c,d),  of the N\'eel-type AF SkL calculated  with the OQMC method when $H$ = 7, $T$ = 0.05.}
\end{figure*}

\begin{figure*}[htb]
\centerline{
 \includegraphics[width=0.45\textwidth,height=6cm,clip=]
 {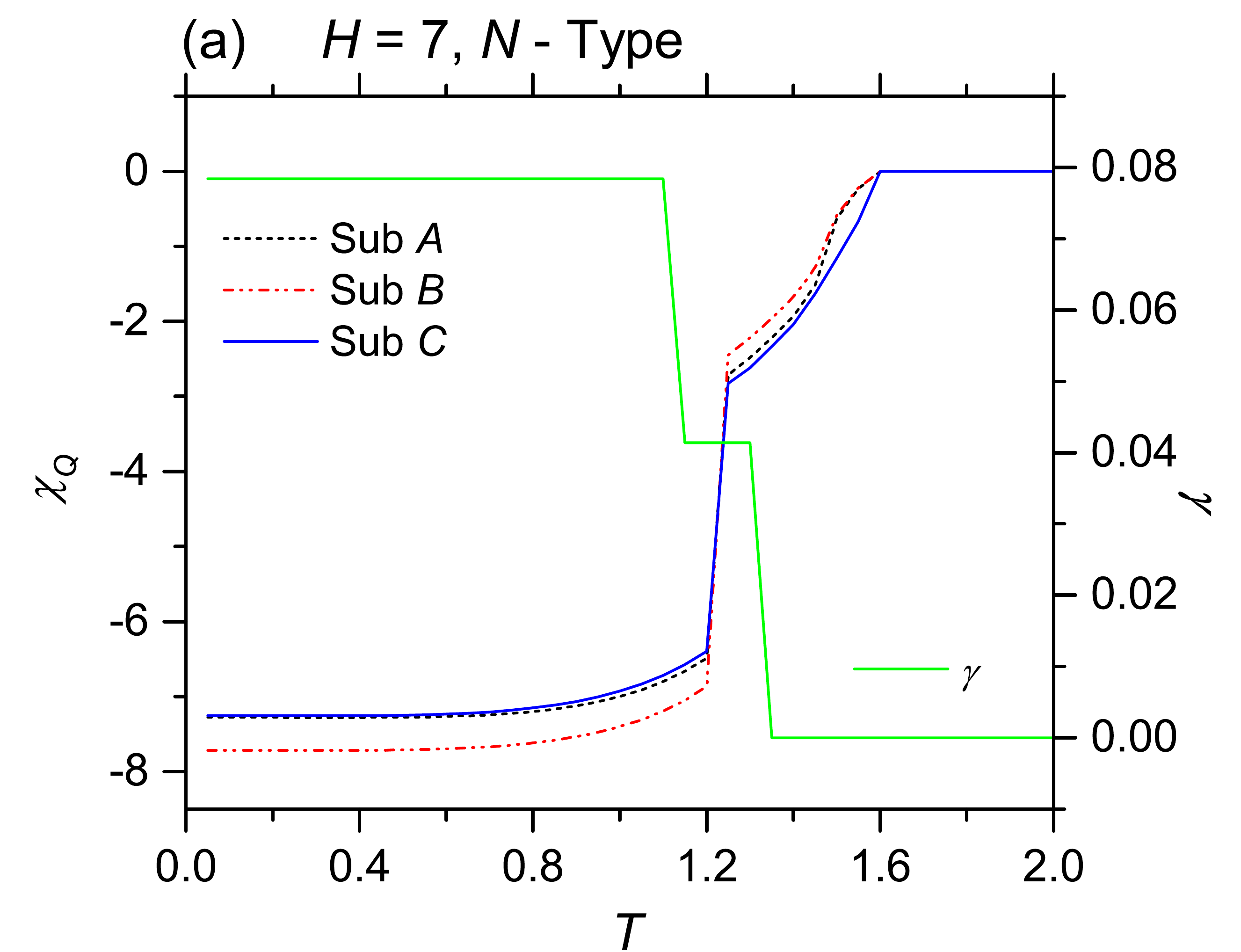}
 \includegraphics[width=0.45\textwidth,height=6cm,clip=]
 {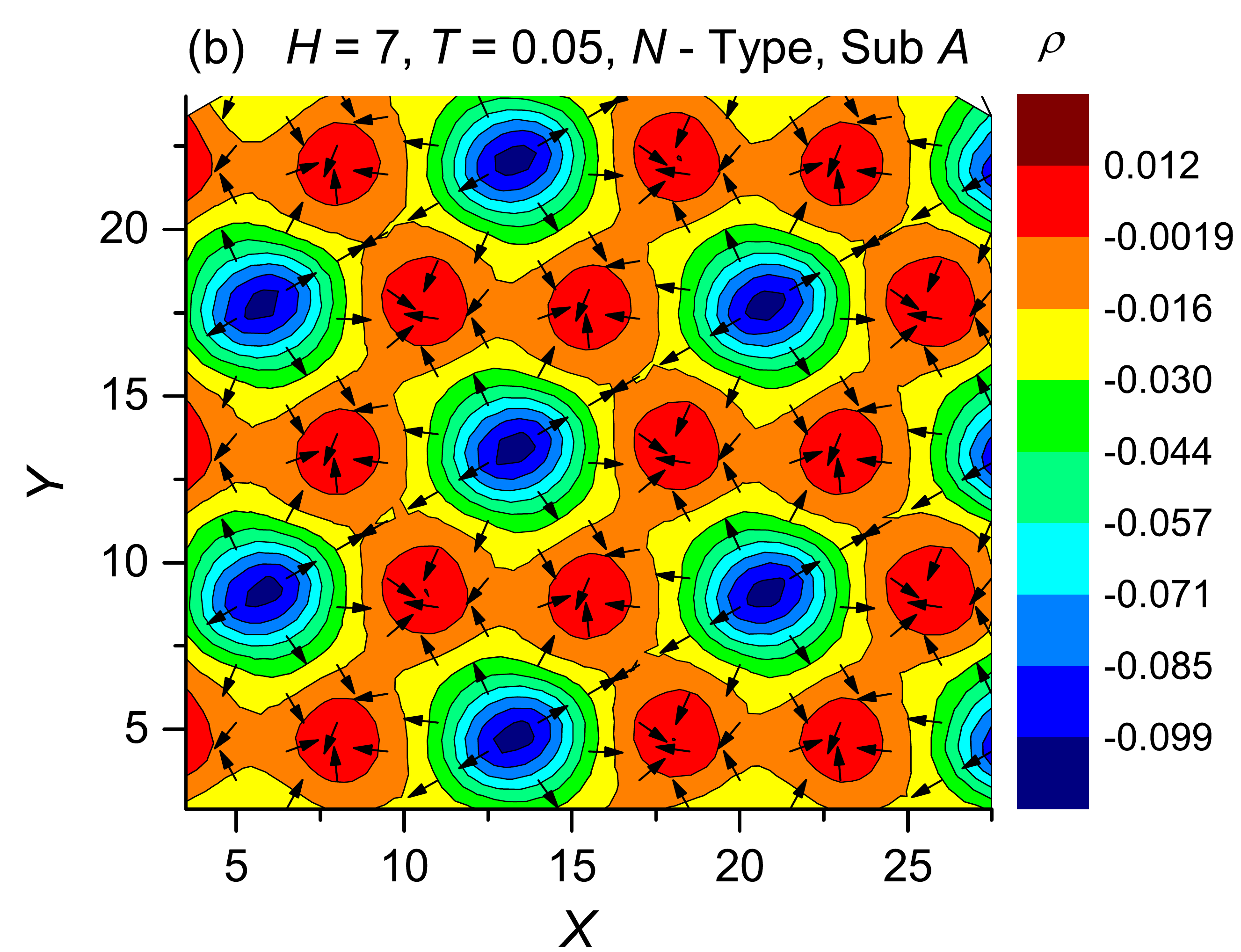}
 }
 \centerline{
 \includegraphics[width=0.45\textwidth,height=6cm,clip=]
 {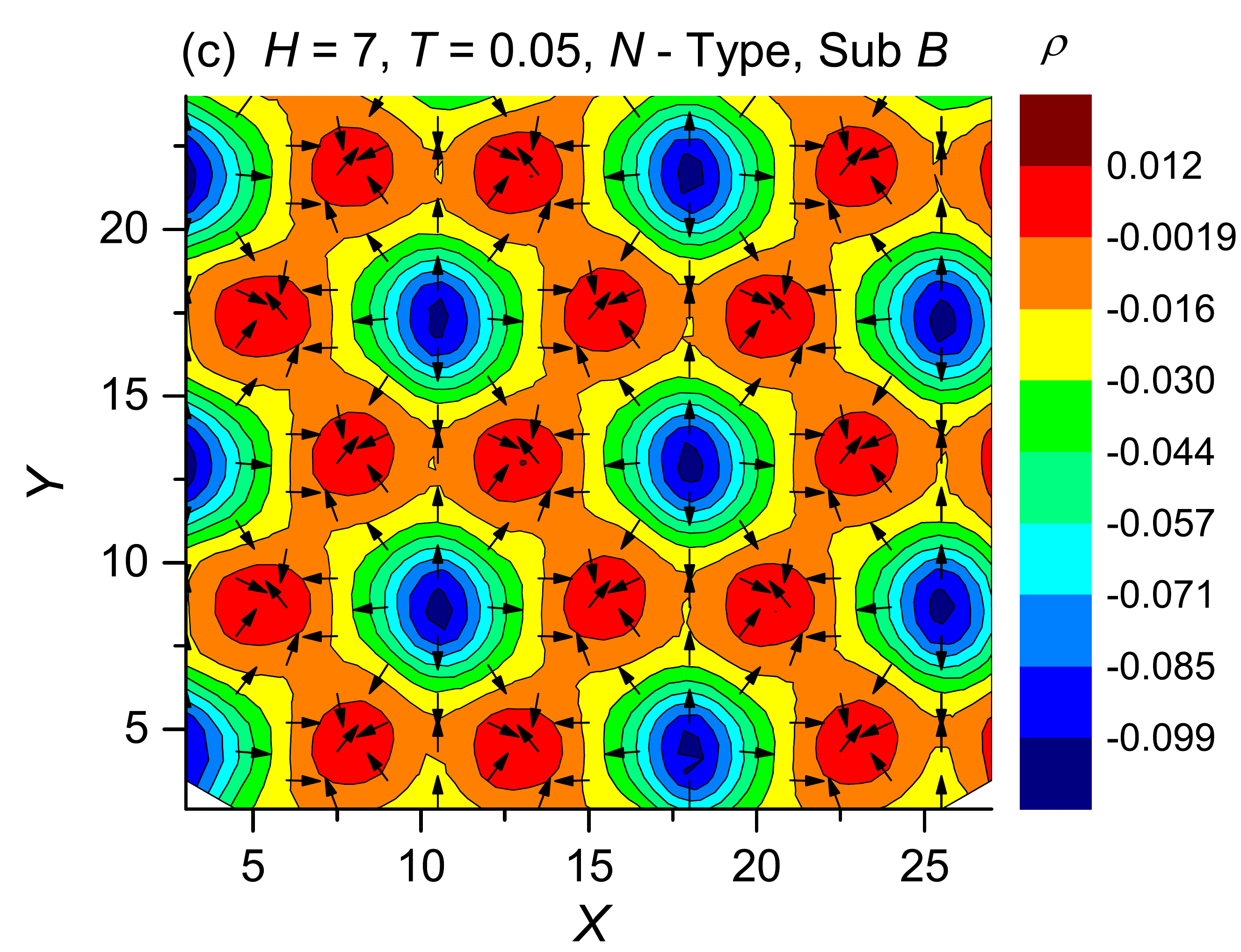}
 \includegraphics[width=0.45\textwidth,height=6cm,clip=]
 {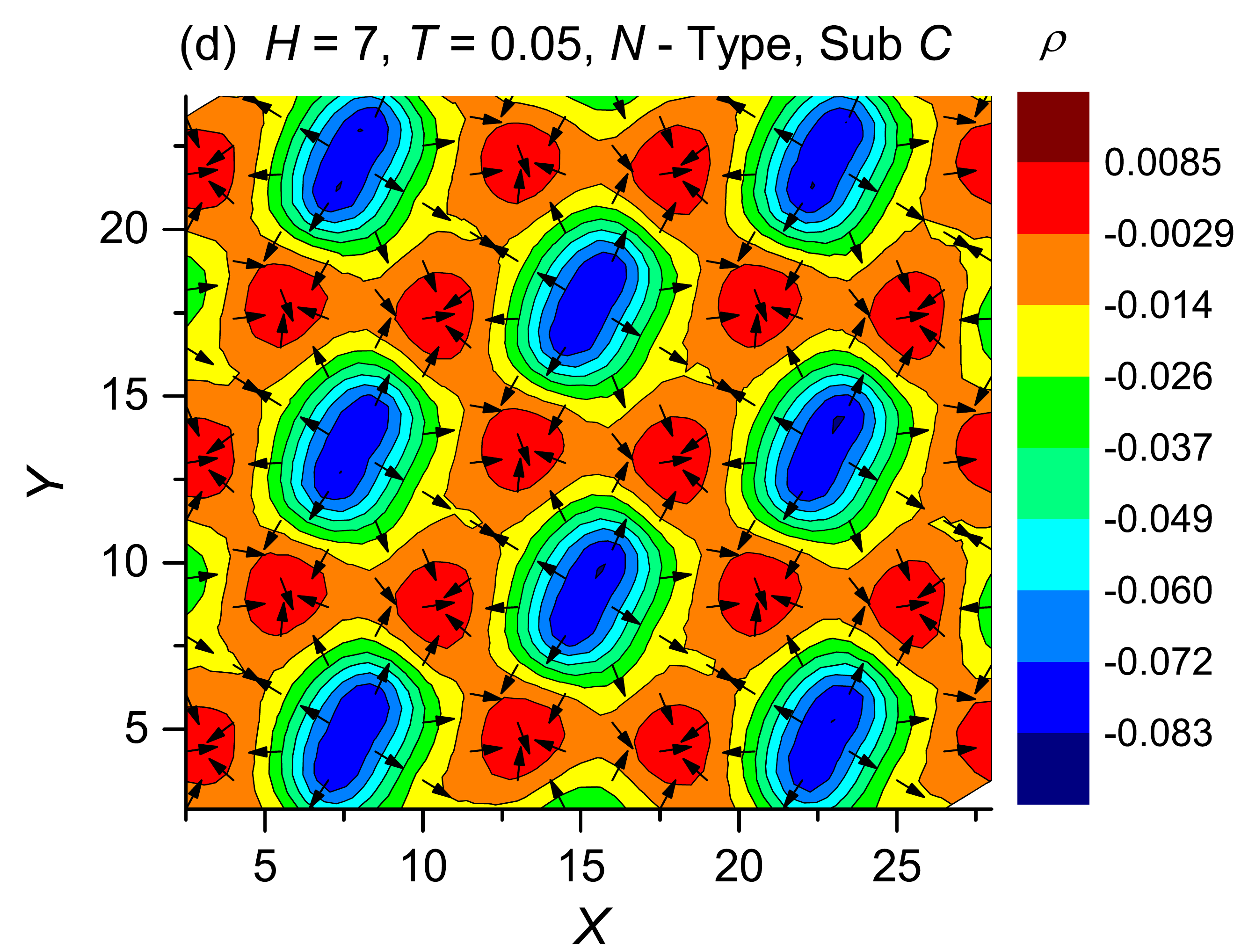}
}
\caption{Total topological charges (a), and  charge density distributions of the three FM sublattices (b,c,d) ,  of the N\'eel-type AFM SkL calculated at $H$ = 7, $T$ = 0.05 using the OQMC method.}
\end{figure*}

As shown in these two figures, this AF SL can also be decomposed to three FM SL, each of them  contains 12 FM skyrmions in the 30 $\times$ 30 lattice, and the topological charge density of each FM sublattice forms a periodical crystal of the same pattern as the spin sublattice.

In comparison with the results shown in Figure 3 and 6, we can see that the two types of  AF SLs obtained with the  same set of parameters exhibit very similar characteristics. So the  AF SLs of the Bloch and N\'eel types form a dual pair.

 \end{document}